\documentclass[english,12pt,a4paper]{article}
\usepackage{authblk}
\usepackage[text={17.0cm,23.7cm}]{geometry}
\usepackage[T1]{fontenc}
\usepackage[utf8]{inputenc}
\usepackage{babel}
\usepackage{hyperref}
\usepackage{cite}
\usepackage{graphicx}
\usepackage{amsmath}
\usepackage{amssymb}
\usepackage[dvipsnames]{xcolor}
\usepackage{yfonts}
\usepackage{ulem}
\usepackage{enumitem}

\newcommand\myshade{80}
\colorlet{mylinkcolor}{ForestGreen}
\colorlet{mycitecolor}{Red}
\colorlet{myurlcolor}{violet}

\hypersetup{
  linkcolor  = mylinkcolor!\myshade!black,
  citecolor  = mycitecolor!\myshade!black,
  urlcolor   = myurlcolor!\myshade!black,
  colorlinks = true
}

\DeclareMathOperator{\artanh}{artanh}

\newcommand{\mdm}{\text{m}_\text{DM}}

\newcommand{\ER}{E_R}
\newcommand{\dER}{\text{d}E_R}
\newcommand{\mNi}{m_{N_i}}

\newcommand{\CSHM}{\text{C}_\text{SHM}}
\newcommand{\Csh}{\text{C}_\text{sh}}

\newcommand{\Nsh}{N_\text{sh}}

\begin{document}
	\title{\vspace{-2cm}
		{\normalsize
			\flushright TUM-HEP 1217/19,\,KIAS-P19046\\}
		\vspace{0.6cm}
		\textbf{Impact of substructure on local dark matter searches}\\[8mm]}

	\author[1,2]{Alejandro Ibarra}
	\author[3]{Bradley J. Kavanagh}
	\author[1]{Andreas Rappelt}
	\affil[1]{\normalsize\textit{Physik-Department T30d, Technische Universit\"at M\"unchen, James-Franck-Stra\ss{}e, 85748 Garching, Germany}}
	\affil[2]{\normalsize\textit{School of Physics, Korea Institute for Advanced Study, Seoul 02455, South Korea}}
	\affil[3]{\normalsize\textit{GRAPPA, Institute of Physics, University of Amsterdam, 1098 XH Amsterdam, The Netherlands}}

	\date{}

\maketitle
\begin{abstract}
	Dark matter substructure can contribute significantly to local dark matter searches and may provide a large uncertainty in the interpretation of those experiments.
	For direct detection experiments, sub-halos give rise to an additional dark matter component on top of the smooth dark matter distribution of the host halo.
	In the case of dark matter capture in the Sun, sub-halo encounters temporarily increase the number of captured particles.
	Even if the encounter happened in the past, the number of dark matter particles captured by the Sun can still be enhanced today compared to expectations from the host halo as those enhancements decay over time.
	Using results from an analytical model of the sub-halo population of a Milky Way-like galaxy, valid for sub-halo masses between $10^{-5}\,M_\odot$ and $10^{11}\,M_\odot$, we assess the impact of sub-halos on direct dark matter searches in a probabilistic way.
	We find that the impact on direct detection can be sizable, with a probability of $\sim 10^{-3}$ to find an $\mathcal{O}(1)$ enhancement of the recoil rate.
	In the case of the capture rate in the Sun, we find that $\mathcal{O}(1)$ enhancements are very unlikely, with probability $\lesssim 10^{-5}$, and are even impossible for some dark matter masses.
\end{abstract}

\section{Introduction}

One of the central assumptions  of modern cosmology is the presence of a non-luminous matter component, commonly known as dark matter (DM) \cite{Bertone:2004pz,Bertone:2018xtm}.
Although the exact nature of dark matter still remains unknown, many proposed models predict a small interaction with Standard Model particles, as in the case of weakly interacting massive particles (WIMPs) \cite{Roszkowski:2017nbc} appearing (for example) in supersymmetric theories \cite{Jungman:1995df}.
This opens up the possibility to search for DM via various possible signatures.
Currently pursued strategies include, among others, collider searches for dark matter production \cite{Kahlhoefer:2017dnp}, indirect searches for the products of dark matter annihilation \cite{Gaskins:2016cha} as well as direct searches for the signatures of dark matter scattering off nuclei and electrons in the lab \cite{Schumann:2019eaa}.
Furthermore, dark matter plays an important role on cosmological scales, driving the formation of structure in the Universe \cite{Primack:1997av}.

In the framework of cold and collisionless dark matter, the growth of structure is hierarchical, with small structures forming first and then later merging into larger structures \cite{Diemand:2009bm,Frenk:2012ph}.
As a consequence of this, we expect that the dark matter halos enclosing galaxies should contain a large number of sub-halos.
Sub-halos of the Milky Way include the known dwarf satellite galaxies, as well as the Small and Large Magellanic Clouds, ranging in mass from roughly $10^7\,M_\odot$ up to $10^{10}\,M_\odot$ \cite{Strigari:2008ib,2015arXiv151103346B}.
However, these massive objects represent only a small fraction of the sub-halos present in the Milky Way.
Many more sub-halos are too small to accrete enough material to form stars and they thus remain dark \cite{Thoul:1996by,Bullock:2000wn,Somerville:2001km,Nadler:2018iux}. These dark sub-halos are expected to exist with masses all the way down to $10^{-6}\, M_\odot$, roughly the mass of the Earth \cite{Zybin:1999ic,Hofmann:2001bi,Berezinsky:2007qu}.
In light of this, a number of studies have explored the properties and distribution of DM sub-halos within Milky Way-like galaxies, see e.g.~Refs.~\cite{Johnston:2001wh,Diemand:2006ey,Springel:2008cc,BoylanKolchin:2009an,Giocoli:2009ie,Gao:2010tn,Kamionkowski:2010mi,Ishiyama:2011af,Sanchez-Conde:2013yxa,Stref:2016uzb,Okoli:2017pqr,Hiroshima:2018kfv} for a non-exhaustive list.
Indeed, a recent study of stellar kinematics using data from the Gaia satellite \cite{GaiaCollaboration2018} suggests that a substantial fraction of the local DM halo may be in substructure \cite{Necib:2018iwb}.
On ultra-local scales, the DM density could be enhanced by DM streams, such as Sagittarius \cite{Freese:2003na} or S1 \cite{OHare2018}.
There has also been recent interest in the possibility that DM forms clusters \cite{Nussinov:2018ofp} or `blobs' \cite{Grabowska:2018lnd}, local overdensities perhaps generated by DM self-interactions.

Dark matter substructure can lead to a number of detectable signatures. 
One possibility is to directly look for the signal of DM annihilation inside sub-halos.
Various studies \cite{Diemand:2006ik, Strigari:2006rd, Pieri:2007ir, Jeltema:2008ax, Ishiyama:2014uoa, Bartels:2015uba} have investigated how this enhances the gamma-ray signal from the host halo and found boost factors up to $\mathcal{O}(10)$. 
Another possibility is to look for evidence of close passages between DM sub-halos and cold stellar streams of the Milky Way \cite{Erkal:2015kqa,Bovy:2016irg,Banik:2018pjp,Banik:2018pal,Bonaca:2018fek}. 
These close passages would be imprinted as gaps and other features in the stellar streams, mapped out by precision astrometric surveys such as Gaia \cite{GaiaCollaboration2018}. 
Furthermore, sub-structures might influence local DM searches by increasing the number of dark matter particles in a finite region, relative to the smooth background.

In this work, we focus on such `local' DM searches: Direct Detection, the search for DM-nucleus scattering in Earth-based detectors; and Solar Capture, the search for neutrino signals from DM which has scattered and been captured in the Sun \cite{Danninger:2014xza}. 
The signal enhancement due to such clustering of dark matter has been studied for direct detection experiments in several works \cite{Stiff:2001dq,Helmi:2002ss,Green:2003ud,Vogelsberger:2007ny,Kamionkowski:2008vw}. 
For example, it has been argued that no individual sub-halo should dominate the scattering rate \cite{Helmi:2002ss} and that the contribution of streams to the local distribution should add in such a way as to be indistinguishable from a smooth halo \cite{Vogelsberger:2007ny}.
However, the impact of sub-structure on direct detection may still be sizeable; for example Refs.~\cite{Stiff:2001dq, Kamionkowski:2008vw} find a percent-level probability of the Earth currently passing through a sub-halo. 
The impact of sub-structure on Solar capture has been less extensively studied; for example, Ref.~\cite{Koushiappas:2009ee} studied a small number of benchmark scenarios for possible DM overdensities.  
For capture in the Sun, we note that sub-halo encounters during the whole Solar lifetime can be important.
Each encounter leads to an enhancement in the number of captured particles, which may persist even after the Sun has traversed the sub-halo, depending on the efficiency of the annihilation process.
It is even possible for these enhancements to accumulate if they persist long enough, highlighting the need for a careful treatment with realistic sub-halo populations.

Here, we revisit the impact of DM sub-structure on local searches.  We describe the potential signal enhancements which could be observed due to individual DM sub-halos in both direct detection and Solar capture. Furthermore, we use state-of-the-art semi-analytic models for Milky Way substructure to derive the probability distributions for these enhancements for different dark matter masses and interaction types. In this way, we are able to assess how the population of sub-halos as a whole may impact local DM searches.

The structure of this article is as follows. In section~\ref{sec:Sub-halos}, we summarize the expected properties of Milky Way sub-halos and derive the velocity distribution of sub-halos and dark matter particles bound to them relative to the Sun or the Earth. We then update the calculation of the impact of sub-halos on direct detection experiments in section~\ref{sec:DD}. In section~\ref{sec:NT}, we build a probabilistic model to determine the boost factor of the neutrino signal due to the annihilation of dark matter particles which have been captured by scattering in the Sun. Finally, we conclude and summarize the results of this work in section~\ref{sec:Conclusions}.

\section{Dark matter flux in the Solar System} \label{sec:Sub-halos}

The Milky Way disk is believed to be embedded in a halo of dark matter particles with density and velocity distributions which are, to a first approximation, stationary and spherically symmetric. For dark matter searches in the Solar System, the so-called Standard Halo Model (SHM) is commonly adopted, where the local density is $\rho^{\rm loc}_{\rm SHM}=0.3\,{\rm GeV}\,{\rm cm}^{-3}$ and the local velocity distribution takes the  Maxwell-Boltzmann form \cite{Green:2011bv}:
\begin{align}
\label{eq:f_MB}
f_\mathrm{SHM} ( \vec{ v } ) = \frac { 1} { ( 2\pi \sigma _ { v } ^ { 2} ) ^ { 3/ 2} N _ { \text{esc} } } \exp \left[ - \frac { | \vec{ v } + \vec{v}_\odot| ^ { 2} } { 2\sigma _ { v } ^ { 2} } \right]\quad \text{for } |\vec v| \leq v_\mathrm{max}\,,
\end{align}
with normalization constant:
\begin{align}
N _ { \text{esc} } = \operatorname{erf} \left( \frac { v _ { \text{esc} } } { \sqrt { 2} \sigma _ { v } } \right) - \sqrt { \frac { 2} { \pi } } \frac { v _ { \text{esc} } } { \sigma _ { v } } \exp \left( - \frac { v _ {\mathrm{esc}} ^ { 2} } { 2\sigma _ { v } ^ { 2} } \right)\,.
\end{align}
Here, $v_\odot \approx 244$ km/s is the local velocity of the Sun with respect to the Galactic frame ~\cite{Xue:2008se,McMillan:2009yr,Bovy:2009dr}, $\sigma_v \approx 156$ km/s is  the velocity dispersion ~\cite{Kerr:1986hz,Green:2011bv} and $v_{\rm max}= v_\mathrm{esc} + v_\odot$ is the maximal velocity of DM particles gravitationally bound to the Galaxy (expressed in the Solar frame), with $v_\mathrm{esc}\approx 544 \,{\rm km}/{\rm s}$ being the escape velocity from the Milky Way~\cite{Smith:2006ym,Piffl:2013mla}. We note that while these values are observationally determined, they have associated uncertainties at the level of 10\% \cite{Green:2011bv,Green:2017odb,Wu:2019nhd}. The value of the local density $\rho^{\rm loc}_{\rm SHM}$ is also known only to within a factor of a few \cite{Read:2014qva} and a number of refinements to the SHM have been proposed (see e.g.~\cite{Bozorgnia:2017brl,Evans:2018bqy,Mandal:2018efq}). In this work, we keep the properties of the smooth DM component fixed and focus not on deviations from the SHM, but on the possible additional contribution of DM substructure.
 
 The Milky Way DM halo is expected to contain a population of sub-halos with different masses and located at different distances from the Galactic Center, which may alter, in a time-dependent way, the local density and velocity distribution. The population of sub-halos can be characterized by the halo mass function (i.e.~the number of sub-halos in a given mass range), their spatial distribution and their velocity distribution. A number of works have attempted to extract sub-halo populations from simulations (e.g.~\cite{Springel:2008cc,Gao:2010tn,Ishiyama:2011af}) or model the processes which govern the evolution and properties of sub-halos (e.g.~\cite{Stref:2016uzb,Hiroshima:2018kfv,Stref:2019wjv,Ando:2019xlm,Ishiyama:2019hmh}).

\begin{figure}[!t]
	\begin{center}
		\hspace{-0.75cm}
		\includegraphics[width=0.49\textwidth]{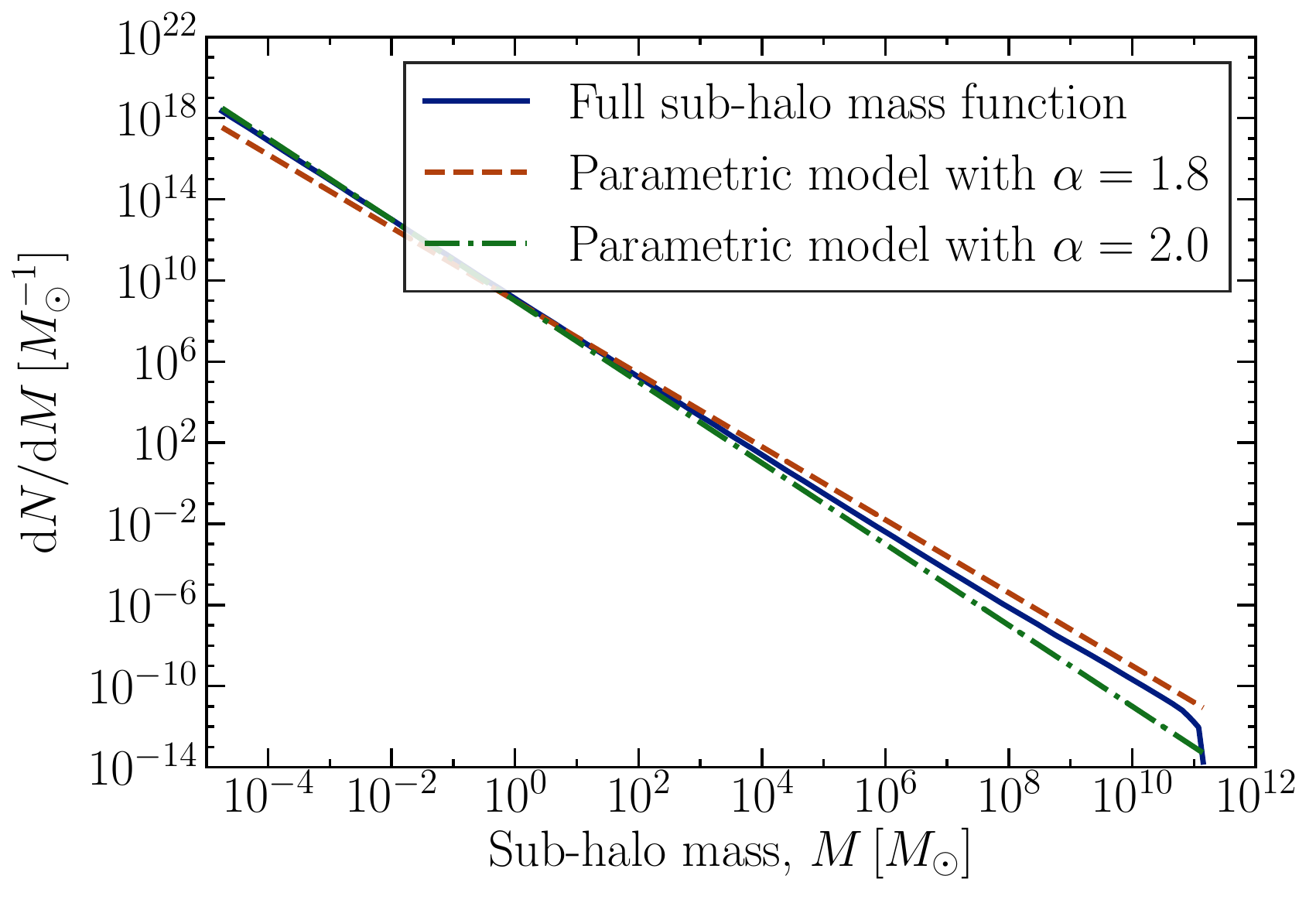}
		\includegraphics[width=0.47\textwidth]{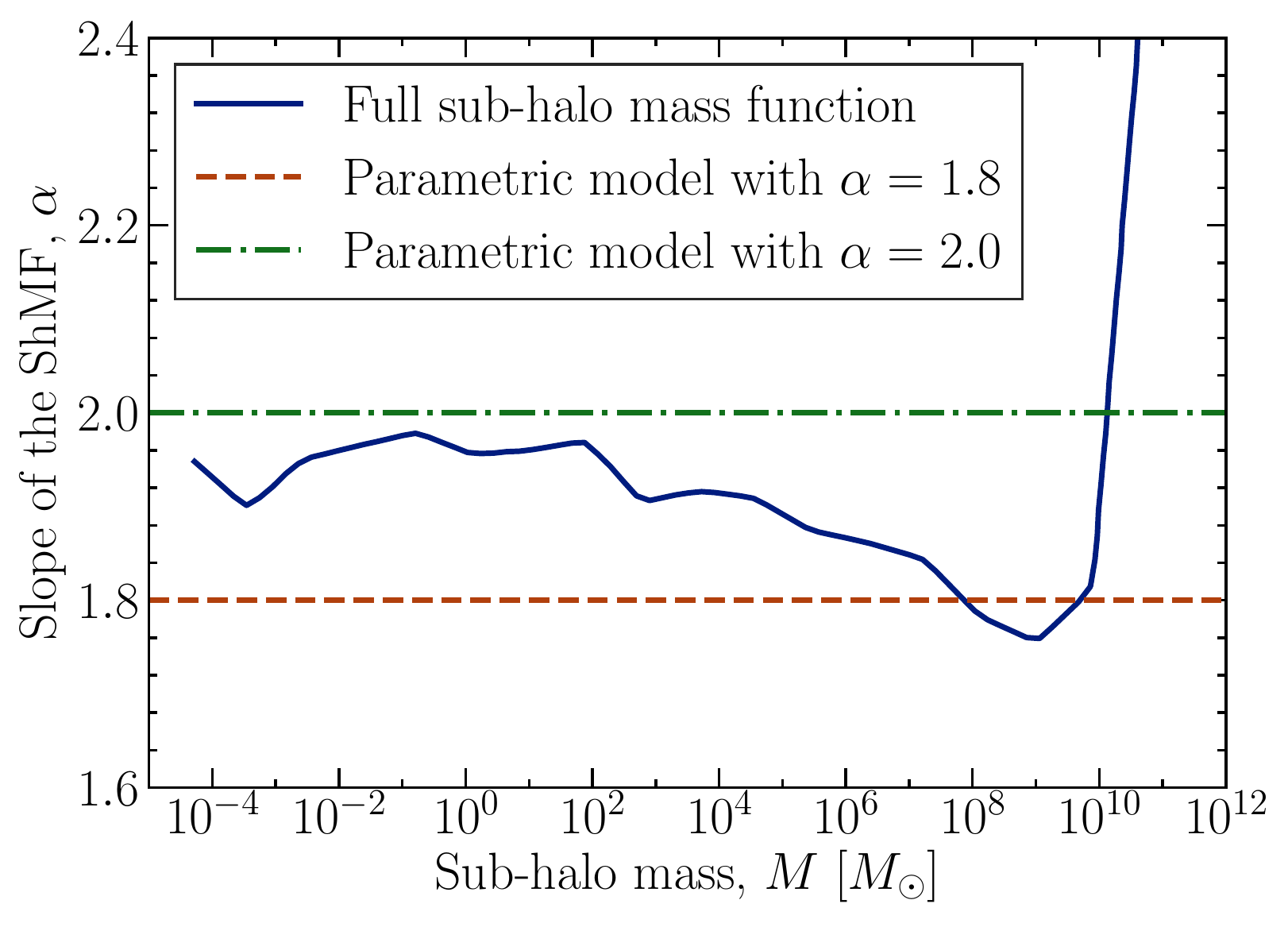}
	\end{center}
	\caption{\small{\it Left plot:} Sub-halo mass function (ShMF) for a DM halo of total mass $1.8 \times 10^{12}\,M_\odot$, taken from Ref.~\cite{Hiroshima:2018kfv}, as well as for two parametric models with ShMF proportional to $M^{-\alpha}$ with $\alpha=1.8$ and $2.0$. {\it Right plot:} Dependence of the slope $\alpha$ of the  ShMF with the sub-halo mass for the cases showed in the left plot.}\label{fig:Distribution}
\end{figure}

In our work we will adopt the halo mass function reported in \cite{Hiroshima:2018kfv}, obtained using a semi-analytic prescription for sub-halo accretion and tidal stripping by the host halo. This prescription was calibrated against the results of $N$-body simulations \cite{Ishiyama:2014gla}, allowing the method to cover sub-halo masses from $\sim 10^{11}~M_\odot$ down to $\sim 10^{-6}~M_\odot$, much smaller than those that can be resolved in the corresponding $N$-body simulations. We use tabulated halo mass functions and sub-halo properties provided by the authors of \cite{Hiroshima:2018kfv}, though we note that fitting functions are also provided in Appendix A of Ref.~\cite{Ando:2019xlm}. The halo mass function at redshift $z=0$ for a DM host halo of total mass $1.8\times 10^{12}\,M_\odot$ is shown for reference in the left panel of figure~\ref{fig:Distribution}. This mass function may be written in the form $\mathrm{d}N/\mathrm{d}M\propto M^{-\alpha(M)}$, with $\alpha(M)$ a mass-dependent parameter which ranges between 1.8 and 2.0 over the relevant mass range, as shown in the right panel. We also show in figure~\ref{fig:Distribution} the halo mass function and $\alpha(M)$ for two parametric models with $\mathrm{d}N/\mathrm{d}M\propto M^{-1.8}$ and $\mathrm{d}N/\mathrm{d}M\propto M^{-2.0}$, where the normalization of the power-law is fit to the results of \cite{Hiroshima:2018kfv}. The low mass sub-halos predicted in all of these models may have a significant impact in the search for dark matter inside the Solar System, as we will analyze in sections \ref{sec:DD} and \ref{sec:NT}. 
 
We will assume, following the results of \cite{Zhu:2015jwa}, that sub-halos in the Milky Way are spatially distributed following an Einasto profile \cite{Einasto1965, Graham:2005xx}
\begin{align}
\label{eq:Einasto}
\ln(n_\text{sh}(r)/n_{-2})=-\frac{2}{\gamma}\left[\left(\frac{r}{r_{-2}}\right)^\gamma -1\right]\,,
\end{align}
with $r_{-2}$ the radius at which the slope of the distribution equals $-2$,  $n_{-2}$ the number density of sub-halos at that location, and $\gamma$  a shape parameter. In our analysis we adopt the values $\gamma=0.854$ and $r_{-2}=245.1$ kpc from  \cite{Zhu:2015jwa}. We choose $n_{-2}$ such that $\int_{V_\text{MW}}\text{d}r^3\,n_\text{sh}(r)=\Nsh$, where $\Nsh$ is the total number of sub-halos, which we calculate from the mass function of \cite{Hiroshima:2018kfv} to be approximately $2.6\times 10^{16}$. It should be borne in mind that the Milky Way disk depletes the number of sub-halos close to the galactic center \cite{Kelley:2018pdy, Richings2018}. While the mass function of \cite{Hiroshima:2018kfv} does not account for this depletion, the spatial distribution we adopt from \cite{Zhu:2015jwa} is derived from hydrodynamical simulations including baryons, so we do not further correct for this effect. Finally, we will also assume that the sub-halo centers of mass move in the Milky Way halo following the same Maxwell-Boltzmann distribution, Eq.~\eqref{eq:f_MB}, as the smooth DM component.

Each sub-halo has an internal structure which is characterized by a density distribution and a velocity distribution. For the density distribution we adopt  an NFW density profile \cite{Navarro:1995iw,Navarro:1996gj} up to a truncation radius $r_t$, with density parameter $\rho_s$ and scale radius $r_s$:
\begin{align}
\rho_\mathrm{sh}(r) = 
\begin{cases}
\displaystyle{\frac {\rho _ { s } } { \left( r / r _ { s } \right) \left( 1+ r / r _ { s } \right) ^ { 2} }} & \text{for } r \leq r_t \,, \\
0 &  \text{for } r > r_t\,.
\end{cases}
\label{eq:rho_sh}
\end{align}
In field halos (those not bound within larger halos), the NFW profile would be truncated at the virial radius \cite{Allgood:2005eu,Drakos:2017gfy,2017MNRAS.472.2694S,2012ARA&A..50..353K}, defined such that the mean density of the halo inside the virial radius is a factor of 200 larger than the critical density (see e.g. \cite{Gao:2010tn}). For sub-halos, tidal stripping leads to a smaller truncation radius $r_t$, which can be determined from the mass loss of the sub-halo within the host \cite{2010MNRAS.406.1290P}. Similarly, the characteristic density and radius $\rho_s$ and $r_s$ at the present epoch may be determined from the corresponding values at accretion by accounting for tidal mass loss. The mass of an individual sub-halo is then given by
\begin{align}
\label{eq:halo_mass}
M = \frac{4\pi}{3} \rho_s r_s^3 f(c_V)\,,
\end{align}
where $c_V$ is the concentration parameter, defined as $c_V = r_t/r_s$, and $f \left( c _ { V } \right) = \ln (1+c _ { V } )- c _ { V } / \left( 1+ c _ { V } \right)$. For field halos, the concentration parameter is expected to follow a log-normal distribution \cite{Bullock:1999he,Wechsler:2001cs,Ishiyama:2011af,Correa:2015dva}, leading to typical values in the range $c_V\simeq 8$ and $c_V\simeq 35$. Instead, for MW sub-halos with masses between $10^{-6}\,M_\odot$ and $10^{11}\,M_\odot$, the concentration parameter typically takes values between $c_V\simeq 1$ and $c_V\simeq 35$, suggesting a larger number of denser, more concentrated sub-halos than those in the field. As for the mass function, we will make use of the distribution of individual sub-halo properties ($\rho_s$, $r_s$, $r_t$) as calculated in Ref.~\cite{Hiroshima:2018kfv}.

\begin{figure}[!t]
	\begin{center}
		\hspace{-0.75cm}
		\includegraphics[width=0.49\textwidth]{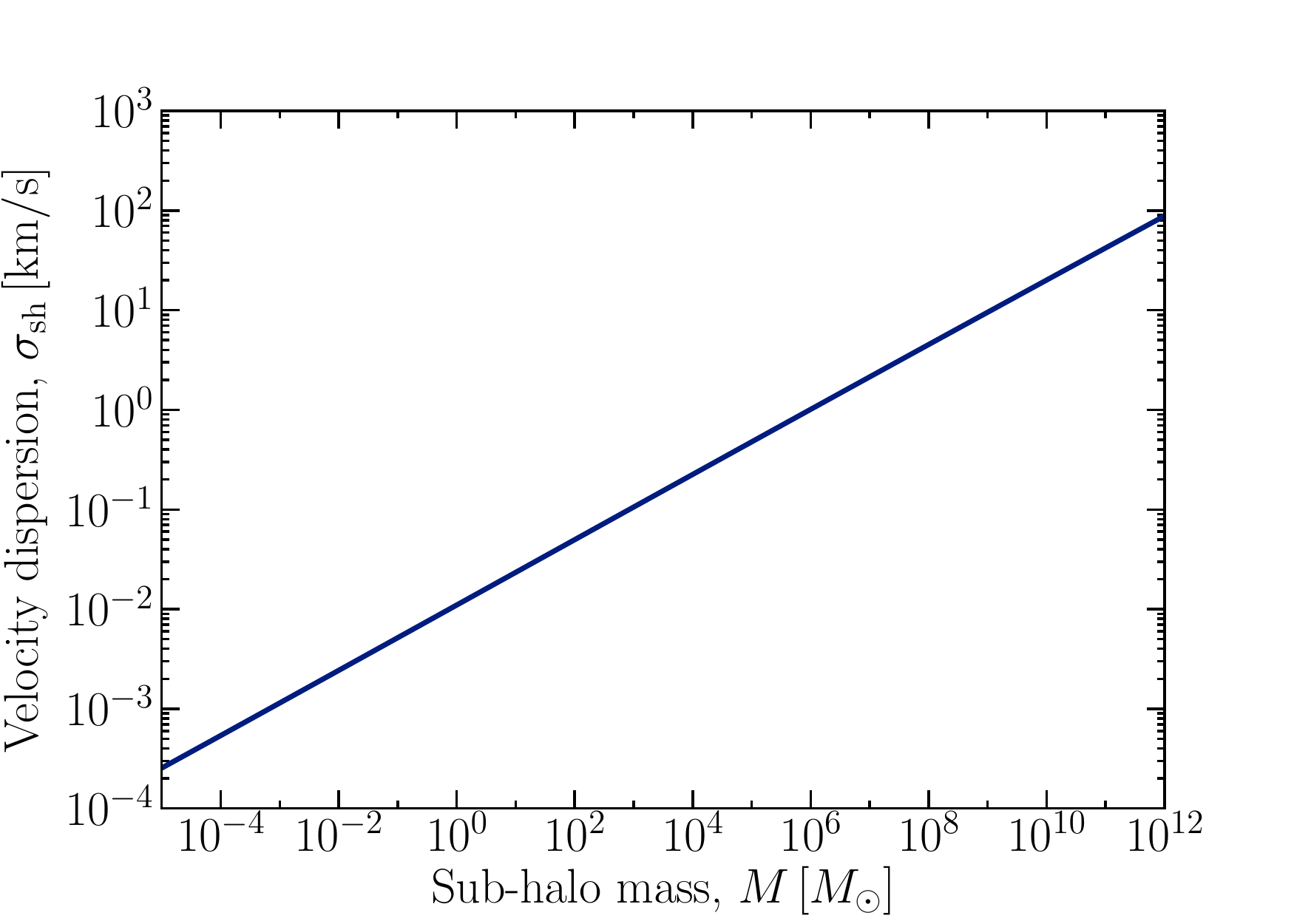}
	\end{center}
	\caption{\small Internal velocity dispersion as a function of the sub-halo mass for concentration parameters within $1\sigma$ from the mean value. }
	\label{fig:HaloDispersion}
\end{figure}

To determine the internal velocity distribution of the individual particles in the sub-halo, we assume that it is virialized. Hence, the DM velocity distribution relative to the center of mass of the sub-halo has a Maxwell-Boltzmann form with dispersion $\sigma_\text{sh}$:
\begin{align}
	\label{eq:MB}
	f_{\rm sh|sh}(\vec v) = \frac{1}{(2 \pi \sigma_{\rm sh}^2)^{3/2}}\exp\left[- \frac{|\vec v|^2}{2\sigma_{\rm sh}^2}\right]\,.
\end{align}
Accordingly the mean-square speed of sub-halo particles is $\langle |\vec v|^2\rangle = 3 \sigma_\text{sh}^2$. To determine the velocity dispersion we use the virial theorem, which relates the potential energy $U$ of the sub-halo to its kinetic energy $T$:
\begin{align}
T = -\frac{1}{2}\,U\,.
\end{align}
The total kinetic energy of a sub-halo of mass $M$  reads $T = \frac{1}{2} M \langle v^2\rangle$, while its potential energy is given by its gravitational binding energy
\begin{align}
U = - 4 \pi G _ { N } \int _ {0} ^ { r_t } \frac { M _ { \mathrm { enc } } ( r ) } { r } r ^ { 2 } \rho_\text{sh} ( r ) \mathrm { d } r\,,
\end{align}
with  $M_\text{enc}(r)$ the enclosed mass of the sub-halo as a function of the distance from the sub-halo center. For $\rho_\mathrm{sh}(r)$ following an NFW profile, as in Eq.~\eqref{eq:rho_sh}, we obtain:
\begin{align}
U &= -8 \pi ^2 G_N \rho_s^2 r_s^5 \left[ 1-2 \frac{\ln (c_V+1)}{(c_V+1)} -  \frac{1}{(c_V+1)^2}\right]\\
&= -\left(\frac{\pi}{6}\right)^{1/3}  \frac{ G_Nc_V}{f(c_V)^2} \bar{\rho}^{1/3} M^{5/3}\left[ 1-2 \frac{\ln (c_V+1)}{(c_V+1)} -  \frac{1}{(c_V+1)^2}\right]\,,
\end{align}
where $\bar{\rho} = 3 f(c_V) \rho_s/c_V^3$ is the mean density inside the sub-halo. Finally, applying the virial theorem, we obtain: 
\begin{align}
\sigma_\text{sh}^2 = -\frac{U}{3 M} = \frac{1}{3}\left(\frac{\pi}{6}\right)^{1/3}  \frac{ G_Nc_V}{f(c_V)^2} \bar{\rho}^{1/3} M^{2/3}\left[ 1-2 \frac{\ln (c_V+1)}{(c_V+1)} -  \frac{1}{(c_V+1)^2}\right]\,.
\end{align}
which depends mostly on the total sub-halo mass, $M$, and has a weak dependence on the concentration parameter, $c_V$. The velocity dispersion is shown in figure~\ref{fig:HaloDispersion} and can be approximated by
\begin{align}
 \sigma_\text{sh}\approx 1.5\,{\rm km}/{\rm s}\,\left(\frac{M}{10^6\,M_\odot}\right)^{\frac{1}{3}}.
\end{align}
The internal density and velocity distributions calculated above are expressed in the sub-halo center-of-mass frame. When expressed in the Solar frame, one finds a time-dependent density distribution as the Sun traverses the sub-halo at a distance $r$ from the sub-halo center, $\rho_{\rm sh}^{\rm loc}(t) \equiv \rho_{\rm sh}^{\rm loc}[r(t)]$. The resulting velocity distribution of DM particles inside the sub-halo reads:
\begin{align}
\label{eq:f_MB_Sol}
f_{{\rm sh},\vec v_{\rm sh}}(\vec v) = \frac{1}{(2 \pi \sigma_{\rm sh}^2)^{3/2}}\exp\left[- \frac{|\vec v+\vec v_{\rm rel}|^2}{2\sigma_{\rm sh}^2}\right]\,,
\end{align}
where the velocity of the sub-halo with respect to the Sun is $\vec{v}_\mathrm{rel} = \vec{v}_\odot - \vec{v}_\mathrm{sh}$ and where the distribution of the individual sub-halo velocities in the Galactic rest frame is $f_{\rm SHM}(\vec v_{\rm sh})$, given in Eq.~\eqref{eq:f_MB}. For small  sub-halos, namely sub-halos with $M\lesssim 10^6 \,M_\odot$, the velocity dispersion is much smaller than the speed of the sub-halo center of mass,  $|\vec v_{\rm sh}|\gg \sigma_{\rm sh}$. Therefore,  the sub-halo behaves effectively as stream of DM particles all moving with the same
velocity. In the Solar frame, the velocity distribution of DM particles from a sub-halo with velocity $\vec v_{\rm sh}$ can thus be approximated by:
\begin{align}
f_{{\rm sh},\vec v_{\rm sh}}(\vec v)\simeq \delta(\vec v+\vec v_{\rm rel})\,.
\label{eq:f_delta}
\end{align}

Finally, we can write down the flux of dark matter particles in the Solar frame, assuming that the Sun travels inside a sub-halo with mass $M$ and concentration parameter $c_V$ which moves with respect to the Galactic frame with velocity $\vec v_{\rm sh}$. This flux contains two components: a flux of  DM particles coming from the smooth Milky Way dark matter halo, characterized by time-independent mass density and velocity  distributions $\rho_{\rm SHM}$ and $f_{\rm SHM}(\vec v)$, and a flux of DM particles from sub-halos, characterized by a time-dependent mass density $\rho_{\rm sh}^{\rm loc}(t)$ and a time-independent velocity distribution $f_{{\rm sh},\vec v_{\rm sh}}(\vec v)$. This flux reads: \footnote{We add the superscript `loc' to emphasize that these quantities are evaluated locally, at the Sun or Earth's position.}
\begin{align}
F(\vec v,t)= v \Big[\frac{\rho^{\rm loc}_{\rm SHM}}{m_{\rm DM}} f_{\rm SHM}(\vec v)+\frac{\rho_{\rm sh}^{\rm loc}(t)}{m_{\rm DM}} f_{{\rm sh},\vec v_{\rm sh}}(\vec v)\Big]\,.
\label{eq:flux}
\end{align}

\section{Impact of sub-halos in direct detection experiments}\label{sec:DD}

We now apply our model for DM sub-structure to determine the expected rate at direct detection experiments. The rate of nuclear recoils induced by scatterings of DM particles traversing a detector at the Earth is given by \cite{Lewin:1995rx,Cerdeno:2010jj}:
\begin{align}
R= \sum_i \int_{0}^\infty \text{d}E_R \, \epsilon_i (E_R) \frac{\xi_i}{m_{A_i} } \int_{v \geq v_{\text{min},i}^{(E_R)}} \text{d}^3 v F(\vec v+\vec v_{\rm Earth}, t_0) \, \frac{\text{d}\sigma_i}{\text{d}E_R}(v, E_R) \,.
\label{eq:scattering_rate}
\end{align}
Here, $F(\vec v+\vec v_{\rm Earth},t_0)$ is the flux of DM received at the Earth at the present time (as described in the previous section), with $\vec v$ the DM velocity expressed in the Solar frame. In addition, $\text{d}\sigma_i/\text{d}E_R$ is the differential cross section for the elastic scattering of DM off a nuclear isotope $i$ with mass $m_{A_i}$ and mass fraction $\xi_i$ in the detector, producing a nuclear recoil with energy $E_R$. Furthermore, $v_{\text{min},i}(E_R) = \sqrt{m_{A_i} E_R/(2 \mu_{A_i}^2)}$ is the minimal velocity necessary for a DM particle to induce a recoil with energy $E_R$ and $\mu_{A_i}$ is the reduced mass of the DM-nucleus system. The efficiency $\epsilon_i(E_R)$ gives the probability to detect a nuclear recoil off the target nucleus $i$. Finally, the total number of expected recoil events at a direct detection experiment reads ${\cal N}=R\cdot \mathcal{E}$, with $\mathcal{E}$ the exposure (i.e.~mass multiplied by live-time) of the experiment. 

We consider the case when the Sun is traversing at the current epoch $t_0$ a sub-halo of mass $M$ and concentration parameter $c_V$, which is moving with velocity $\vec v_{\rm sh}$ with respect to the Galactic frame. The scattering rate can be readily calculated inserting Eq.~(\ref{eq:flux}) into Eq.~(\ref{eq:scattering_rate}). For low mass sub-halos, for which the velocity distribution in the Solar frame can be approximated by Eq.~(\ref{eq:f_delta}), one finds a increment in the scattering rate
\begin{align}
{\cal I}_R\equiv \frac{R}{R_{\rm SHM}}-1\simeq \frac{\rho^{\rm loc}_{\rm sh}[r(t_0)]}{\rho_{\rm SHM}}
\frac{R^{\rho^{\rm loc}_{\rm SHM}}_{\delta(\vec v +\vec{v}_\text{rel})}}{R_{\rm SHM}}\,.
\label{eq:def_IR}
\end{align}
Here, $\vec{v}_\text{rel}=\vec v_\odot+\vec v_{\rm Earth}-\vec v_{\rm sh}$ is the velocity of the sub-halo with respect to the Earth and $\rho^{\rm loc}_{\rm sh}[r(t_0)]$ is the contribution of the sub-halo to the total dark matter density (with the Sun located a distance $r(t_0)$ from the center of the sub-halo).

\begin{figure}[!t]
	\begin{center}
		\hspace{-0.75cm}
		\includegraphics[width=0.49\textwidth]{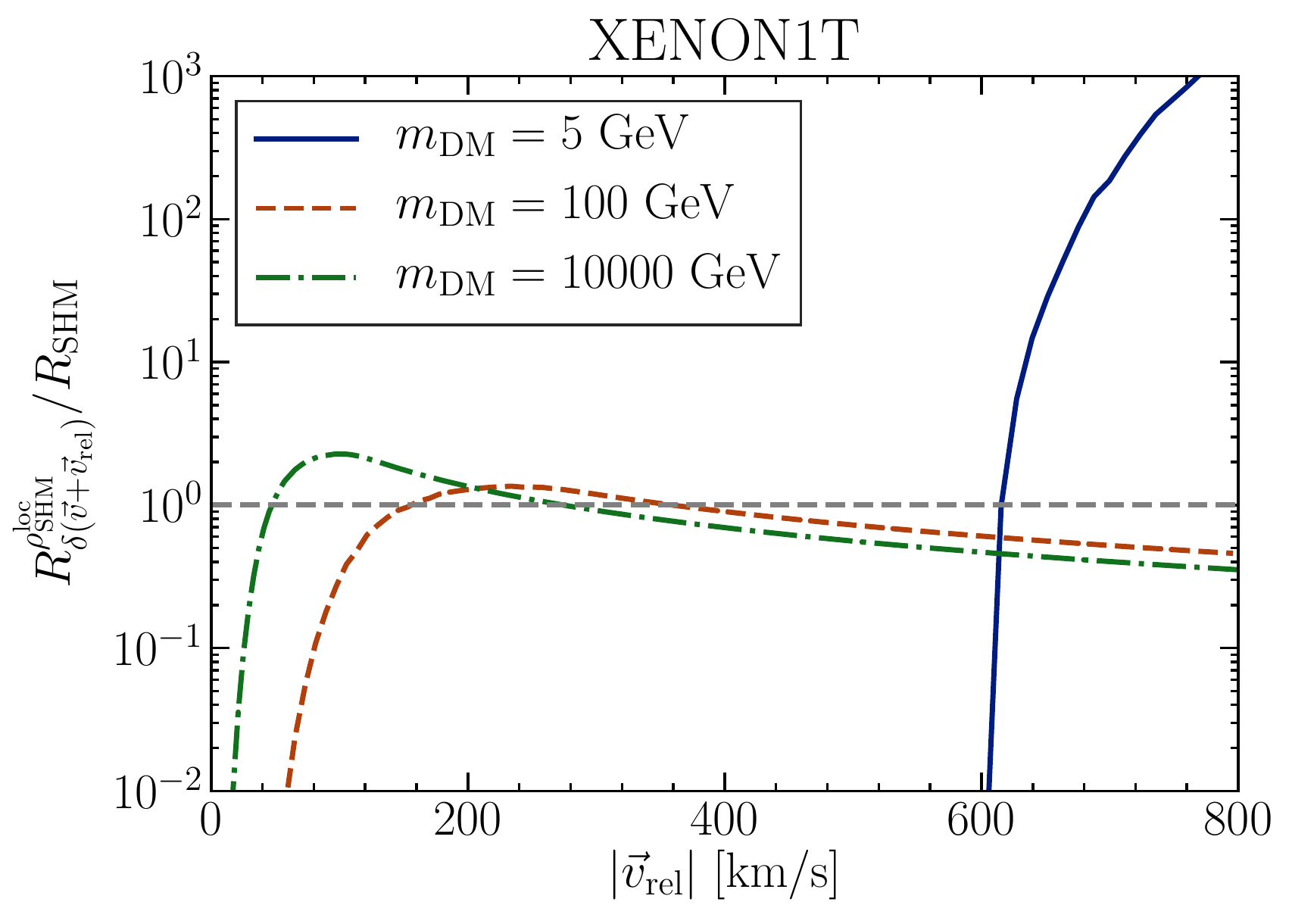}
		\includegraphics[width=0.49\textwidth]{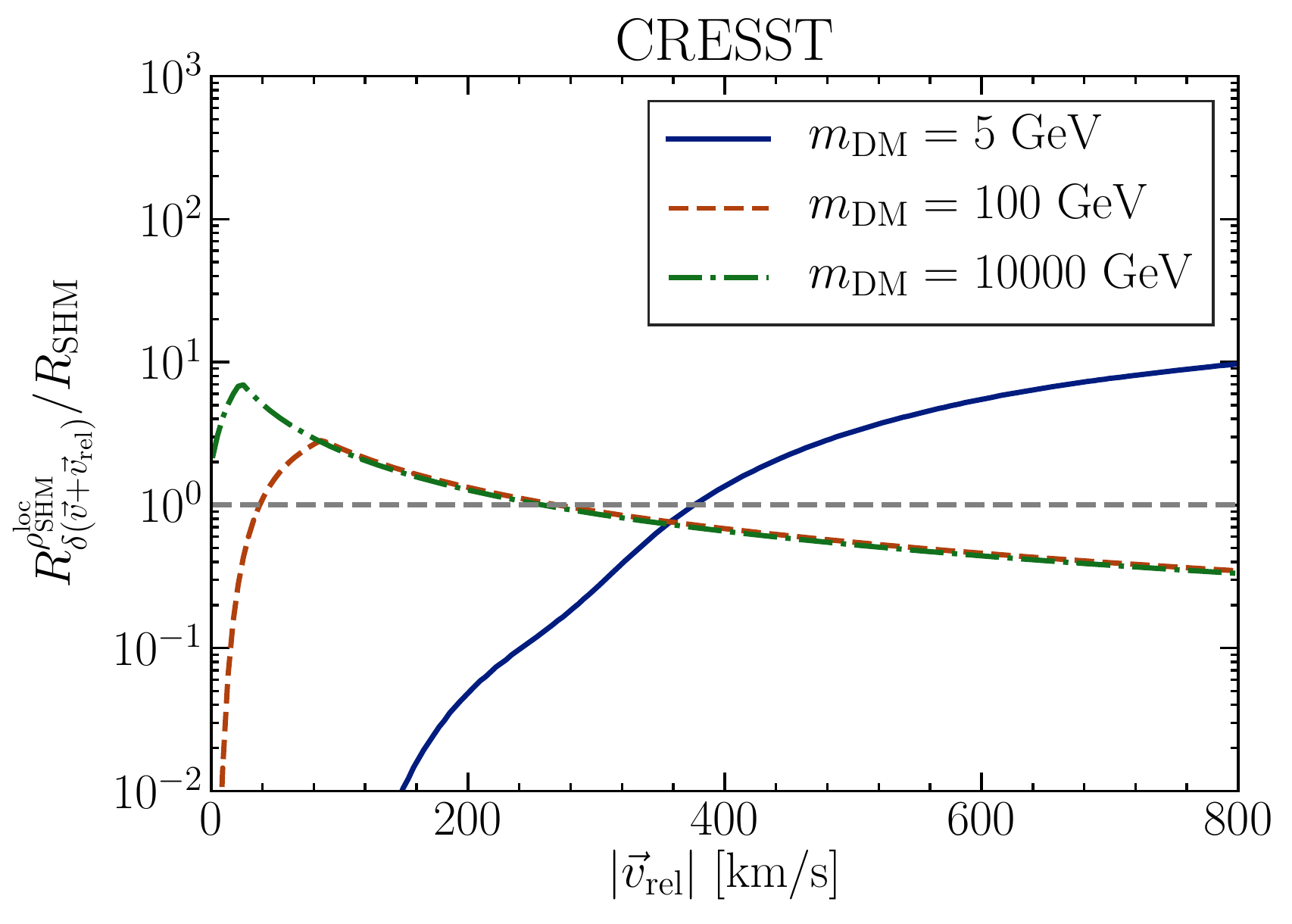}
	\end{center}
	\caption{\small Scattering rate induced at XENON1T (left plot) and CRESST (right plot) for dark matter streams with velocity  $|\vec{v}_\mathrm{rel}|$, relative to the scattering rate assuming the Standard Halo Model, for dark matter masses $m_{\rm DM}=5$, 100 and 10000 GeV.}
	\label{fig:R_streams}
\end{figure}

\begin{figure}[!t]
	\begin{center}
		\hspace{-0.75cm}
		\includegraphics[width=0.49\textwidth]{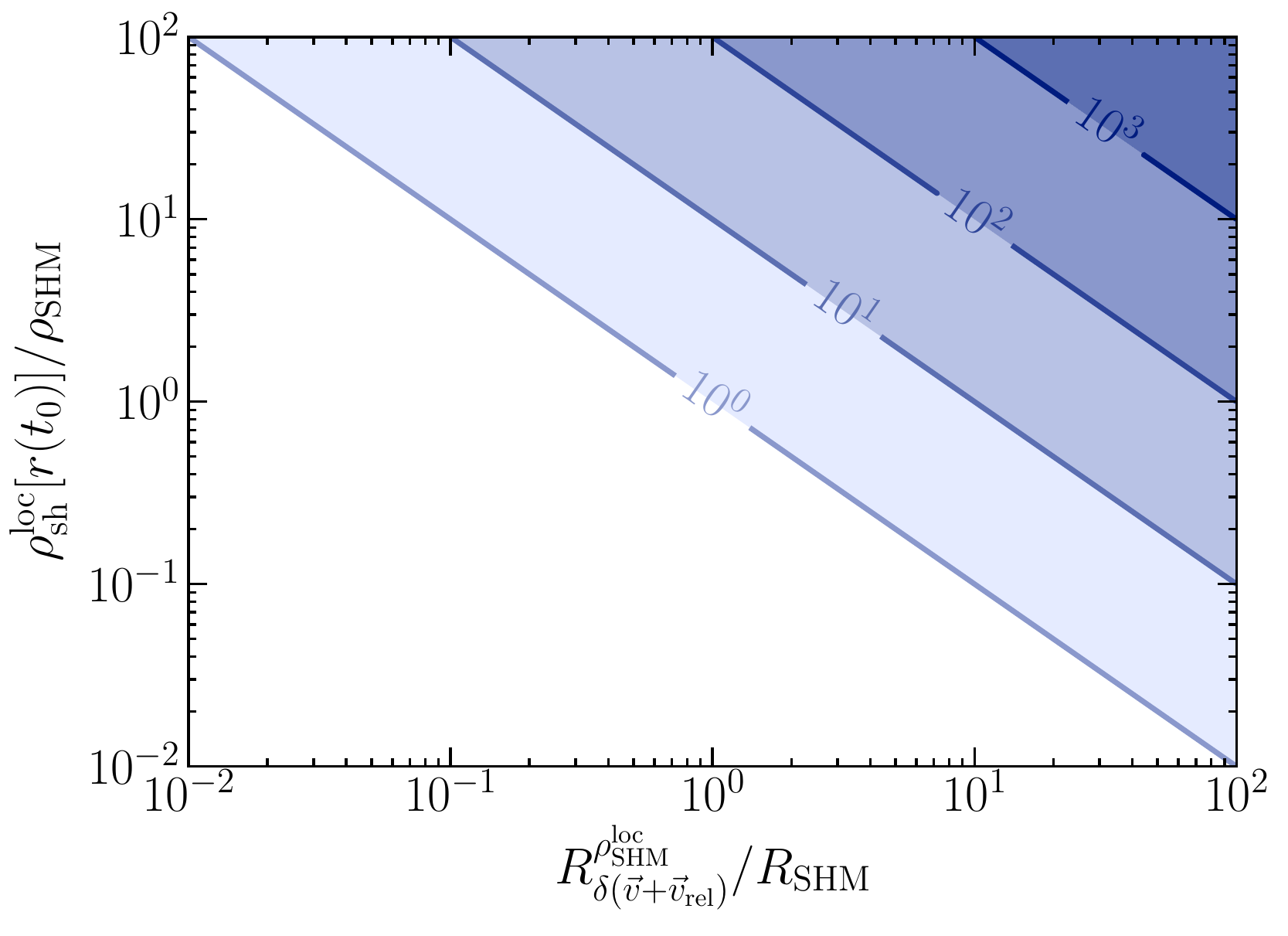}
	\end{center}
	\caption{\small Increment in the scattering rate $\mathcal{I}_R$ relative to the expectations from the Standard Halo Model in the parameter space spanned by $\rho^{\rm loc}_{\rm sh}[r(t_0)]/\rho_{\rm SHM}$ (the contribution from sub-halos to the local density, relative to the SHM value) and $R^{\rho^{\rm loc}_{\rm SHM}}_{\delta(\vec v +\vec{v}_\text{rel})}/R_{\rm SHM}$ (the capture rate of dark matter particles from the sub-halo, relative to the capture rate of dark matter particles in the smooth halo).} 	
	\label{fig:IR}
\end{figure}

The factor $R_\mathrm{SHM}$ is the event rate for the smooth SHM described at the start of section~\ref{sec:Sub-halos}, while  $R^{\rho^{\rm loc}_{\rm SHM}}_{\delta(\vec v +\vec{v}_\text{rel})}$ is the scattering rate for a stream of dark matter particles with density $\rho_{\rm SHM}$ and velocity distribution $f(\vec v)=\delta(\vec v +\vec{v}_\text{rel})$. This rate is illustrated in figure~\ref{fig:R_streams} for XENON1T \cite{Aprile:2018dbl} (left panel) and for CRESST \cite{Angloher:2015ewa} (right panel), for three representative dark matter masses.~\footnote{To calculate the detector responses we used the DDCalc package \cite{Workgroup:2017lvb, Athron:2018hpc}.} For large dark matter masses, the scattering rate for streams (relative to the SHM) is ${\cal} O(0.1-1)$ for generic values of the relative velocity, and can be much larger for small dark matter masses and large relative velocities. Accordingly, there can in principle be a large enhancement in the scattering rate at these two experiments if the sub-halo also gives an ${\cal O}(1)$ contribution to the local dark matter density, as shown in Fig.~\ref{fig:IR}.

From the current body of knowledge about the sub-halo population in the Milky Way (summarized in section \ref{sec:Sub-halos}), one can estimate the probability of having an increment in the event rate ${\cal I}_R$, defined in Eq.~\eqref{eq:def_IR}. Assuming that we are currently inside a sub-halo, the probability that the sub-halo moves with velocity $\vec v_{\rm sh}$ with respect to the Galactic frame is given straightforwardly by the Maxwell-Boltzmann distribution in Eq.~\eqref{eq:f_MB}. We must then determine the probability of finding ourselves at the current time inside a sub-halo, along with the distribution of density enhancements $ \rho_{\rm sh}^{\rm loc}[r(t_0)] \equiv \tilde \rho$.

The probability  $P(\tilde \rho , M, c_V)$ that we find ourselves inside a sub-halo of mass $M$ and concentration $c_V$ such that the total DM density receives an additional contribution $\tilde \rho$ can be written:
\begin{align}
P(\tilde \rho , M, c_V) = P(D , M, c_V) \left|\frac{\mathrm{d}\,\tilde\rho(D, M, c_V)}{\mathrm{d}D}\right|^{-1} \qquad \text{for } D < r_t(M, c_V)\,,
\label{eq:P_of_rho}
\end{align}
where $\tilde{\rho}(D, M, c_V)$ is the sub-halo density a distance $D$ from the center of the sub-halo, Eq.~\eqref{eq:rho_sh}, and  $P(D , M, c_V)$ is the probability of finding the center of such a sub-halo at the distance $D$ from Earth. This in turn can be calculated from the probability of finding a sub-halo with mass $M$, $P(M)$; the probability of finding a concentration parameter $c_V$ for that sub-halo mass $P(c_V|M)$; and the probability of finding a sub-halo with a given mass and concentration at the distance $D$ from the Earth $P(D|M,c_V)$:
\begin{align}
P(D,M,c_V)= P(M) P(c_V|M) P(D|M,c_V) \,.
\label{eq:PDMcV}
\end{align}
More concretely, the probability of finding a sub-halo of mass $M$ is $P(M)=\frac{1}{N_{\rm sh}}\mathrm{d}N/\mathrm{d}M$, with $\mathrm{d}N/\mathrm{d}M$ shown in figure~\ref{fig:Distribution}, and $\Nsh$ the total number of sub-halos, given in \cite{Hiroshima:2018kfv}. The distribution of concentrations $c_V$ for sub-halos of mass $M$, $P(c_V|M)$, is obtained using the tabulated results of Ref.~\cite{Hiroshima:2018kfv}.
Lastly, we assume for simplicity that the spatial distribution of sub-halos is independent of their mass and concentration. Hydrodynamical simulations suggest that baryonic disruption should preferentially deplete the most massive sub-halos at small galactocentric radii \cite{Zhu:2015jwa}. At larger radii (relevant in this context), simulations suggest instead only a weak dependence of the spatial distribution on the sub-halo properties. The probability of finding a sub-halo center at a distance $D$ from the Earth can then be estimated as:
\begin{align}
P(D) \equiv \frac{1}{N_\mathrm{sh}}\frac{\mathrm{d}N_{\rm sh}(D)}{\mathrm{d}D}=\frac{4\pi D^2 \bar{n}_\mathrm{sh}(D)}{N_\mathrm{sh}}\,,
\end{align}
where $N_\mathrm{sh}(D)$ is the number of sub-halos that can be found in the spherical shell located at distance $D\rightarrow D + \mathrm{d}D$ from the Earth, with $\bar{n}_\mathrm{sh}(D)$ the average number density of sub-halos at that distance:
\begin{align}
\bar{n}_\mathrm{sh}(D) = \frac{1}{2}\int_{-1}^1 n_\mathrm{sh}[r(D, \psi)]\,\mathrm{d}\cos\psi\,.\label{eq:nbar}
\end{align}
Here, $n_\mathrm{sh}(r)$ is the number density of sub-halos as a function of the galactocentric radius (given in  Eq. \eqref{eq:Einasto}), which can be expressed in terms of the distance of the Sun to the Galactic center using $r(D,\psi) = \sqrt{D^2 + r_\odot^2 - 2 r_\odot D \cos\psi}$, with $r_\odot \approx 8.5\,\mathrm{kpc}$ and $\psi$ the angular separation between the Galactic center and the sub-halo center. 

\begin{figure}[!t]
	\begin{center}
		\hspace{-0.75cm}
		\includegraphics[width=0.49\textwidth]{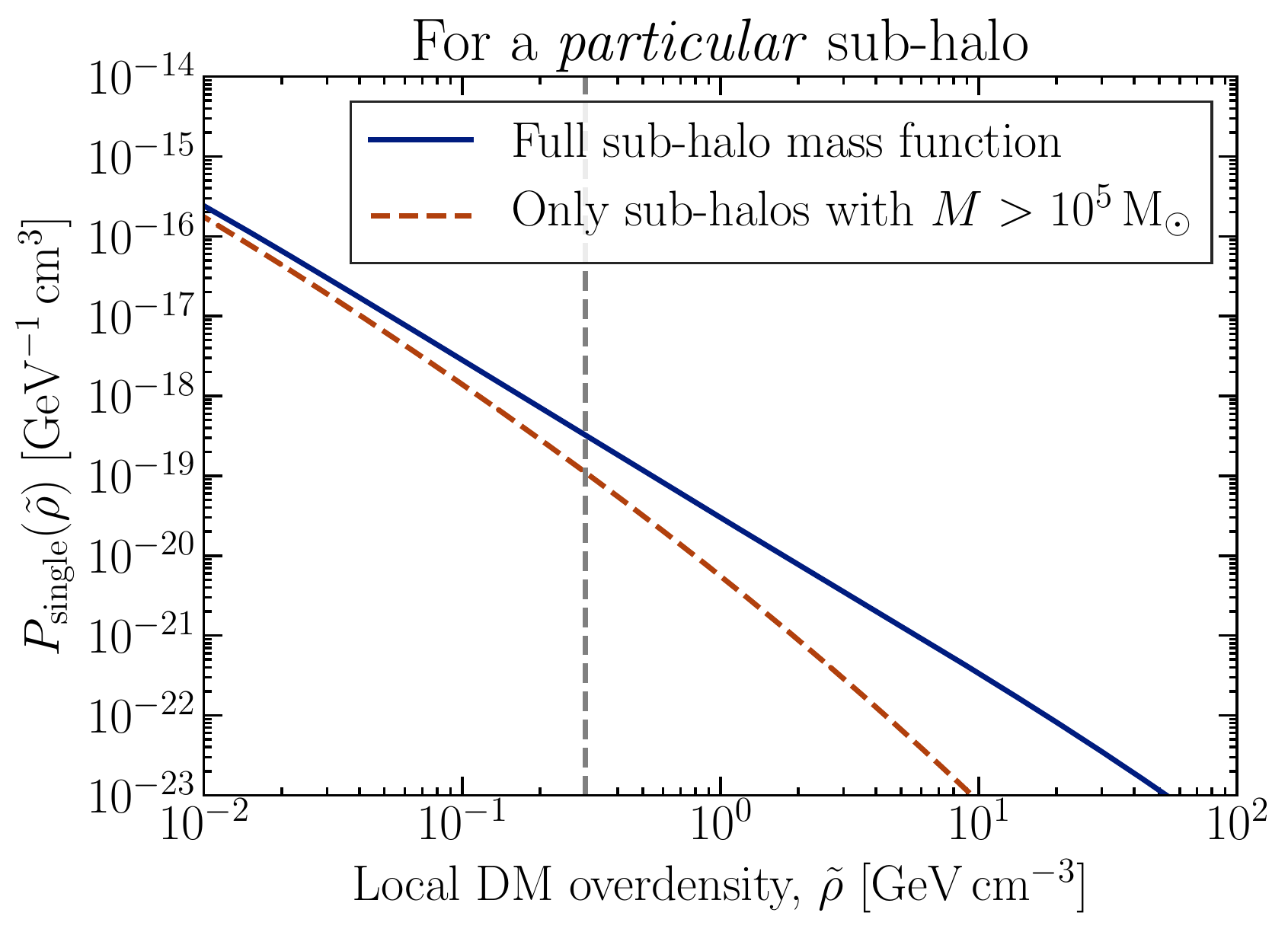}
		\includegraphics[width=0.49\textwidth]{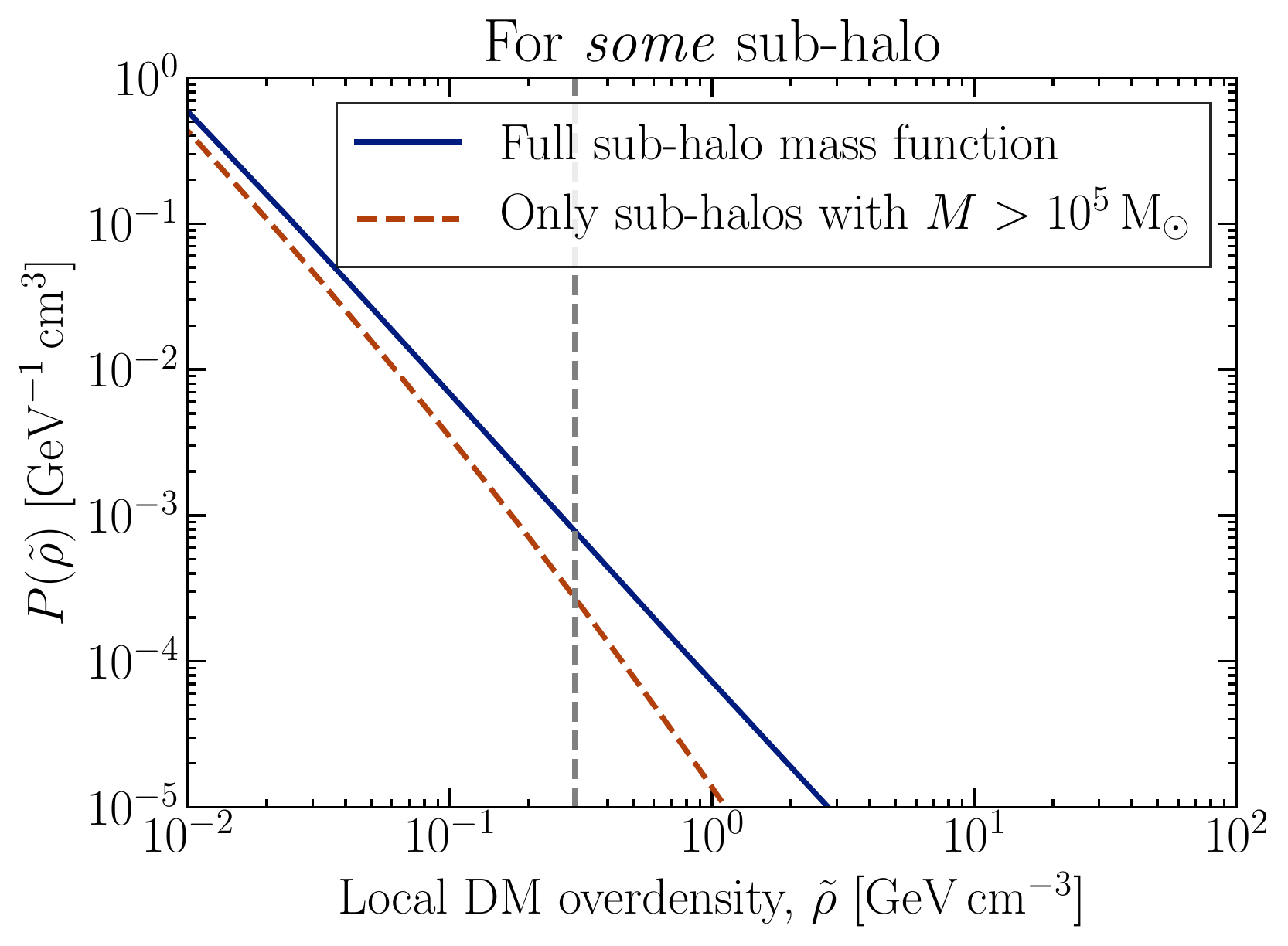}
	\end{center}
	\caption{\small Probability distribution of producing an overdensity $\tilde \rho$ at the position of the Earth due to the passage through one particular sub-halo (left plot) and due to the passage through some sub-halo (right plot) assuming the ShMF calculated in Ref.~\cite{Hiroshima:2018kfv}. The figure also shows for comparison the corresponding probability distributions when considering only sub-halos with mass greater than $10^5\,M_\odot$, as resolved by numerical N-body simulations.}	
	\label{fig:SingleDD}
\end{figure}

Finally, one obtains the probability distribution for finding a contribution $\tilde \rho$ to the local density (due to a \textit{single} sub-halo) by integrating Eq.~\eqref{eq:P_of_rho} over all sub-halo masses and concentrations:
\begin{align}
P_{\rm single} (\tilde \rho)=
	\int_{M_{\rm min}}^{M_{\rm max}} {\rm d}M 
	\int _{0}^{\infty} {\rm d}c_V  
	P(\tilde \rho, M, c_V)\,.
\end{align}
This probability distribution is shown in the left panel of figure~\ref{fig:SingleDD}, assuming the sub-halo mass function of \cite{Hiroshima:2018kfv}. We take into account sub-halos with masses ranging from $M_\text{min}=10^{-6}\,M_\odot$ up to $M_\text{max}=10^{12}\,M_\odot$.
As apparent from the plot, the probability that a single sub-halo dominates the local density is small. This is to be expected, as the fraction of the Milky Way volume occupied by a given individual sub-halo is also small. Indeed, the probability of having an enhancement of any size in the local dark matter density due to a {\it particular single} sub-halo, which we denote as $p_\text{single}$, is $p_\text{single}\equiv \int {\rm d}\tilde{\rho}\, P_\text{single}(\tilde{\rho}) \approx 5 \times 10^{-17}$. Nevertheless, due to the large number of sub-halos present in the galaxy, there is a fairly high probability that the Earth is currently immersed in a sub-halo.
The probability of being immersed in \textit{any one} sub-halo today is obtained from the Binomial distribution:
\begin{align}
p_1 &= \Nsh\,p_\text{single}\,(1-p_\text{single})^{\Nsh-1}\approx \Nsh\,p_\text{single}\,(1-p_\text{single})^{\Nsh} \sim 12\%\,, 
\end{align}
while the probability that the Earth is {\it not} immersed in any sub-halo is\footnote{In practice, we have used the approximation $\log(1+x) \approx x$ for $0 < x \ll 1$ to evaluate the probabilities.}
\begin{align}
p_0 = (1-p_\text{single})^{\Nsh}\approx \exp(-\Nsh\cdot p_\text{single}) = 88\%\,.
\end{align}

In the right panel of figure~\ref{fig:SingleDD}, we plot the probability $P(\tilde{\rho}) \equiv (p_1/p_\mathrm{single}) \times P_\mathrm{single}(\tilde{\rho}) $ that some sub-halo enhances the local density by $\tilde \rho$, which is the relevant quantity for the purposes of direct detection at Earth. It follows from the plot that there is a 0.1\% probability  ($10^{-3}\%$ probability) of finding a contribution to the local dark matter density from sub-halos which is equal to $\rho_{\rm SHM}^{\rm loc}$ ($10 \rho_{\rm SHM}^{\rm loc}$). The plots also show for reference the probability distributions considering only sub-halos with mass $M\geq 10^5 M_\odot$, namely those that can be resolved and identified with current $N$-body simulations. Including also the lighter sub-halos, as done in this work, clearly enhances the probability of finding a larger DM density in the Solar System, owing to their larger number density. More concretely, the low mass sub-halos increase by a factor $\sim 3$ ($\sim 10$) the probability of finding a contribution from the sub-halos to the total local DM  density which is a factor $\sim 1$ ($\sim 10$) larger than $\rho_{\rm SHM}^{\rm loc}$.

\begin{figure}[!tp]
	\begin{center}
		\hspace{-0.75cm}
		\includegraphics[width=0.49\textwidth]{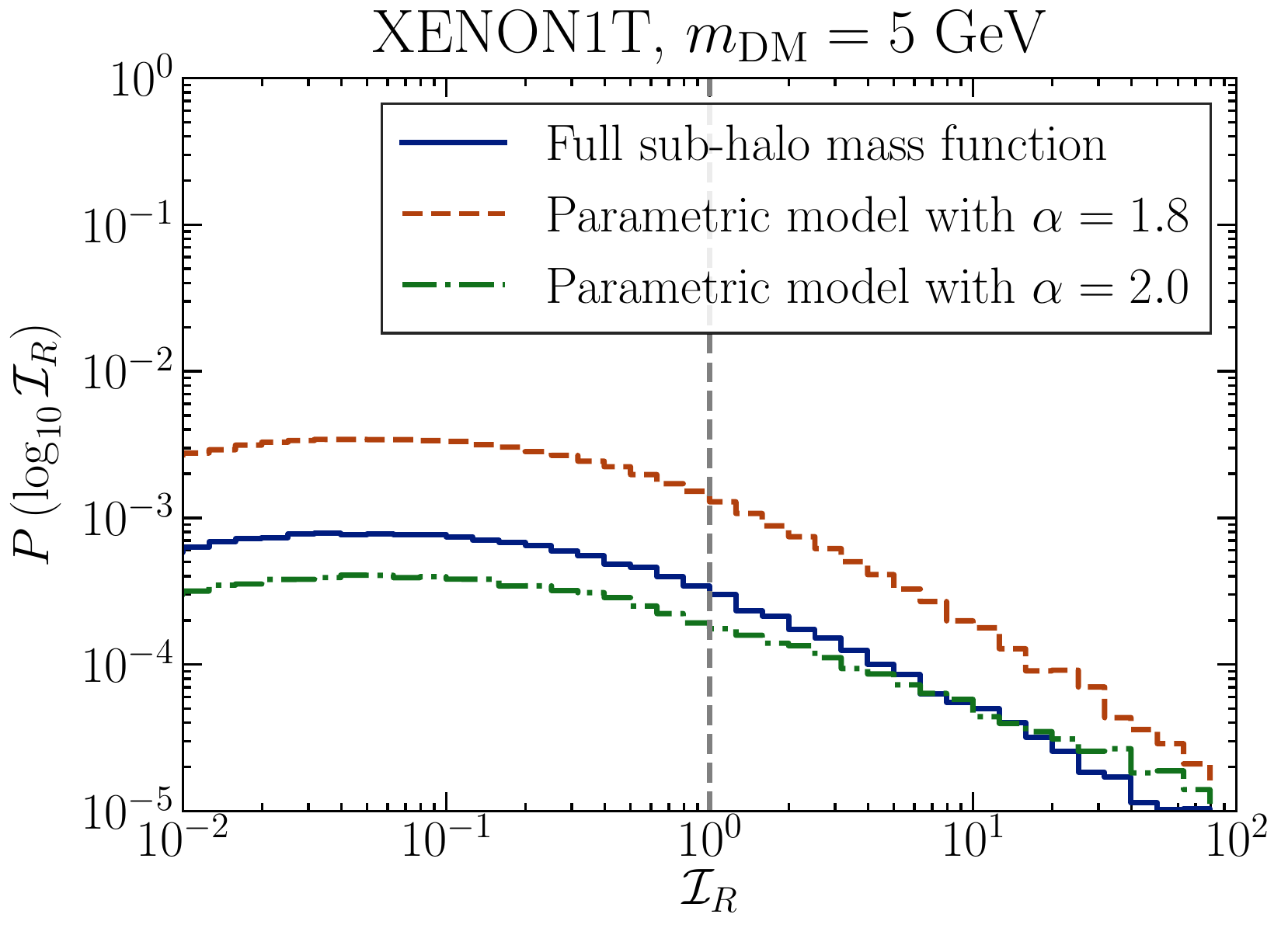}
		\includegraphics[width=0.49\textwidth]{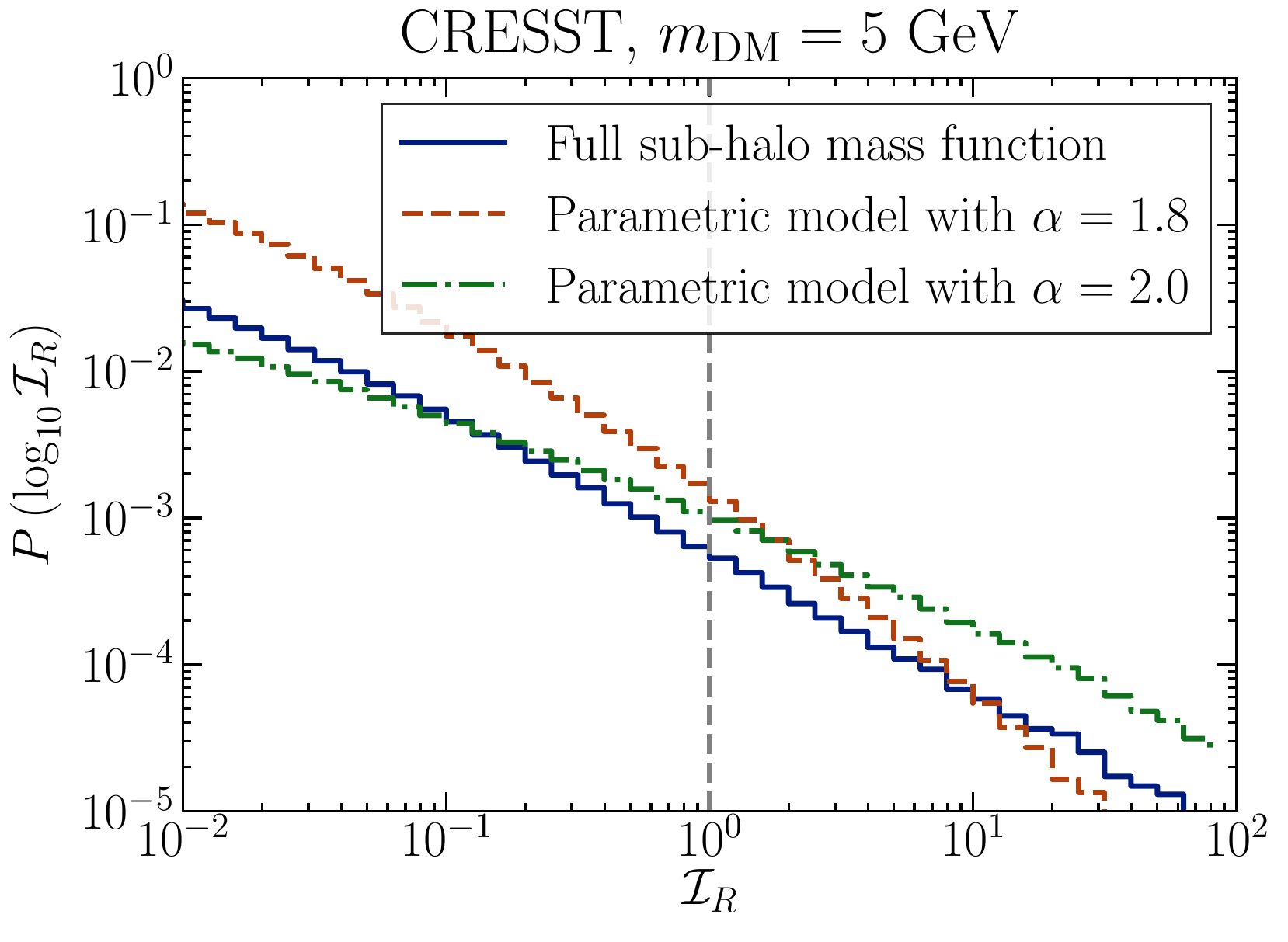}\\
		\hspace{-0.75cm}	
		\includegraphics[width=0.49\textwidth]{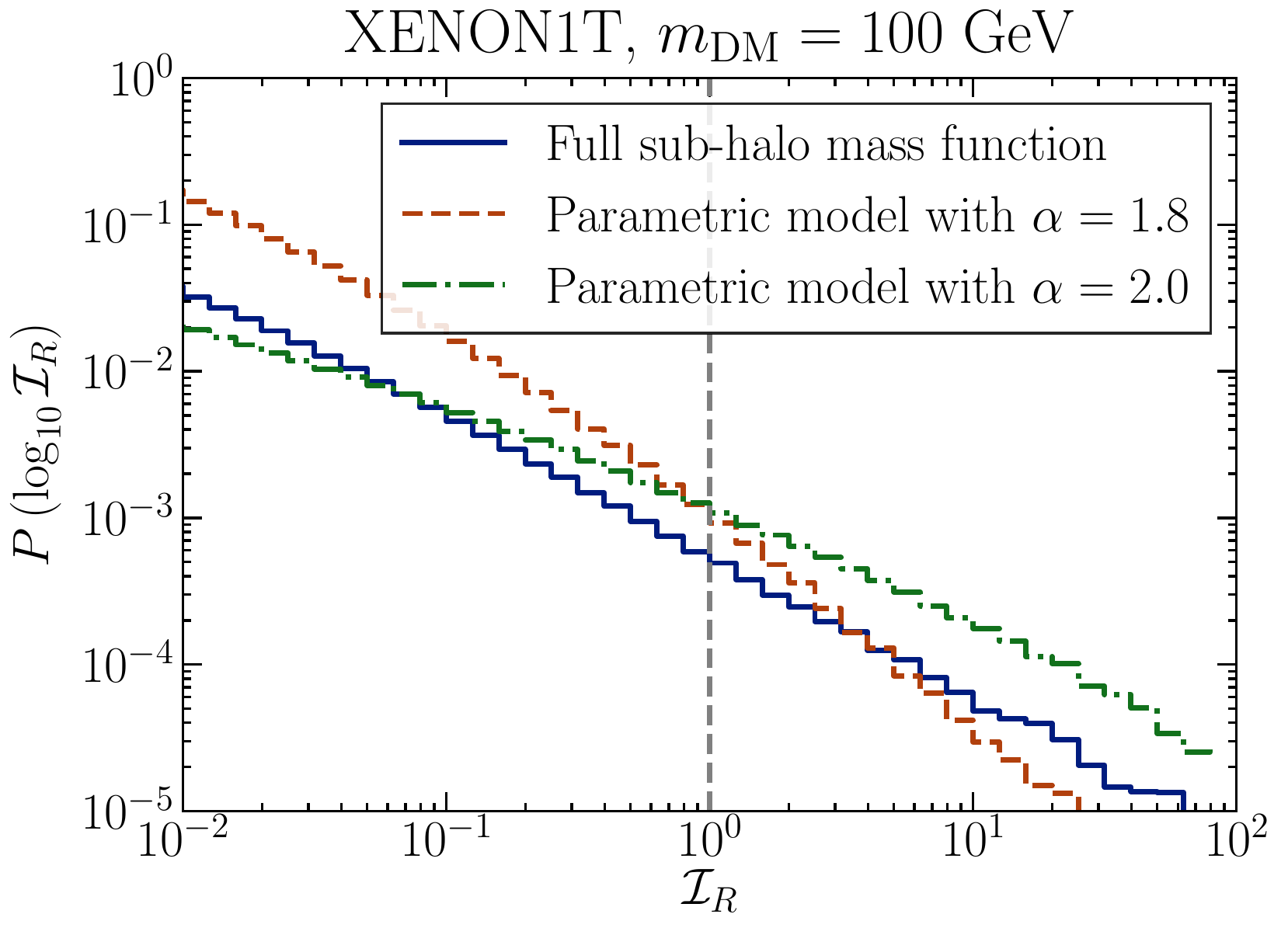}
		\includegraphics[width=0.49\textwidth]{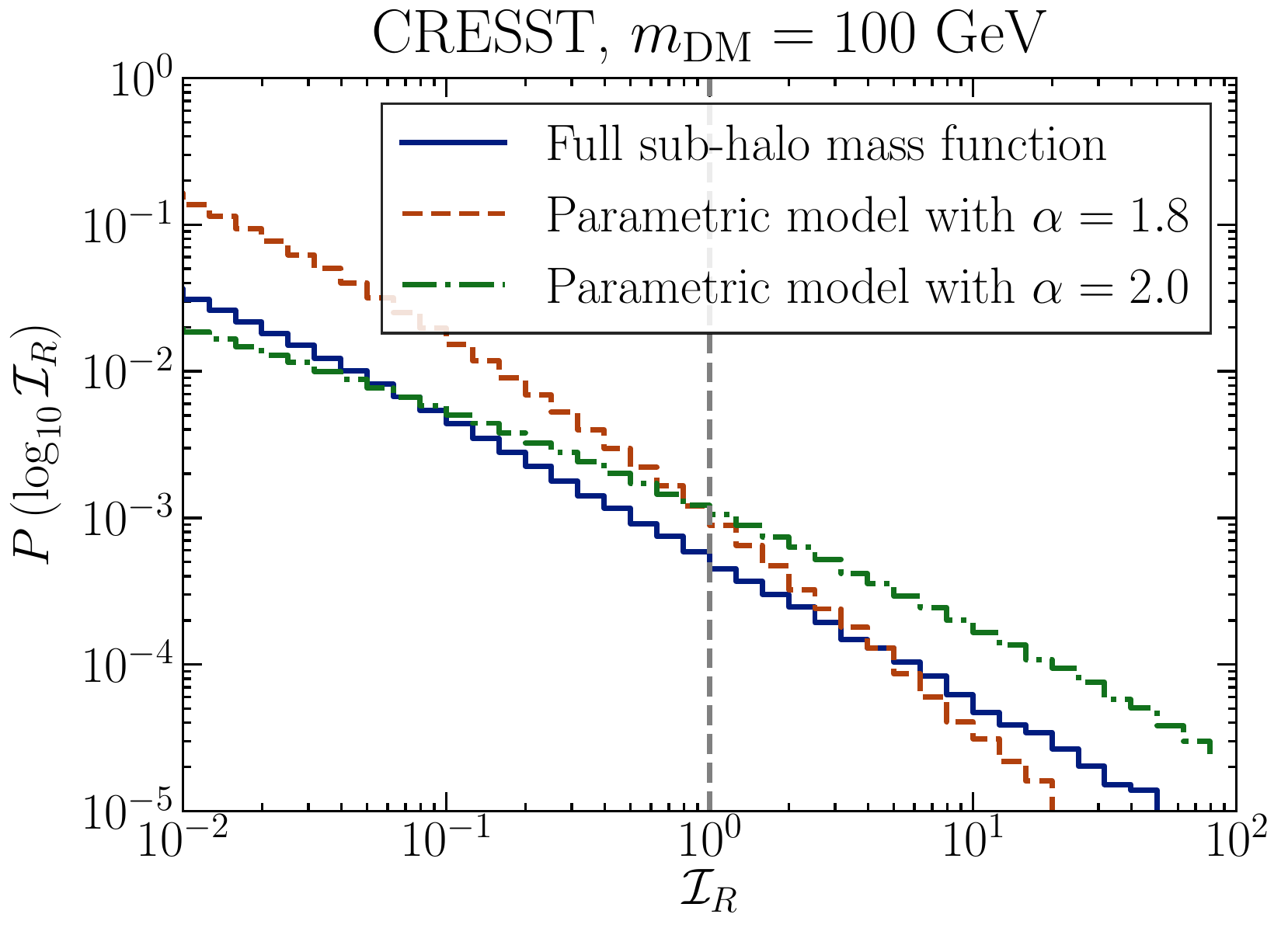}\\
		\hspace{-0.75cm}
		\includegraphics[width=0.49\textwidth]{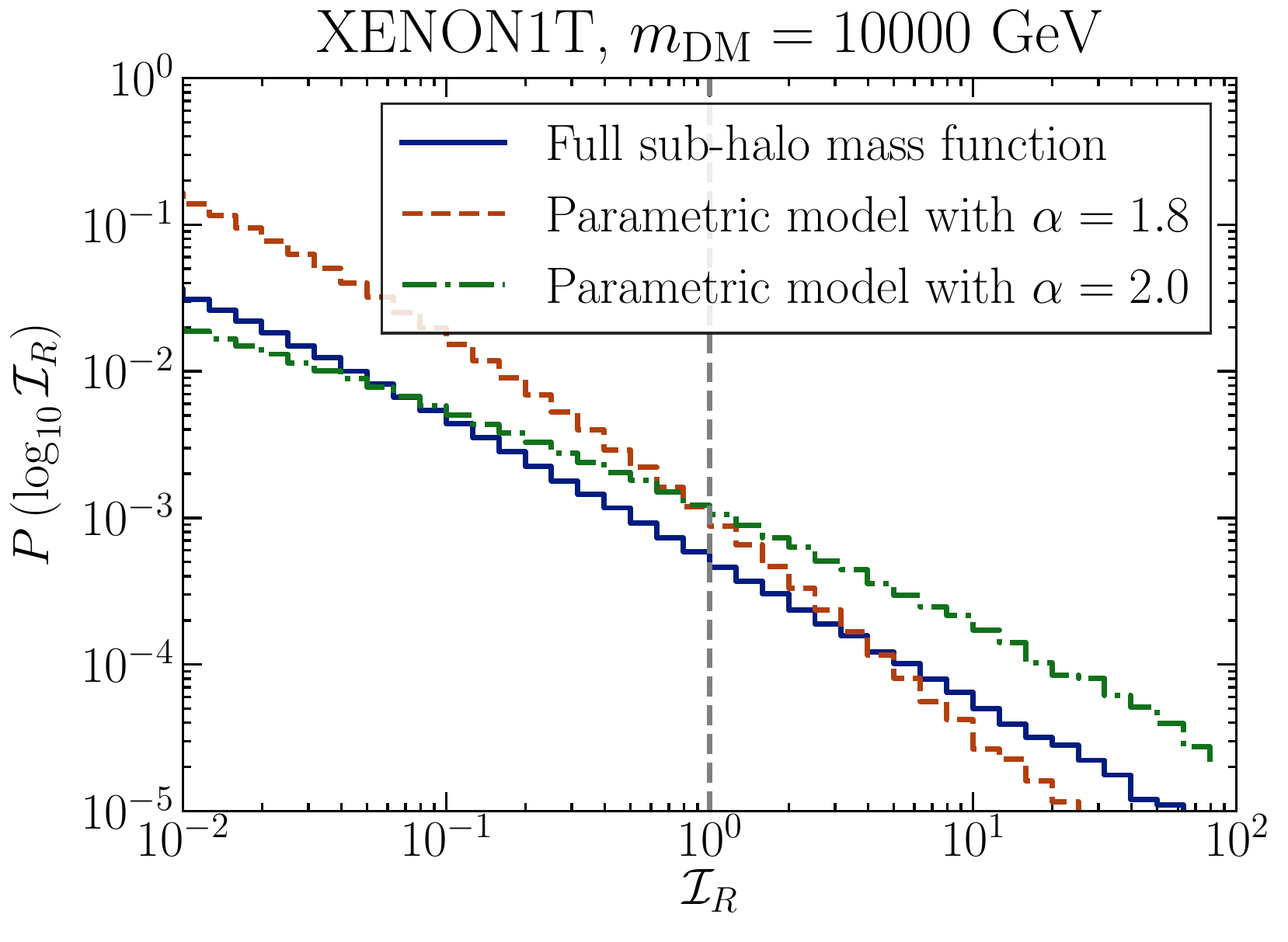}
		\includegraphics[width=0.49\textwidth]{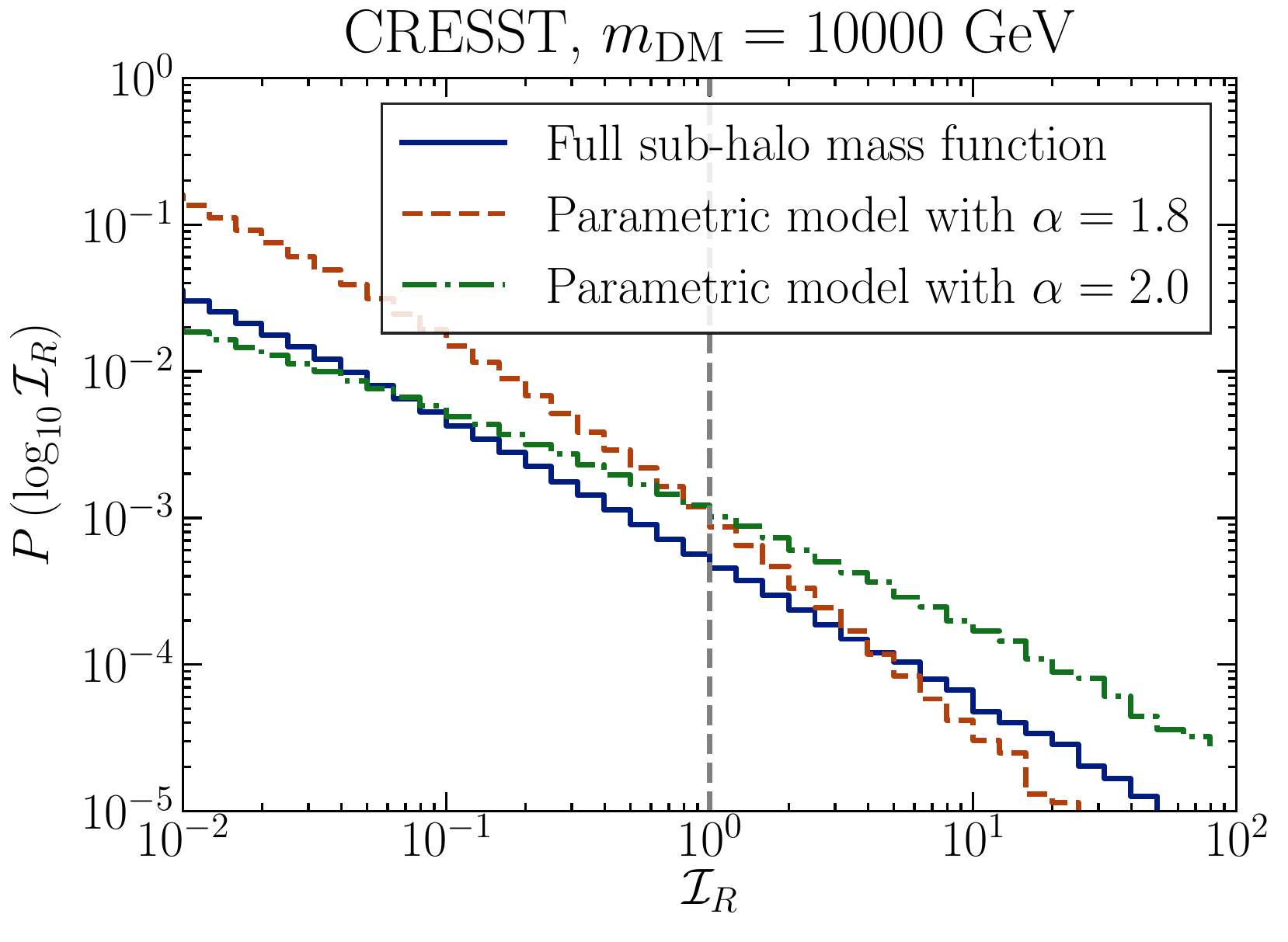}
	\end{center}
	\caption{\small Probability distributions of the increment in the scattering rate at XENON1T (left panel) and CRESST (right planel), for a dark matter mass of 5 GeV (top panel), 100 GeV (middle panel) and 10000 GeV (bottom panel).}
	\label{fig:DDboost}
\end{figure}

After determining the probability distributions for the sub-halo relative velocity and for finding an enhancement $\tilde \rho$ in the local dark matter density, it is straightforward to calculate the probability distribution of finding an increment ${\cal I}_R$ in the scattering rate at a given direct detection experiment. These probability distributions are shown as solid blue lines in figure~\ref{fig:DDboost}, for XENON1T (left panels) and CRESST (right panels), and for dark matter masses $m_{\rm DM}=5$ GeV (top panels), 100 GeV (middle panels) and 10 TeV (bottom panels). For both experiments, we find that there is a probability $\lesssim 7\times 10^{-4}$ of increasing the event rate by an ${\cal O}(1)$ factor. 

To assess the impact of our assumptions on the probability distributions, we have also analyzed the cases where the mass function follows a simpler power law $\propto M^{-1.8}$ or $\propto M^{-2.0}$, namely the upper and lower limits on $\alpha(M)$ in the right panel of figure~\ref{fig:Distribution}. For these cases, we find that the probability of finding a $\sim 1$ ($\sim 10)$ enhancement on the rate is at most a factor $\sim 3$ ($\sim 4$) larger than the one obtained with the full sub-halo mass function from Ref.\cite{Hiroshima:2018kfv}. We should also be careful that we are not `over-counting' the density enhancement by allowing multiple sub-halos to contribute to the density at the present time. We find that this physically invalid configuration (of multiple overlapping sub-halos) has only a small probability and so can be neglected:
 \begin{align}
p_2 &= \begin{pmatrix}\Nsh\\2\end{pmatrix}\,(p_\text{single})^2\,(1-p_\text{single})^{\Nsh-2}\approx\frac{1}{2}(\Nsh\,p_\text{single})^2\,p_0=0.7\%\,.
 \end{align}
 Lastly, let us note that the increment in the rate could be modified if the density profile of disrupted sub-halos is not described by an NFW profile, but rather by, for instance, an ``exponential profile'' of the form
\begin{align}
\rho(r)=\frac{\rho_0}{r^\gamma}\exp\left(-\frac{r}{\text{R}_b}\right),
\end{align}  
with parameters $\rho_0$, $\gamma$ and $\text{R}_b$, as advocated in~\cite{Hooper:2016cld, Kazantzidis:2003hb}. Repeating our analysis using a different halo profile would require, for consistency, the use of the corresponding halo-mass function calculated along the lines of \cite{Hiroshima:2018kfv}. Unfortunately, this analysis is not available in the literature. Accordingly, investigating how the form of the halo profile affects the increment in the scattering rate is beyond the scope of our work.

\section{Impact of sub-halos on the neutrino flux from the Sun}\label{sec:NT}

The idea that sub-structure can influence the capture rate of DM in the Sun was first introduced in \cite{Koushiappas:2009ee}. Here, we extend this idea and build a probabilistic model based on realistic sub-halo distributions to assess the boost factor of the neutrino signal from the Sun. We note that the same rationale we present here applies also for DM capture inside the Earth and the possibility of observing a high-energy neutrino flux from the center of the Earth \cite{Gould:1987ir,Green:2018qwo}.

Dark matter particles crossing the Sun can scatter with Solar matter and lose energy, occasionally becoming gravitationally bound to the Sun. Once captured, and due to subsequent scatterings, these particles will sink to the Solar core and generate a DM overdensity where the annihilation rate is enhanced, possibly leading to observable signals at neutrino telescopes \cite{Silk:1985ax}.
The capture rate of DM in the Sun can be determined from  \cite{Gould:1987ir, Gould:1991hx}:
\begin{align}
C(t)=\sum_i \int_0^{R_\odot} 4\pi\,r^2\,\text{d} r \, \eta_i(r)\,&\int_{v \leq v_{\text{max},i}^{\text{(Sun)}}(r)} \text{d}^3 v \, \frac{ F (\vec{v},t)}{ v^2}\,w^2(r) \int_{\mdm v^2 /2}^{2 \mu_{N_i}^2 w^2(r)/\mNi} \dER \, \frac{\text{d} \sigma_i}{\dER}(w(r), \ER) \,.
\label{eq:general_formula_capture_rate}
\end{align}
This rate is in general time-dependent, due to the time-dependent dark matter flux at the position of the Sun $F(\vec v,t)$ (as described by Eq.~\eqref{eq:flux}). In this expression, $w(r)\equiv\sqrt{v^2+v_\text{esc}(r)^2}$ is the speed of the DM particle at a distance $r$ from the Sun's center, given an asymptotic velocity $v$. The escape velocity from the Sun at radius $r$ is $v_\text{esc}(r)$, and $v_{\rm max}$ is the maximum velocity for which capture is kinematically possible from scatterings with a nucleus $i$ of mass $m_{N_i}$. In addition, $\eta_i(r)$ denotes the number density of the nuclei $i$ inside the Sun, for which we take the Solar model AGSS09 \cite{Serenelli2009}, and $\text{d} \sigma_i/\dER$ denotes the differential scattering cross section with a nucleus species $i$. This cross section depends on a nuclear form factor $F(\ER)$; for spin-independent interactions we adopt the Helm form factors from Ref.~\cite{Helm:1956zz,Lewin:1995rx} and for spin-dependent interactions, the form factors from Ref.~\cite{Catena:2015uha}.

The time evolution of the number of captured dark matter particles inside the Sun is described by the following differential equation
\begin{align}
\frac{\mathrm{d}N(t)}{\mathrm{d}t}=\text{C}(t)- C_A N(t)^2,\label{eq:DiffEqNsh}
\end{align}
where the effects of evaporation can be neglected for dark matter heavier than a few GeV \cite{Busoni:2017mhe}. Here, $C_A$ characterizes the rate at which the number of dark matter particles is depleted due to annihilation, and which we calculate following Refs.~\cite{Gould:1987ir, Griest:1986yu}. 

We begin by neglecting the effect of the sub-halos, in which case $C(t)=\CSHM$, the time-independent capture rate of DM from the smooth Milky Way dark matter distribution, described by the Standard Halo Model. Assuming that $N(0)=0$ at $t=0$, which we take as the time of formation of the Sun, the solution for the number of captured particles reads,
\begin{align}
	\label{eq:SolSHM}
	N(t)=\sqrt{\frac{\CSHM}{C_A}}\,\tanh\left(\frac{t}{\tau}\right)=N_\text{SHM}\,\tanh\left(\frac{t}{\tau}\right)\,,
\end{align}
which tends to a constant $N_\text{SHM}=\sqrt{\CSHM/C_A}$ when $t\gg \tau$.
Here, $\tau\equiv 1/\sqrt{\CSHM~C_A}$ denotes the equilibration time.
In this case, 
the annihilation rate inside the Sun today is given by:
\begin{align}
	\label{eq:AnnhRate}
	\Gamma_A=\frac{1}{2}\,C_A\,N(t_0)^2=\frac{\CSHM}{2}\,\tanh^2\left(\frac{t_0}{\tau}\right)\,,
\end{align}
where $t_0$ is the time elapsed since the formation of the Sun, $t_0\simeq 4.6~\text{Gyr}$. Assuming  $t_0\gg\tau$, the annihilation rate reduces to
\begin{align}
	\label{eq:AnnhEq}
	\Gamma_A=\frac{\CSHM}{2}\,.
\end{align}
That is, the annihilation rate is entirely determined by the capture rate, and ultimately to the interaction strength of DM particles with the nuclei in the Solar interior. This tight relation between annihilation rate and capture rate is often used to translate the non-observation of an excess of high energy neutrinos from the Sun into constraints on the spin-independent and spin-dependent cross sections.\footnote{In some frameworks, however, the condition $t_0\gg \tau$ may not hold (e.g.~in scenarios where the annihilation rate is p-wave suppressed \cite{Ibarra:2013eba}) and accordingly the annihilation rate gets suppressed by a factor $\tanh^2(t_0/\tau)$. }

In this work we focus on the possibility of a time-varying DM flux at the Sun induced by the passage through a sub-halo. The capture rate is therefore also time-varying and has the form:
\begin{align}
\text{C}(t)=\CSHM+\Csh(t)\Theta(t-t_-)\Theta(t_+-t)\,,
\end{align}
where $\Csh(t)$ describes the enhancement of the capture rate due to the Sun's passage through a single sub-halo between the times $t_-$ and $t_+$. 

To simplify the discussion, we approximate the time-dependent capture rate during the passage by its average value, namely
\begin{align}
\Csh(t)\simeq \langle \Csh(t)\rangle\equiv\frac{1}{\Delta t}\int_{t_-}^{t_+}{\rm d}t \, \Csh(t)\,,
\end{align} 
where $\Delta t = t_+ - t_-$.
Under this approximation, the solution of Eq.~(\ref{eq:DiffEqNsh}) reads:

\begin{align}
N(t)=\begin{cases}
\displaystyle{\sqrt{\frac{C_{\rm SHM}}{C_A}} \tanh \Big( \frac{t}{\tau}\Big)} & {\rm if}~t\leq t_- \,,\\
\displaystyle{
	\sqrt{\frac{C_{\rm SHM}}{C_A}} \frac{ \tanh \Big( \frac{t_-}{\tau}\Big)+\sqrt{1+\frac{\langle C_{\rm sh}(t)\rangle}{\CSHM}} \tanh \Big\{\sqrt{1+\frac{\langle C_{\rm sh}(t)\rangle}{\CSHM}} \frac{(t-t_-)}{\tau}\Big\}}
	{1+ \Big(\sqrt{1+\frac{\langle C_{\rm sh}(t)\rangle }{\CSHM}}\Big)^{-1}\tanh \Big( \frac{t_-}{\tau}\Big) \tanh \Big\{\sqrt{1+\frac{\langle C_{\rm sh}(t)\rangle }{\CSHM}} \frac{(t-t_-)}{\tau}\Big\}}
}& {\rm if}~t_-<t\leq t_+ \,,\\
\displaystyle{\frac{N(t_+)+\sqrt{\frac{C_{\rm SHM}}{C_A}}\tanh\Big(\frac{t-t_+}{\tau}\Big)}{1+N(t_+)\Big(\sqrt{\frac{C_{\rm SHM}}{C_A}}\Big)^{-1}\tanh\Big(\frac{t-t_+}{\tau}\Big)
		\Big]}
}& {\rm if}~t>t_+\,,
\end{cases}
\label{eq:analytic_solution}
\end{align}
where $\tau$ is the equilibration time in the absence of time-dependent contributions. We note that the impact of the passage of the Sun through the sub-halo can be characterized by only three dimensionless quantities: the duration of the passage relative to the equilibration time, $\Delta t/\tau$, the time elapsed since the Sun abandoned the sub-halo relative to the equilibration time $(t_0-t_+)/\tau$, and the time-averaged increment in the capture rate due  to the passage through the sub-halo, $\langle C_{\rm sh}(t)\rangle/C_{\rm SHM}$.

\begin{figure}[!t]%
	\begin{center}
		\hspace{-0.75cm}
		\includegraphics[width=0.49\textwidth]{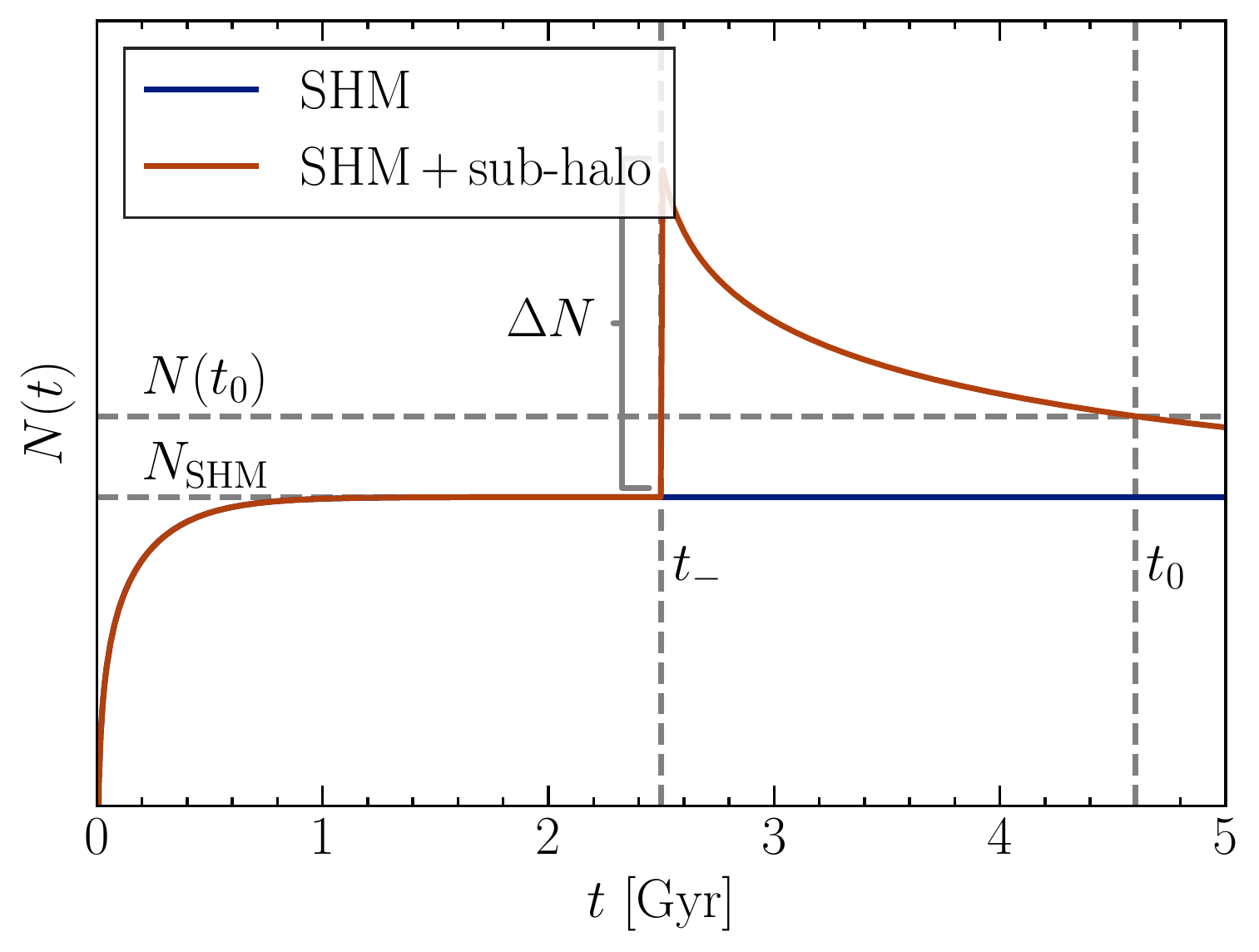}
		\includegraphics[width=0.49\textwidth]{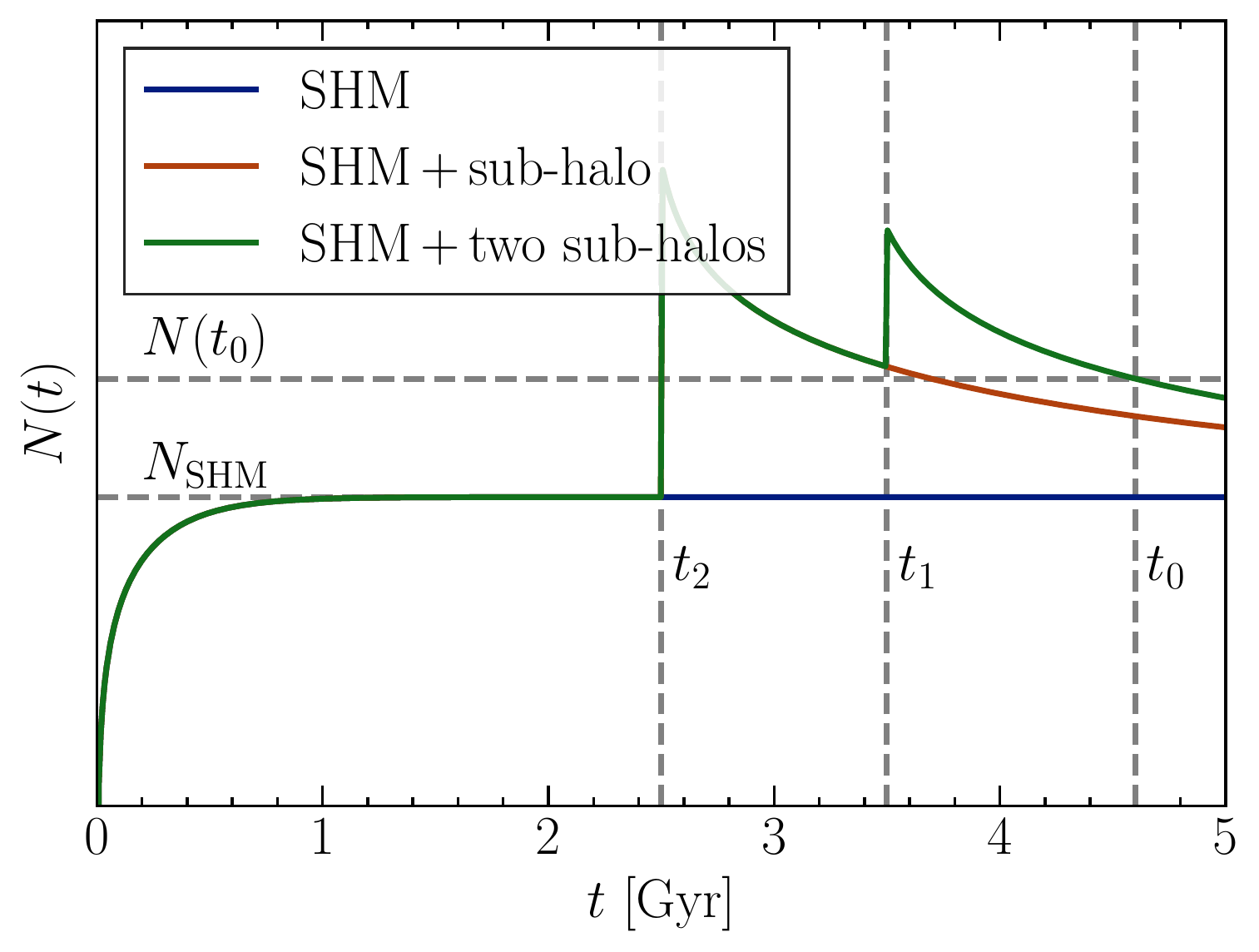}
	\end{center}
	\caption{\small Sketch of the time evolution of the number of captured dark matter particles $N(t)$, and its value at the present time $N(t_0)$, assuming a passage of the Sun through a single sub-halo at the time $t_-$ (left plot) and assuming successive passages through two sub-halos at times $t_2$ and $t_1$ (right plot). For comparison, we also show the corresponding values of $N(t)$ and $N(t_0)$ for the Standard Halo Model.}	
	\label{fig:sketch_Nt}
\end{figure}

The two relevant parameters $\Delta t/\tau$ and $\langle C_{\rm sh}(t)\rangle/C_{\rm SHM}$ are in turn calculable  given a particle DM framework, given the characteristics of the smooth halo and sub-halo components, and given the impact parameter of the Sun when entering the sub-halo. More concretely, for an impact parameter $L$, and under the approximation that the Sun moves inside the sub-halo following a straight line with constant speed $v_{\rm rel}$, it can be shown that the Sun travels  in a complete passage a distance $\Delta d=2\sqrt{r_t^2-L^2}$ in a time $\Delta t=2\sqrt{r_t^2-L^2}/v_{\rm rel}$. Therefore,  $\Delta t/\tau\leq 2r_t/v_{\rm rel}\tau$. For $v_{\rm rel}={\cal O}(100)$ km/s, and taking conservatively $\tau\gtrsim 10^6$ years, one obtains $v_{\rm rel}\tau\gtrsim 30$ pc. Assuming that the sub-halos crossed by the Sun are much smaller than 30 pc \footnote{This is typically true for the most abundant sub-halos, with masses below $M \lesssim 10^{3}\,M_\odot$.}, so that $\Delta t/\tau\ll 1$, the solution simplifies to:
\begin{align}
\begin{split}
N(t)  &\simeq N_{\rm SHM}
\displaystyle{
	\Big[\frac{(N(t_-)+\Delta N)/N_\text{SHM}+\tanh\frac{t-t_-}{\tau}}{1+(N(t_-)+\Delta N)/N_\text{SHM}\cdot\tanh\frac{t-t_-}{\tau}}\Big]}\,.
\end{split}
\end{align}
Here, $N(t_-)$ is the number of captured particles just before entering the sub-halo and $N_\text{SHM} = \sqrt{C_\mathrm{SHM}/C_\mathrm{A}}$ is the equilibrium number of captured particles in the SHM. We also define $\Delta N$ as
\begin{align}
\Delta N\equiv \langle \Csh(t)\rangle\,\Delta t\,,
\end{align}
which is interpreted as the increase in the number of DM particles captured during the passage: $\Delta N\equiv N(t_+)-N(t_-)$. The time evolution of the number of captured dark matter particles is sketched in figure~\ref{fig:sketch_Nt}. From this it is apparent that the number of captured particles, and accordingly the neutrino flux from annihilations, can be significantly enhanced if the passage through the sub-halo was recent, $(t_0-t_-)/\tau\ll 1$, and if the number of particles captured during the passage was large $\Delta N\gg N(t_-)$.  A recent passage through the sub-halo suggests that capture and annihilation were in equilibrium inside the Sun before entering the sub-halo, therefore, it is plausible that $N(t_-)=N_{\rm SHM}$. 

\begin{figure}[!t]
	\begin{center}
		\hspace{-0.75cm}
		\includegraphics[width=0.49\textwidth]{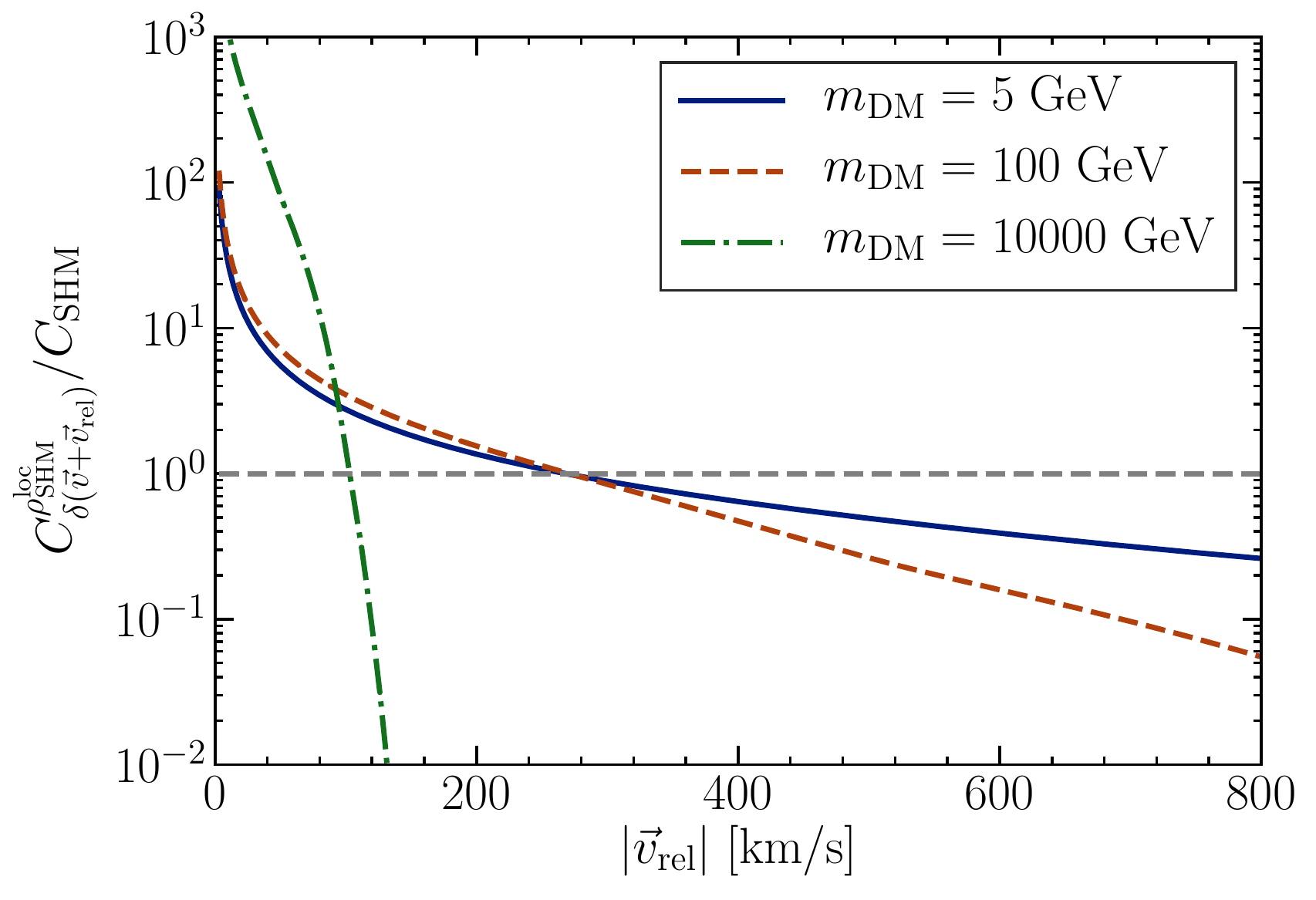}
	\end{center}
	\caption{\small Capture rate at the Sun for dark matter streams with velocity  $|\vec{v}_\mathrm{rel}|$, relative to the capture rate assuming the Standard Halo Model, for dark matter masses $m_{\rm DM}=5$, 100 and 10000 GeV.}
	\label{fig:C_streams}
\end{figure}

The time-averaged increment in the capture rate can be calculated inserting Eq.~(\ref{eq:flux}) into Eq.~(\ref{eq:general_formula_capture_rate}). For small mass sub-halos, for which the velocity distribution in the Solar frame can be approximated by Eq.~(\ref{eq:f_delta}), one finds
\begin{align}
\Delta N\equiv  C^{\rho_{\rm SHM}}_{\delta(\vec v+\vec v_{\rm rel})}\frac{\langle \rho^{\rm loc}_{\rm sh}[r(t)]\rangle}{\rho_{\rm SHM}}\Delta t\,,
\end{align}
where $C^{\rho_{\rm SHM}}_{\delta(\vec v+\vec v_{\rm rel})}$ denotes the capture rate for a DM stream with density $\rho_{\rm SHM}$ and velocity $\vec v_\text{rel}=(\vec v_\odot-\vec v_{\rm sh})$ with respect to the Solar frame, shown in figure~\ref{fig:C_streams} relative to $\CSHM$. The time-averaged quantity $\langle \rho^{\rm loc}_{\rm sh}[r(t)]\rangle$ is related to the average mass density observed by the Sun along its path. For impact parameter $L$, and assuming that the Sun moves inside the sub-halo with constant velocity $v_{\rm rel}$, one finds
 \begin{align}
\langle \rho^{\rm loc}_{\rm sh}[r(t)]\rangle  \simeq 
\frac{1}{\Delta t} \int_{t_-}^{t_+}{\rm d}t\,\rho^{\rm loc}_{\rm sh}[r(t)]=
\frac{1}{2 \sqrt{r_t^2-L^2}}\int_{-\sqrt{r_t^2-L^2}}^{\sqrt{r_t^2-L^2}}{\rm d}x\,\rho^{\rm loc}_{\rm sh}[\sqrt{x^2+L^2}]\,,
\label{eq:av_rho}
\end{align}
where we have used that $\Delta t=2\sqrt{r_t^2 -L^2}/v_{\rm rel}$. For a sub-halo with an NFW profile, Eq.~(\ref{eq:rho_sh}), we find,
\begin{align}
\langle \rho^{\rm loc}_{\rm sh}[r(t)]\rangle=\frac{\rho_s }{(1-c_V^2 z^2)}\left[
\frac{2}{c_V\sqrt{1-z^2}\sqrt{1-c_V^2 z^2}}
\artanh\left(\frac{\sqrt{1-z^2}\sqrt{1-c_V^2 z^2}}{(1+z)(1+c_V z)}\right)-\frac{1}{1+c_V}\right]\,,
\label{eq:av_rho2}
\end{align}
with $z\equiv L/r_t$. In certain limiting cases, this expression can be simplified to:
\begin{align}
\langle \rho^{\rm loc}_{\rm sh}[r(t)]\rangle  \simeq 
\begin{cases}
\displaystyle{
\frac{\rho_s}{c_V(1+c_V)^2}\left[1+\frac{2}{3}\left(\frac{1+3c_V}{1+c_V}\right)\left(1-\frac{L}{r_t}\right)\right]}
	 &~{\rm for~}r_t>L\gg r_s \,,\\
\displaystyle{-\frac{\rho_s}{c_V}\left[\log\frac{L}{r_s}+\frac{c_V}{1+c_V}-\log\frac{c_V}{1+c_V}-\log 2 \right]}
		 &~{\rm for~}L\ll r_s\,.
\end{cases}
\label{eq:av_rho3}
\end{align}
The mean density diverges for $L\rightarrow 0$, as expected for a cuspy profile, which in turn translates into a large enhancement in the capture rate if the Sun traverses the innermost part of the sub-halo. On the other hand, typical sub-halos have a size larger than the scale radius, since the concentration parameter takes values $c_V\sim 1-35$. Therefore, in most instances one expects $L\gg r_s$.

\begin{figure}[!t]
	\begin{center}
		\hspace{-0.75cm}
		\includegraphics[width=0.49\textwidth]{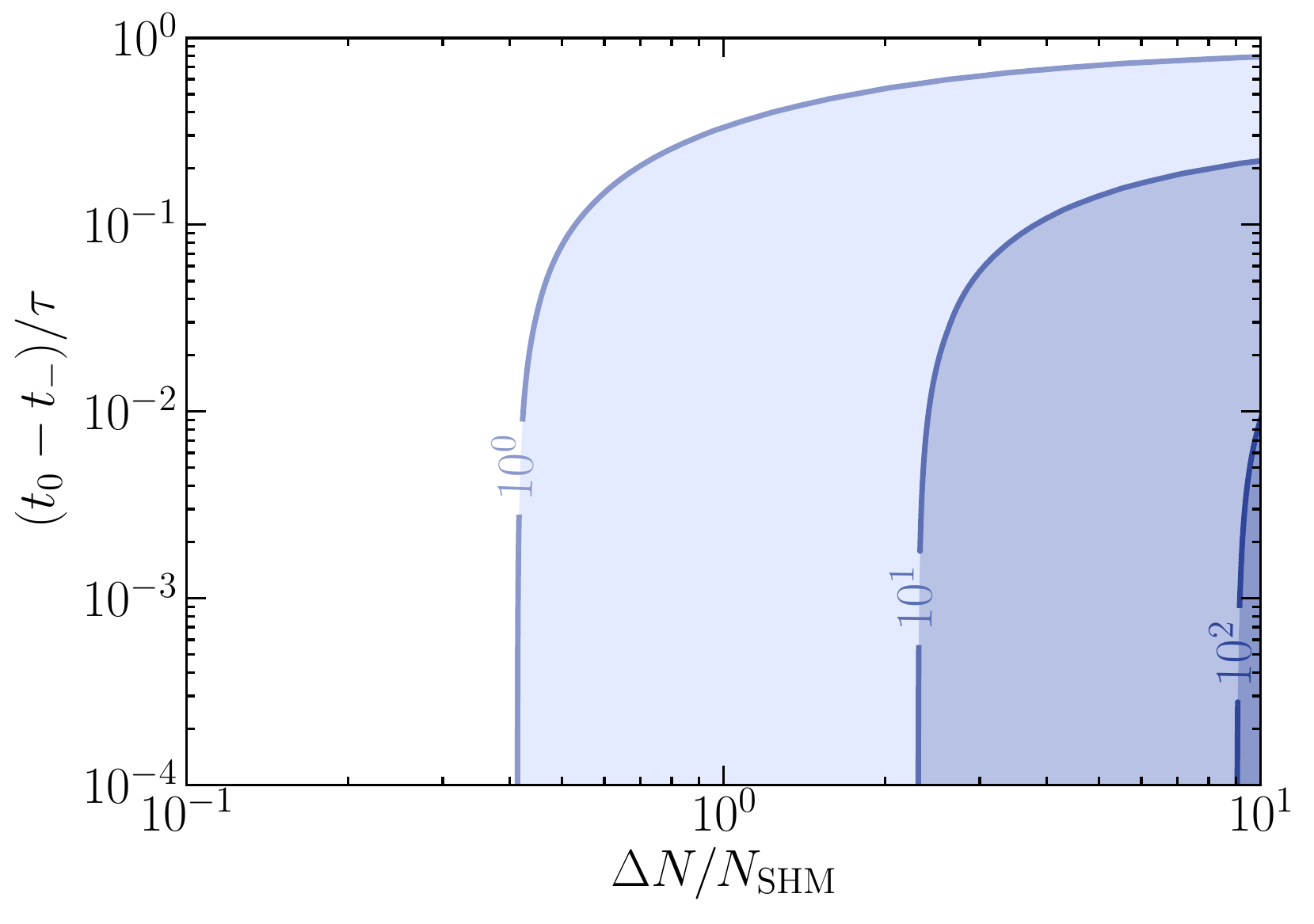}
	\end{center}
	\caption{\small Increment in the annihilation rate $\mathcal{I}_\Gamma$ relative to the expectations from the Standard Halo Model in the parameter space spanned by $(t_0-t_{-})/\tau$ (the time elapsed since the passage of the Sun through the sub-halo, relative to the equilibration time) and $\Delta N/N_\text{SHM}$ (the number of particles in the sub-halo captured during the passage, relative to the number of particles in the SHM when equilibration is reached).
	} 
	\label{fig:IGamma}
\end{figure}

Finally, mirroring our discussion in section~\ref{sec:DD},  we define the increment in the annihilation rate inside the Sun  (and in the neutrino flux) due to the passage through the sub-halo as:
\begin{align}
{\cal I}_\Gamma&=\frac{\Gamma(t_0)}{\Gamma_{\rm SHM}(t_0)}-1\simeq
\displaystyle{
\Big[\frac{(N(t_-)+\Delta N)/N_\text{SHM}+\tanh\frac{t_0-t_-}{\tau}}{1+(N(t_-)+\Delta N)/N_\text{SHM}\tanh\frac{t_0-t_-}{\tau}}\Big]^2\,\tanh\left(\frac{t_0}{\tau}\right)^{-2}-1}\,,
\label{eq:def_IGamma}
\end{align}
where the annihilation rate expected from the smooth Milky Way dark matter distribution is given by $\Gamma_\text{SHM}(t_0)=\frac{1}{2}\CSHM\tanh(t_0/\tau)$.
We show in figure \ref{fig:IGamma} contours of ${\cal I}_\Gamma$ in the parameter space of $\Delta N$ and $(t_0-t_{-})/\tau$, assuming for simplicity that capture and annihilation were in equilibrium before entering the sub-halo, so that $N(t_{-})=N_\text{SHM}$ and $\Gamma_{\rm SHM}(t_{-})=\frac{1}{2}\CSHM$.
We furthermore assume equilibrium for the smooth Milky Way dark matter halo, i.e. $\Gamma_\text{SHM}(t_0)=\frac{1}{2}\CSHM$.
As expected from our previous discussion, the largest enhancement occurs for recent passages and for $\Delta N\gg N(t_-)$. This occurs especially if the sub-halo is very dense; if the impact parameter is small; or if the speed of the Sun relative to the sub-halo is small. We remark that for DM scenarios where the number of captured DM particles from the smooth halo is below the equilibrium value ({\it i.e.} when $\tau\gg t_0$), the passage of the Sun through one or more sub-halos can significantly enhance the total number of captured DM particles.

As in section \ref{sec:DD}, one can estimate the probability of having a certain increment ${\cal I}_\Gamma$ in the annihilation rate inside the Sun.
The probability for finding a certain value of $\Delta N$ from the passage of the Sun through a sub-halo of mass $M$, concentration $c_V$, and velocity $\vec v_{\rm rel} = \vec v_\odot - \vec v_\mathrm{sh}$ is related to the probability of a passage with impact parameter $L$ by
\begin{align}
P(\Delta N,M,c_V,v_{\rm rel})=P(L ,M,c_V,v_{\rm rel}) \left|\frac{{\rm d} \Delta N(L, M, c_V,v_{\rm rel})}{{\rm d}L}\right|^{-1}~~~{\rm for~} L<r_t\,.
\end{align}
Note that $\Delta N$ depends on $L$ through $\langle \rho^{\rm loc}_{\rm sh}[r(t)]\rangle$, given in Eqs.~(\ref{eq:av_rho},\ref{eq:av_rho2}), and through $\Delta t\equiv 2\sqrt{R_{\rm vir}^2 -L^2}/v_{\rm sh}$. The probability for the Sun traversing a sub-halo with impact parameter $L$ can be calculated using a similar rationale as in section \ref{sec:DD} (c.f.~Eq.~\eqref{eq:PDMcV}): 
\begin{align}
P(L,M,c_V,v_{\rm rel})= P(M) P(c_V|M) P(v_{\rm rel}) P(L|M,c_V,v_{\rm rel})\,.
\label{eq:PLMcV}
\end{align}
Here, P$(M)$ and  $P(c_V|M)$ were already discussed in section \ref{sec:DD},  and $P(v_{\rm rel})$ is given by the Maxwell-Boltzmann distribution in Eq.~\eqref{eq:f_MB_Sol} \footnote{With the caveat that we are not interested in the \textit{direction} of the sub-halo velocity, only its magnitude, so that in practice we draw $|\vec v_\mathrm{rel}|$ from the distribution $f(v) = v^2 \int f_\mathrm{SHM}(\vec{v})\,\mathrm{d}\Omega_v$.}. To determine $P(L|M,c_V, v_{\rm rel})$ we note that the probability of the Sun passing through a sub-halo with impact parameter $L$ is equal to the probability of finding a sub-halo center at a perpendicular distance $L$ from the Sun's path during the crossing time $\Delta t$.  The number of sub-halos in the cylindrical shell of width ${\rm d}L$ located at a perpendicular distance $L$ from the Sun and with length equal to $v_{\rm rel} \Delta t$ is $\mathrm{d}N_{\rm sh}(L,M,c_V,v_{\rm rel})\simeq 2 \pi L \,\mathrm{d}L \,v_{\rm rel} \Delta t \, \bar n_{\rm sh}(L)$. Assuming, as we did in section \ref{sec:DD}, that the spatial distribution of sub-halos is independent of their mass and concentration, one then obtains
\begin{align}
P(L|M,c_V,v_{\rm rel})\equiv \frac{1}{N_{\rm sh}}\frac{\mathrm{d}N_{\rm sh}(L,M,c_V,v_{\rm rel})}{\mathrm{d}L}\simeq \frac{2\pi L v_{\rm rel} \Delta t\,\bar n_{\rm sh}(L)}{N_{\rm sh}}\,.
\end{align}
Here, $N_{\rm sh}$ the total number of sub-halos and $\bar n_{\rm sh}(L)$ the average number of sub-halos at a perpendicular distance $L$, written in terms of the galactocentric radius $r$ as:\footnote{Note the slightly different definition compared to Eq.~\eqref{eq:nbar}, due to the different geometry.}
\begin{align}
\bar{n}_\mathrm{sh}(L) = \frac{1}{2\pi}\int_{0}^{2\pi} n_\mathrm{sh}\left[r(L, \psi)\right]\,\mathrm{d}\psi\,.
\end{align}

Finally, one obtains the probability distribution of finding a certain $\Delta N$ from the passage through a single sub-halo by integrating over all possible sub-halo masses, concentration parameters and velocities:
\begin{align}
P_{\rm single}(\Delta N)&=\int_{M_{\rm min}}^{M_{\rm max}} \mathrm{d}M
\int_{0}^{\infty} \,\mathrm{d}c_V
\int_0^{v_{\rm max}} \,\mathrm{d}v_{\rm rel} P(\Delta N, M,c_V, v_{\rm rel})
\end{align}

\begin{figure}[!pth]
	\begin{center}
		\hspace{-0.75cm}
		\includegraphics[width=0.49\textwidth]{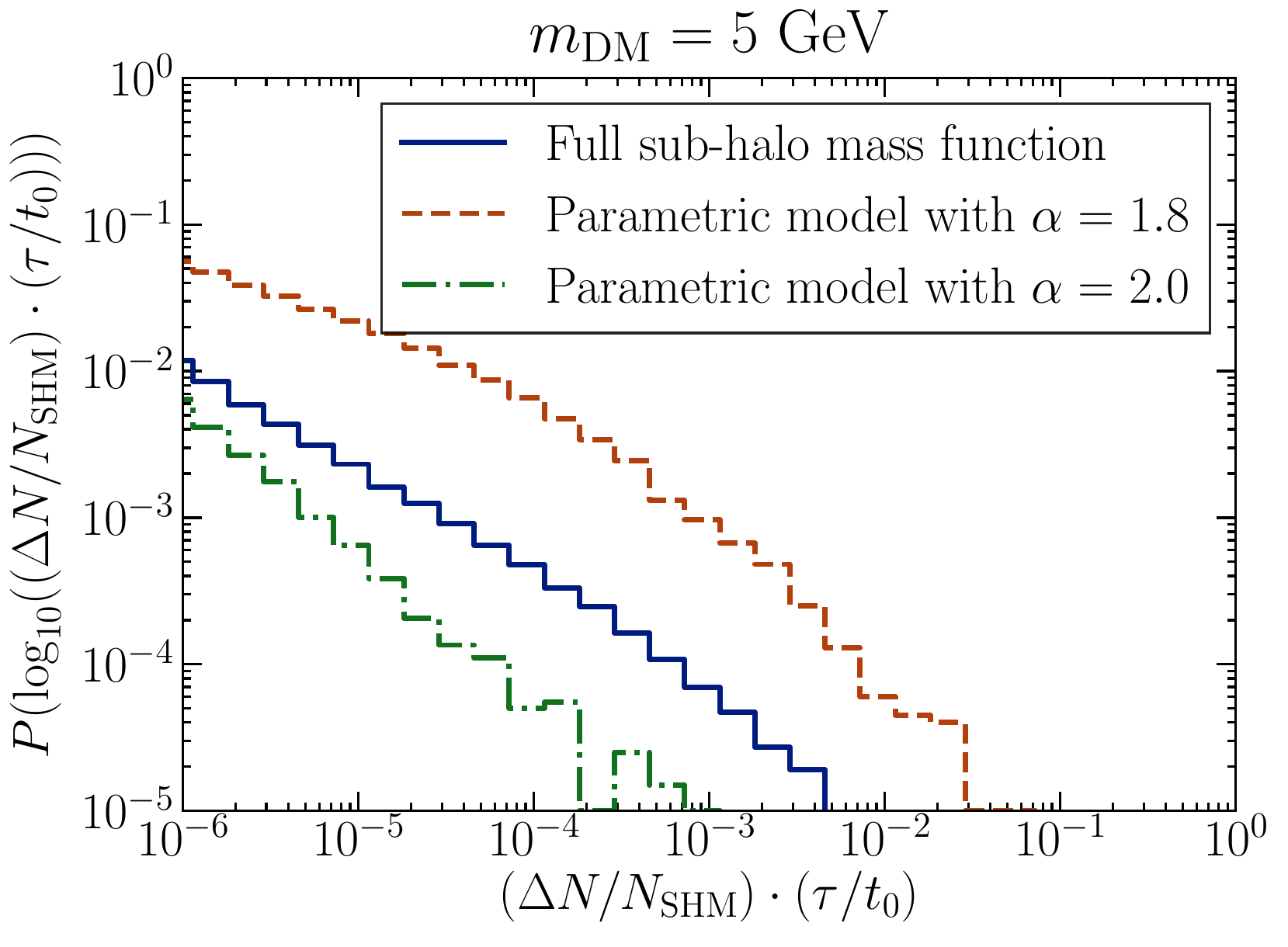}
		\includegraphics[width=0.49\textwidth]{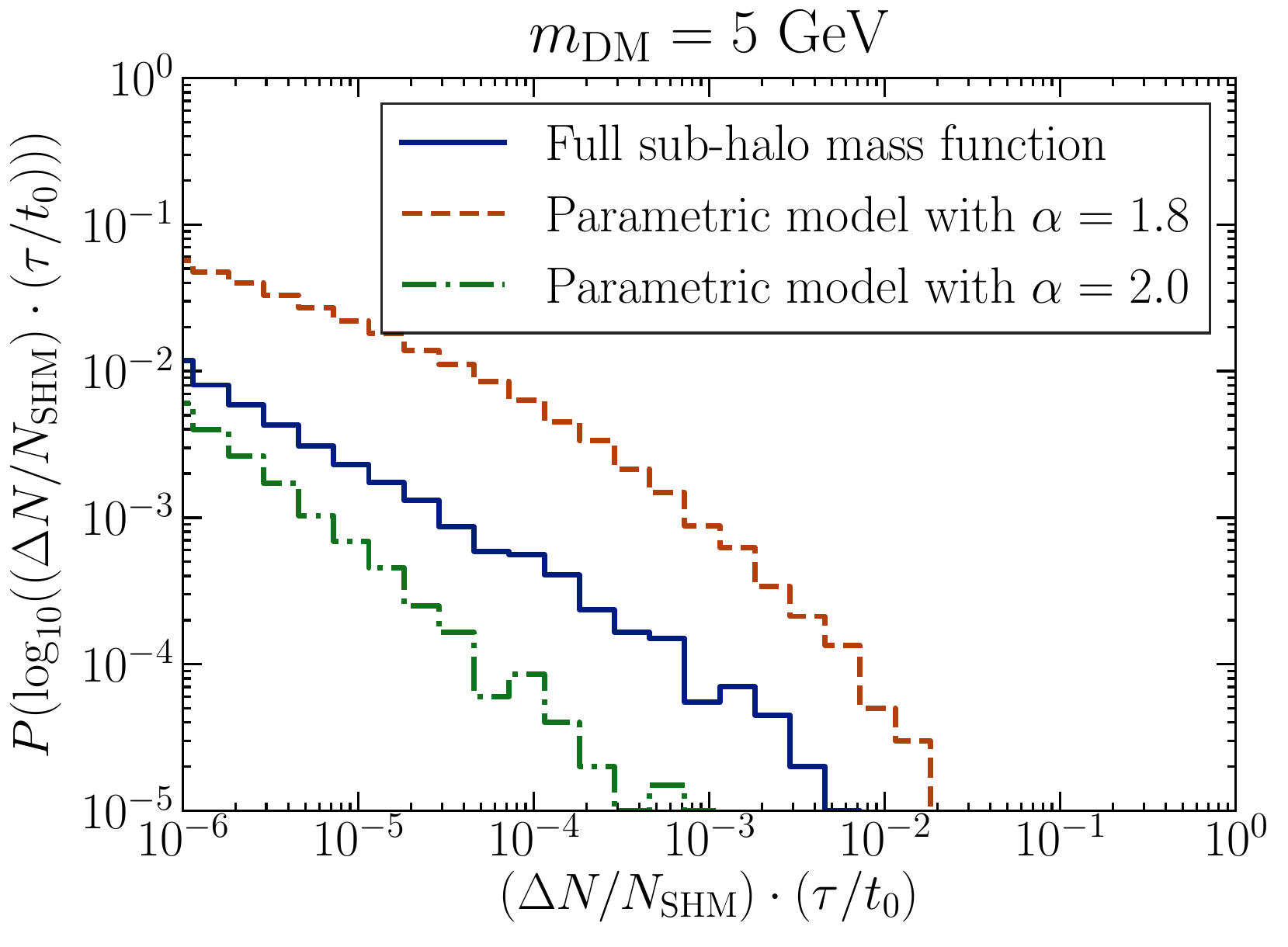}\\
		\hspace{-0.75cm}
		\includegraphics[width=0.49\textwidth]{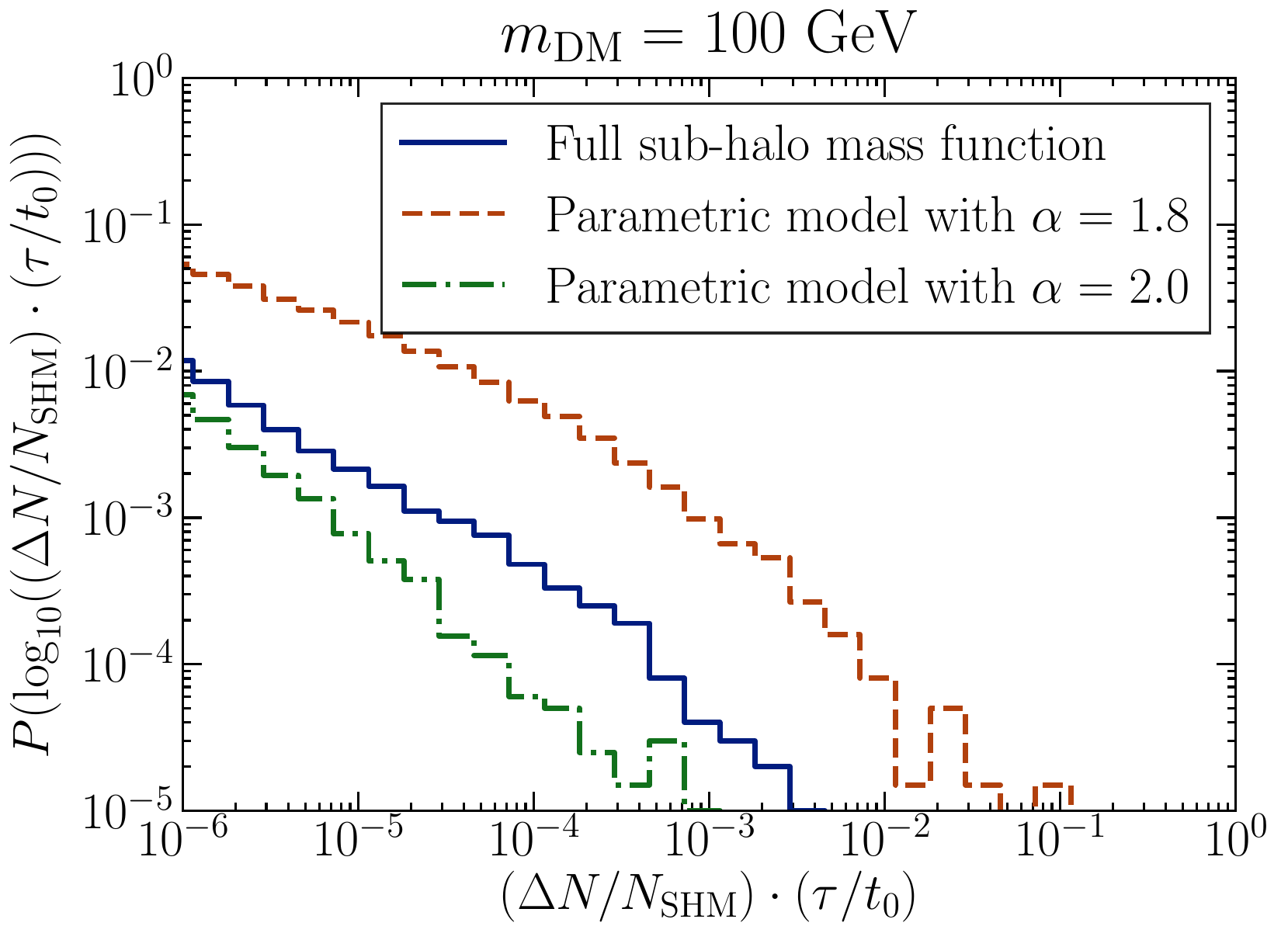}
		\includegraphics[width=0.49\textwidth]{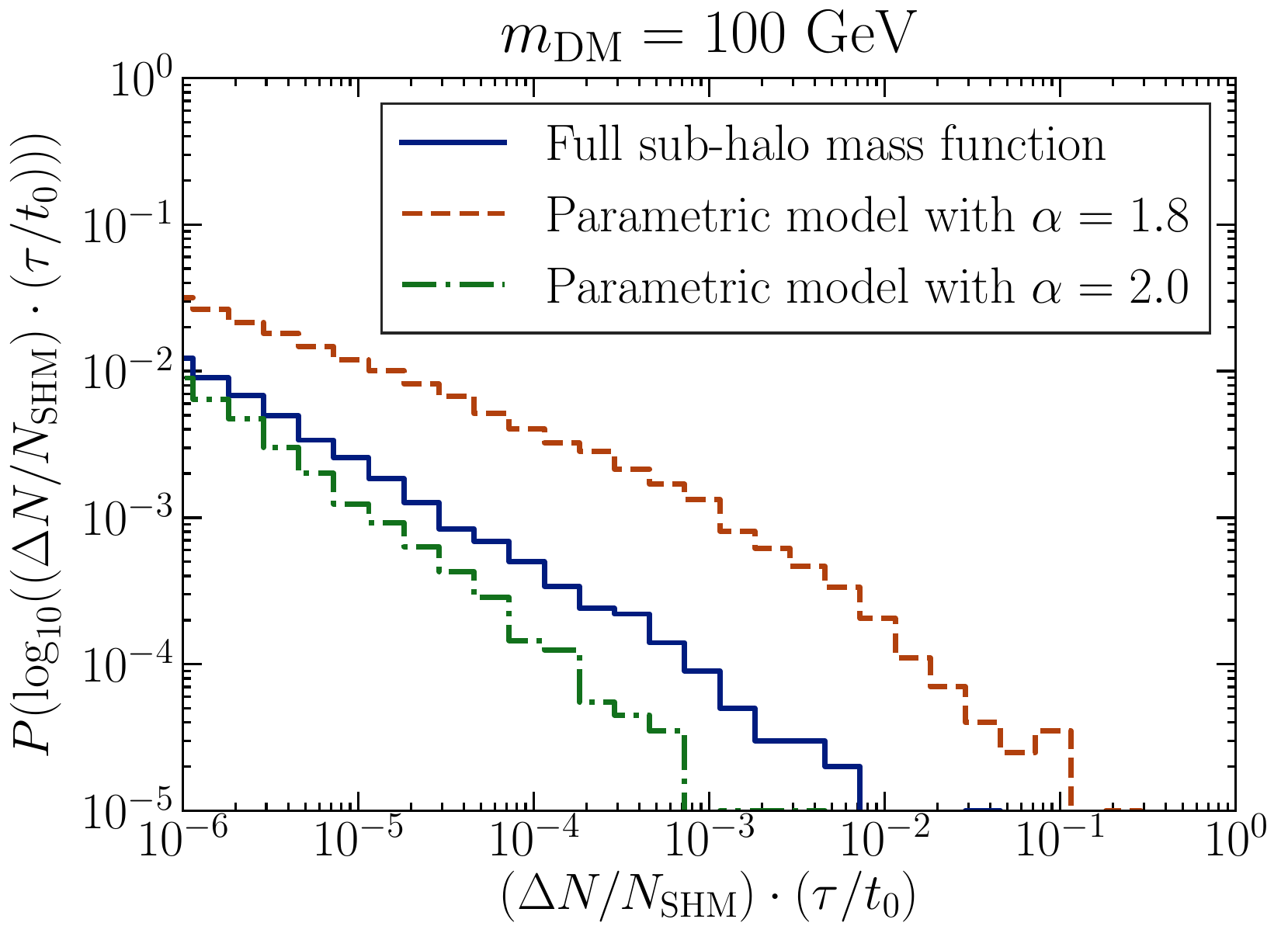}\\
		\hspace{-0.75cm}
		\includegraphics[width=0.49\textwidth]{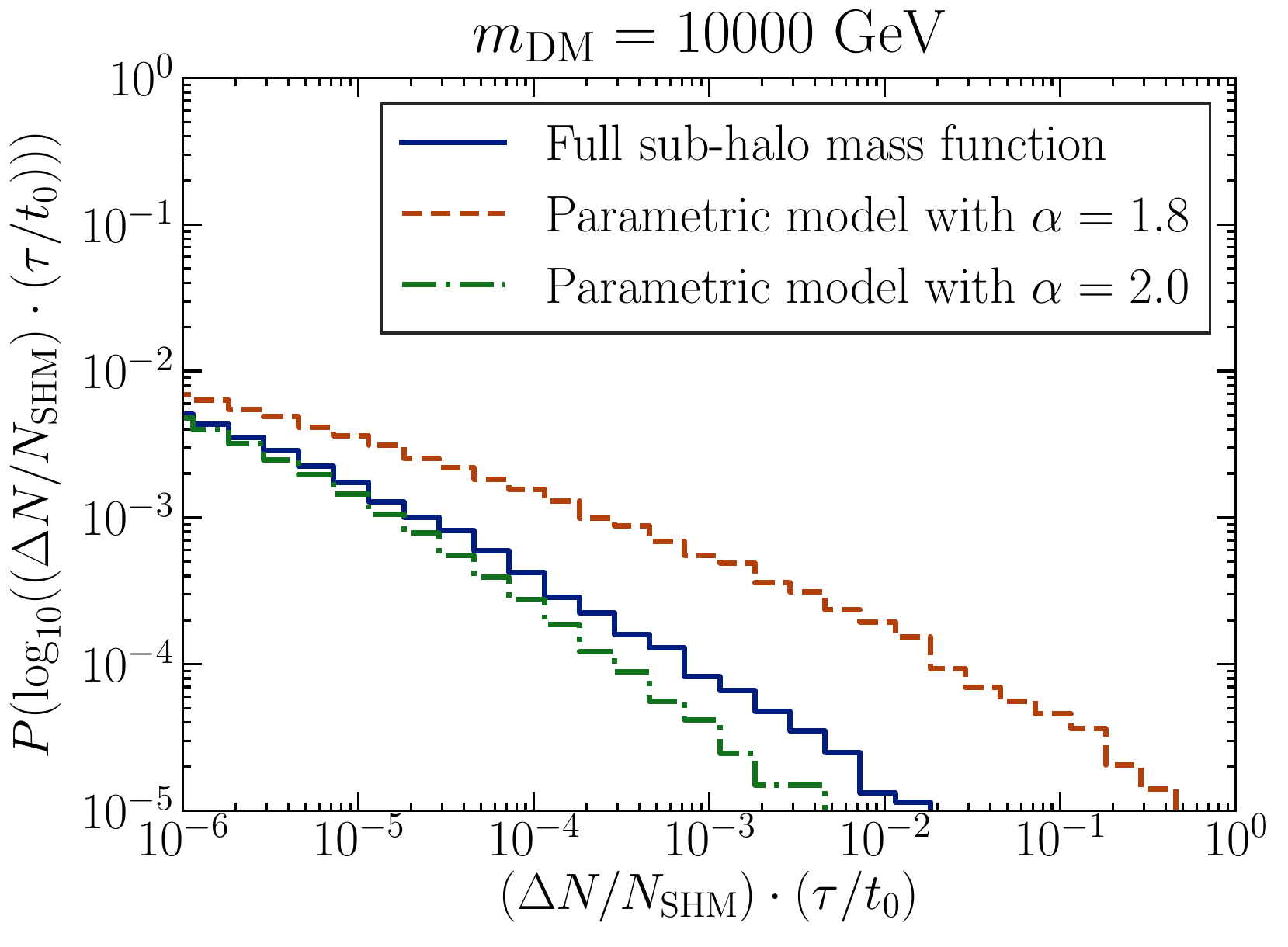}
		\includegraphics[width=0.49\textwidth]{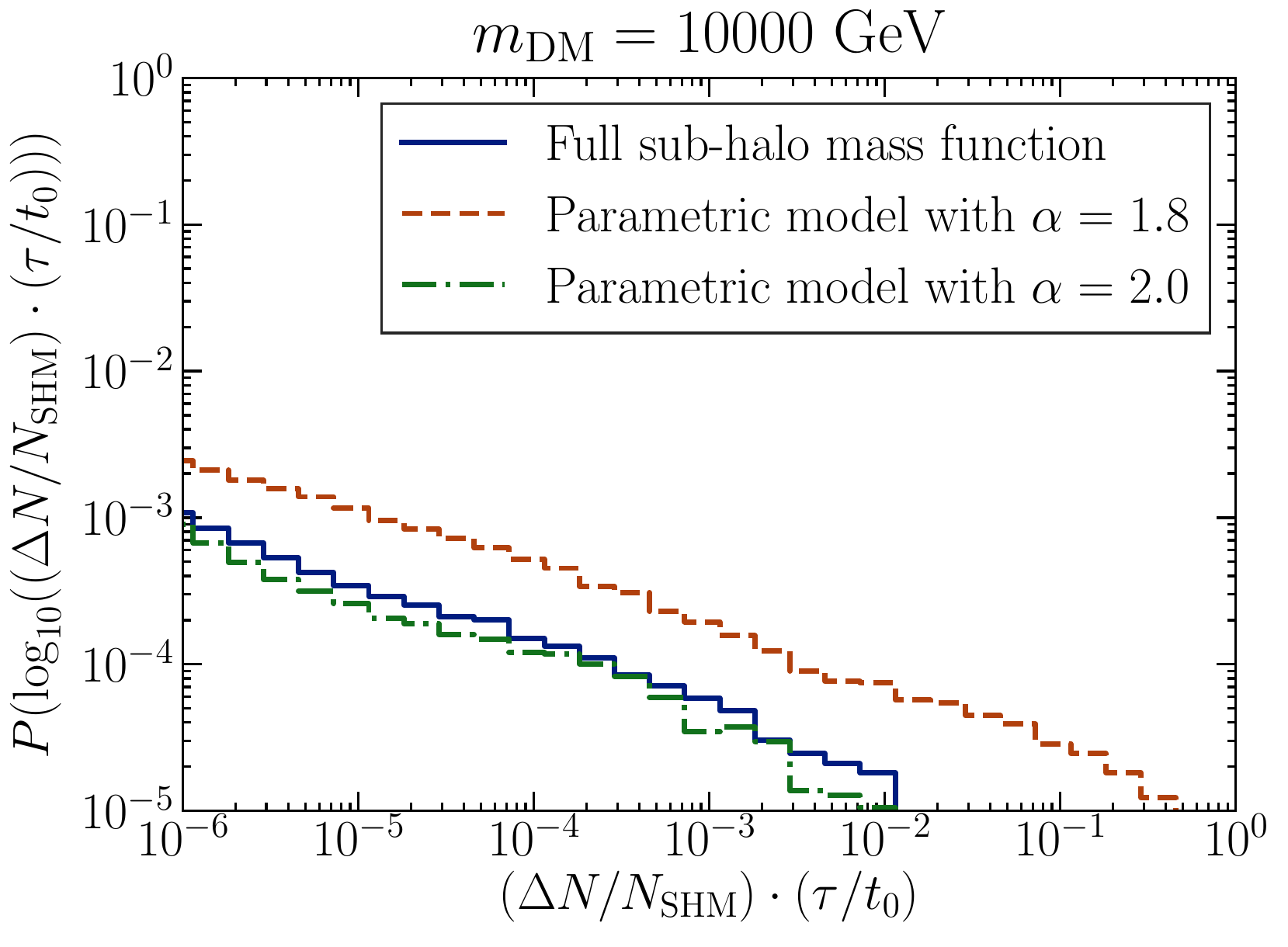}
	\end{center}	
	\caption{\small Probability distribution of the dimensionless and cross section independent combination of parameters $(\Delta N/N_\text{SHM})\cdot(\tau/t_0)$, assuming the spin-independent interaction (left panel) or the spin-dependent interaction (right-panel) for a dark matter mas of 5 GeV (top panel) 100 GeV (middle panel) and 10000 GeV (bottom panel).}
	\label{fig:BFPDF}
\end{figure}

The resulting distributions are shown in figure~\ref{fig:BFPDF} for capture through the spin-independent (left panels) or the spin-dependent (right panels) interactions, for DM masses $m_{\rm DM}=5$ GeV (top panels), 100 GeV (middle panels) and 10 TeV (bottom panels). We show the probability distributions for $\Delta N/N_\mathrm{SHM}$ multiplied by a factor $\tau/t_0$ in order to emphasize the impact of equilibration. If DM particles annihilate efficiently, then the equilibrium number of captured particles $N_\mathrm{SHM}$ is suppressed. This means that if we decrease the equilibration time $\tau$, the number of particles captured from the sub-halo $\Delta N$ is increased relative to the equilibrium number. However, a short equilibration time also means that enhancements in the number of captured particles are quickly annihilated away. The factor $t_0/\tau$ counts the number of equilibration times over the lifetime of the Sun, meaning that the factor $\tau/t_0$ roughly corrects for the effects of equilibration after the sub-halo passage.
 In order to assess the impact of varying the halo mass function, we additionally show the results for the power-law mass functions $\mathrm{d}N/\mathrm{d}M \propto M^{-\alpha}$ with fixed exponents $\alpha=1.8$ and $\alpha=2.0$.
For the shallowest sub-halo mass function with $\alpha=1.8$, we observe the largest $\Delta N$ as we have more heavy sub-halos than for the other two. Conversely, we find the smallest enhancement for $\alpha=2.0$. Finally, we find that the distributions are flatter and reach farther towards large $\Delta N$ for heavy dark matter. The reason for this is that the maximal velocity at which a DM particle can be captured by the Sun is small in this case.
Most of the sub-halos therefore lead to no additional DM particles being captured in the Sun. However, if the stream is slow enough, $\Delta N/N_\text{SHM}$ can be large as the capture from the smooth Milky Way DM distribution is small. Such sub-halo crossings contribute to the flat tail of the distribution.

In the case of direct detection, we demonstrated that the largest contribution to a local overdensity comes from low mass sub-halos (see Figure~\ref{fig:SingleDD}). In the case of solar capture, it is more challenging to identify which sub-halos are the most relevant. This is because the size of a sub-halos contribution depends not only on its mass and concentration, but also on its impact parameter $L$ and relative velocity with respect to the Earth $v_\mathrm{rel}$. As a result, sub-halos with the same mass and concentration parameter can give rise to very different increments in the annihilation rate, $\mathcal{I}_\Gamma$.

So far we have studied the effect on the annihilation rate of the crossing through a single sub-halo, concluding that the effects of the sub-halo passage remain significant for a long period of time $\sim \tau$. It is then plausible that the Sun could have traversed more than one sub-halo during this time, resulting in an enhanced accumulation of dark matter particles inside the Sun, and in a larger probability of an increment in the annihilation rate.

\begin{figure}[bth]
	\begin{center}
		\hspace{-0.75cm}
		\includegraphics[width=0.49\textwidth]{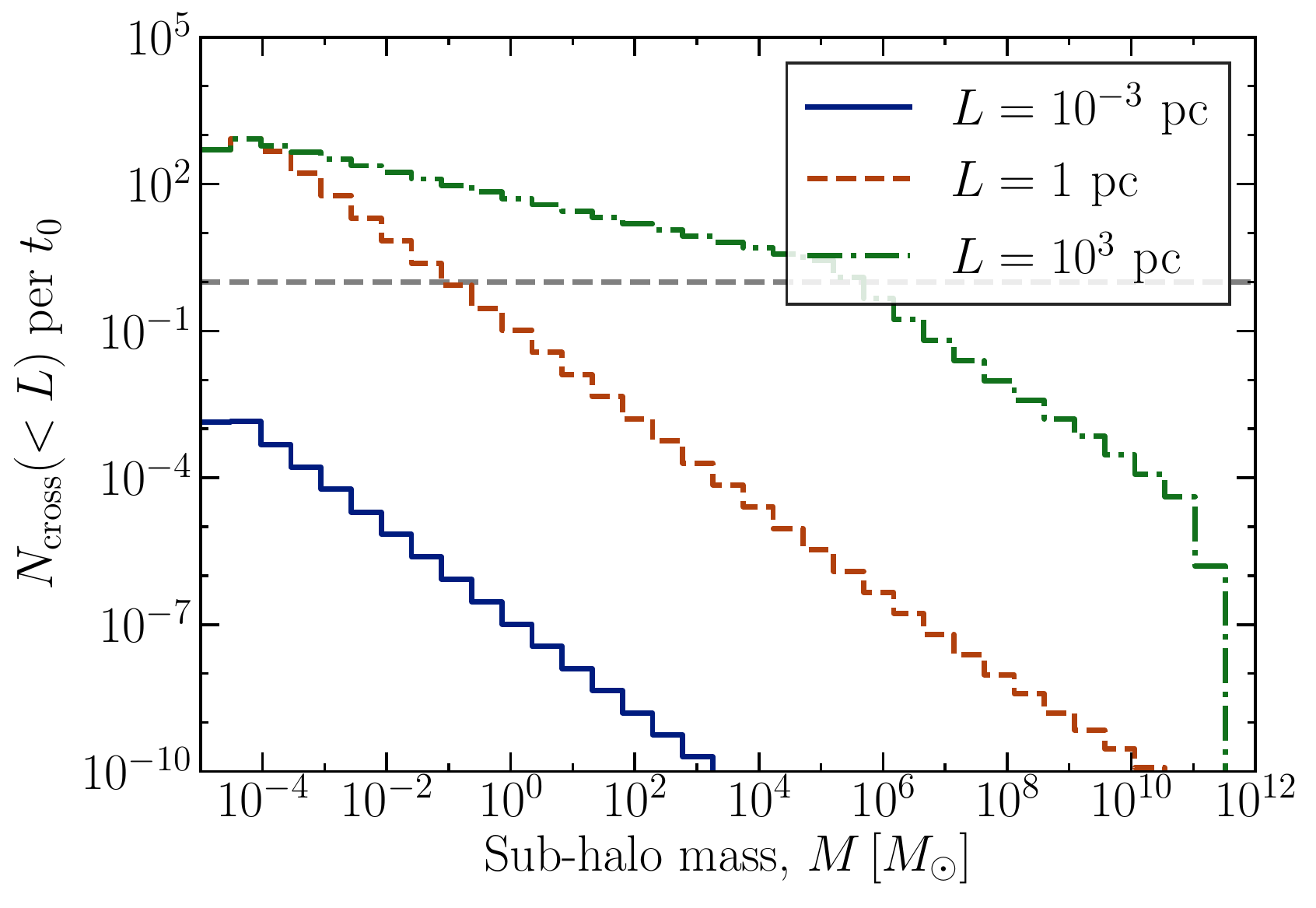}
	\end{center}
	\caption{\small Number of crossings of halos with mass $M$ (per decade in sub-halo mass) with close passage distance \textit{less than} $L$. The horizontal dashed line corresponds to a single crossing during the age of the Sun $t_0$. }
	\label{fig:ShEnc}
\end{figure}

We consider that at the time $t=0$ the number of captured dark matter particles in the Sun is $N(0)=0$, and that there have been ${\cal N}$ sub-halo passages at the times $t_i$, $i=1...{\cal N}$ (ordered such that $0<t_{\cal N}<...<t_1<t_0$)  with duration $\Delta t_i/\tau \ll 1$, each producing an `instantaneous' change in the number of captured DM particles $\Delta N_i$, as sketched in figure~\ref{fig:sketch_Nt}.  The total number of sub-halo encounters $\mathcal{N}$ in a total time $t_0$ can be calculated as:
\begin{align}
\mathcal{N} = \frac{t_0}{t_\mathrm{orb}}N_\mathrm{sh}\int_0^\infty \mathrm{d}\Delta N \,P_\mathrm{single}(\Delta N)\,,
\end{align}
where the integral corresponds to the probability of a crossing a single sub-halo during one Solar orbit around the Galaxy. With an orbital period of $t_\mathrm{orb} = 250\,\mathrm{Myr}$, the factor of $t_0/t_\mathrm{orb} \approx 18$ simply counts the number of Galactic orbits during the Sun's lifetime. We find the expected number of sub-halo crossings to be $\mathcal{N} \sim 100$, which we assume are uniformly distributed in time. Figure~\ref{fig:ShEnc} shows the number of sub-halo crossings as a function of sub-halo mass and impact parameter $L$. Crossing events are dominated by small sub-halos due to their larger abundance, while close passages are very rare. Indeed, crossings at sub-parsec distances from the sub-halo center are expected to occur less than once during the age of the Sun, justifying our assumption that $L \gg r_s$ for the sub-halos of interest (see Eq.~\eqref{eq:av_rho3}).

The number of captured dark matter particles today is then:
\begin{align}
N(t_0) \simeq N_{\rm SHM}
\displaystyle{
	\Big[\frac{(N(t_1)+\Delta N_1)/N_\text{SHM}+\tanh\frac{t_0-t_1}{\tau}}{1+(N(t_1)+\Delta N_1)/N_\text{SHM}\cdot\tanh\frac{t_0-t_1}{\tau}}\Big]}\,,
\end{align}
with 
\begin{align}
N(t_i) \simeq N_{\rm SHM}
\displaystyle{
	\Big[\frac{(N(t_{i+1})+\Delta N_{i+1})/N_\text{SHM}+\tanh\frac{t_0-t_{i+1}}{\tau}}{1+(N(t_{i+1})+\Delta N_{i+1})/N_\text{SHM}\cdot\tanh\frac{t_0-t_{i+1}}{\tau}}\Big]}~~~~{\rm for~} i=1,...,{\cal N}\,.
\end{align}
Finally, the increment in the annihilation rate reads
\begin{align}
{\cal I}_\Gamma&=\frac{\Gamma(t_0)}{\Gamma_{\rm SHM}(t_0)}-1\simeq
\left(\frac{N(t_0)}{N_{\rm SHM}\,\tanh\left(\frac{t_0}{\tau}\right)}\right)^2-1
\end{align}

\begin{figure}[!p]
\begin{center}
	\hspace{-0.75cm}
	\includegraphics[width=0.49\textwidth]{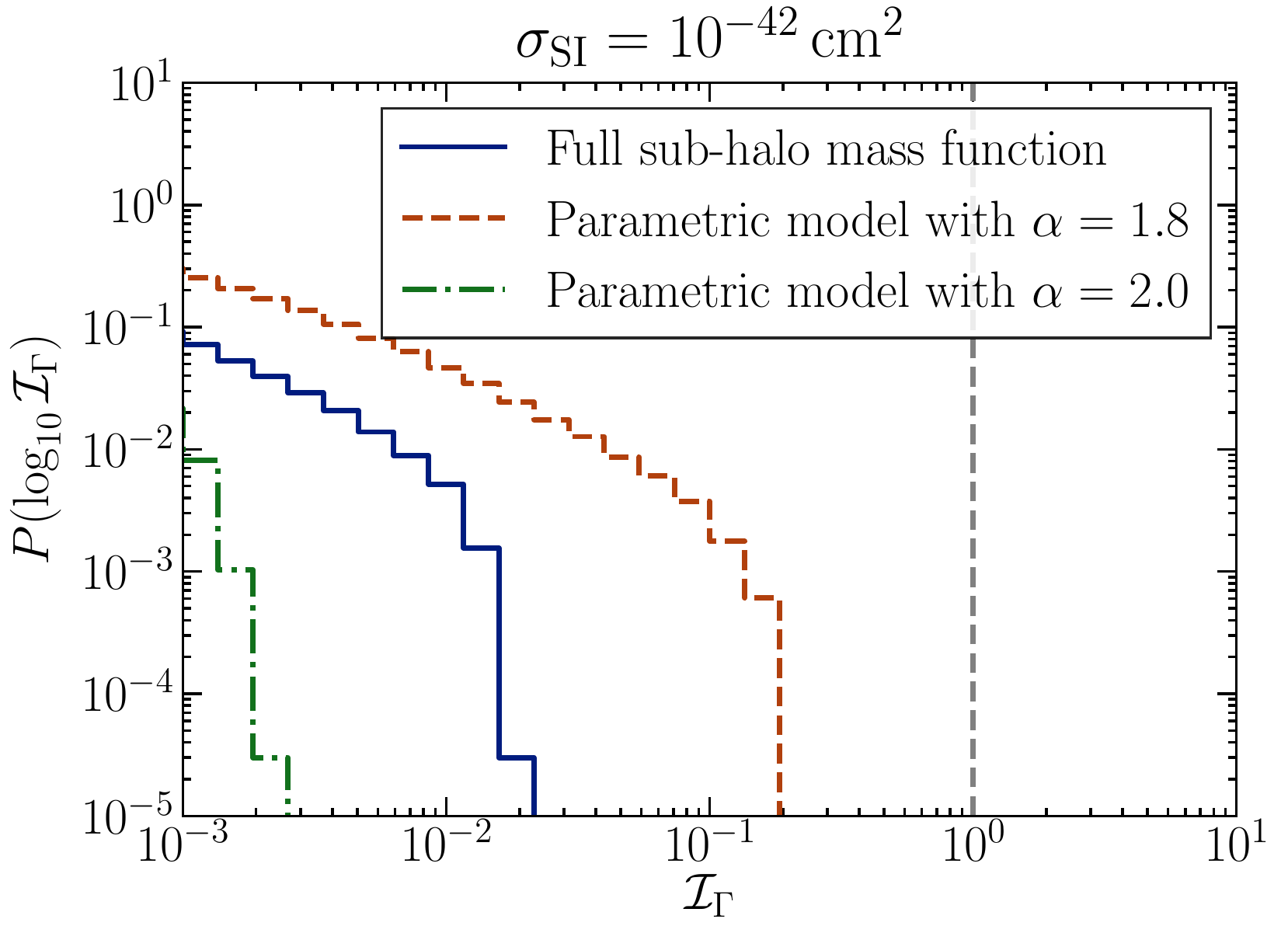}
	\includegraphics[width=0.49\textwidth]{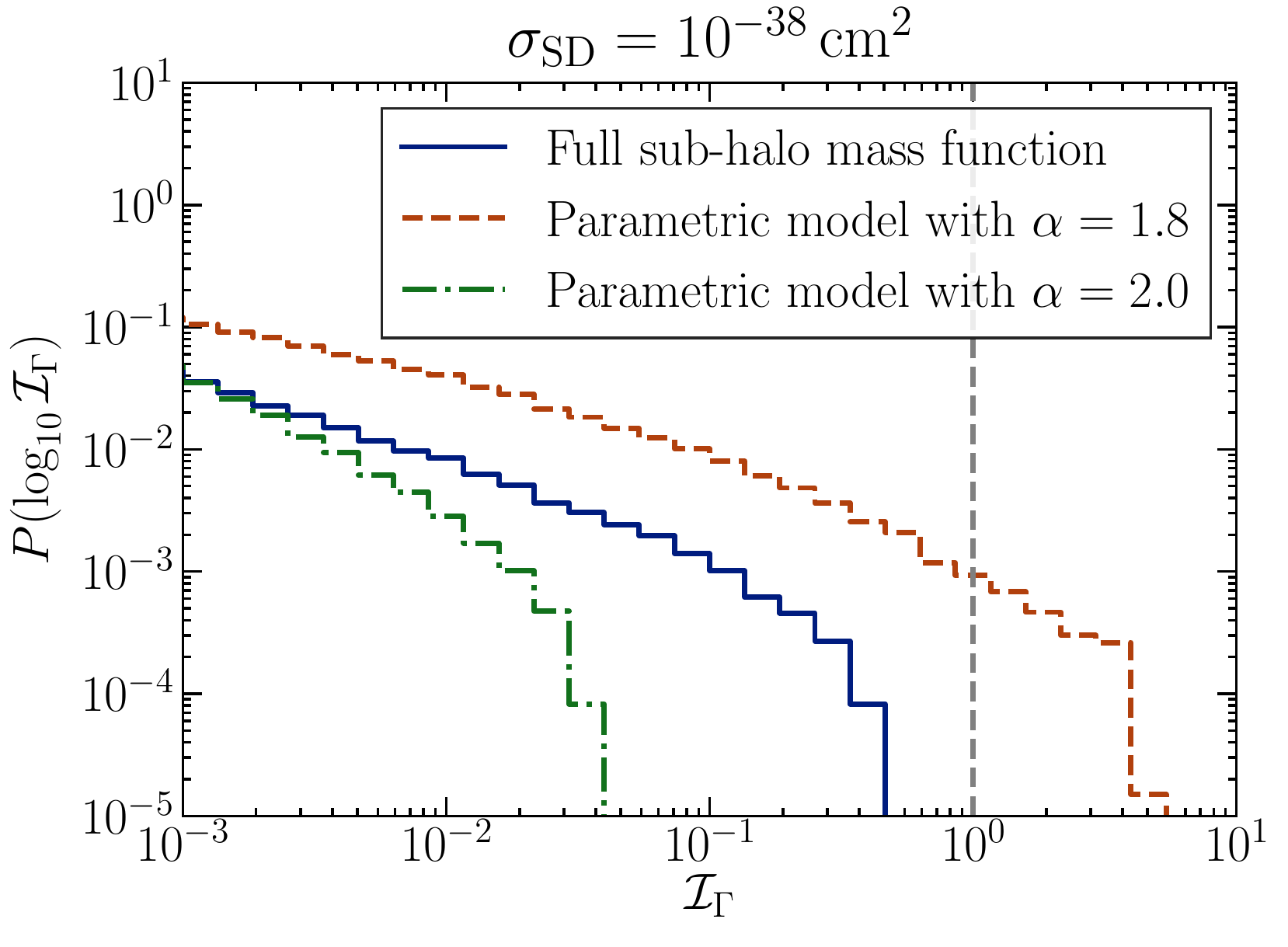}\\
	\hspace{-0.75cm}
	\includegraphics[width=0.49\textwidth]{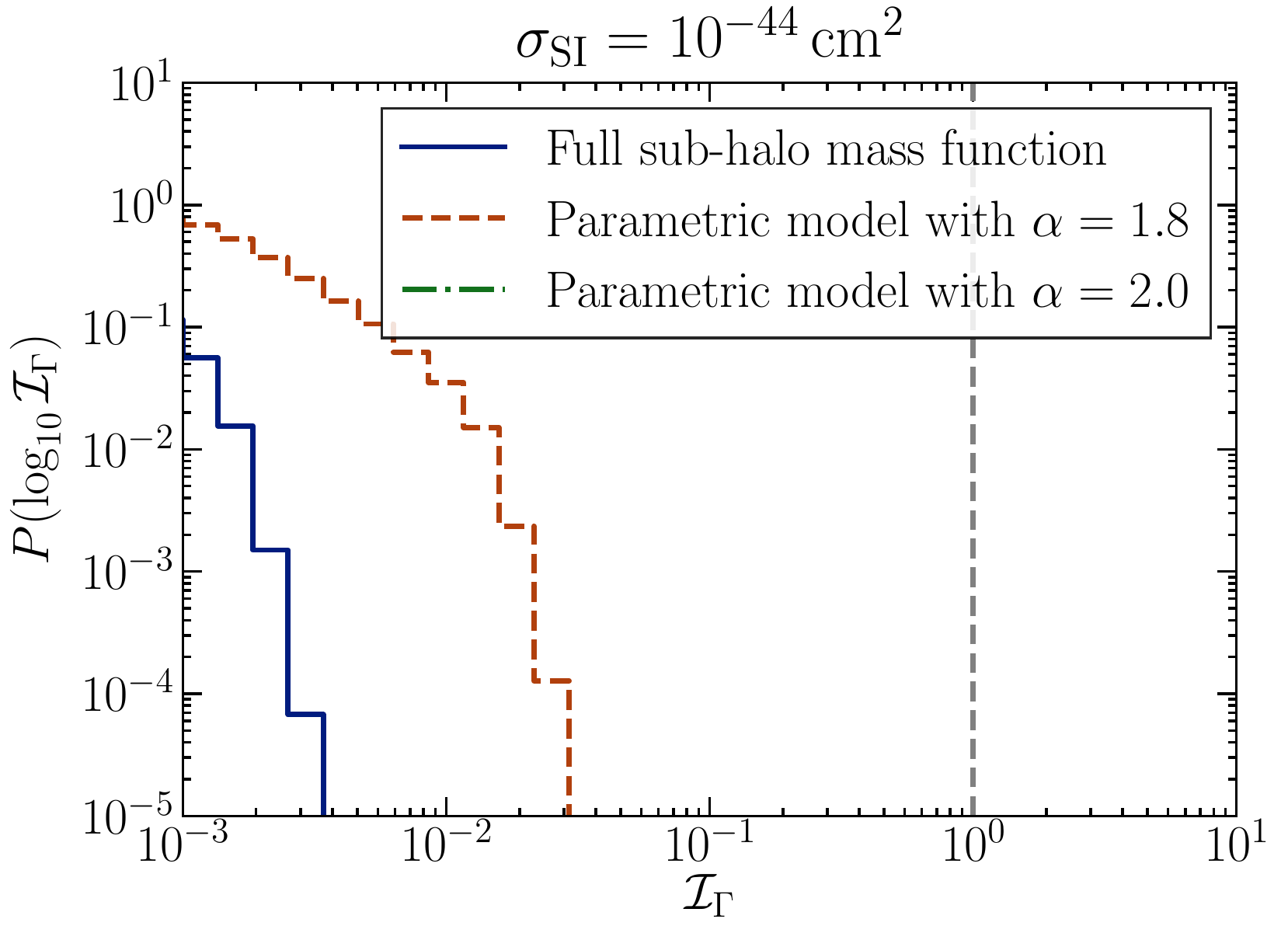}
	\includegraphics[width=0.49\textwidth]{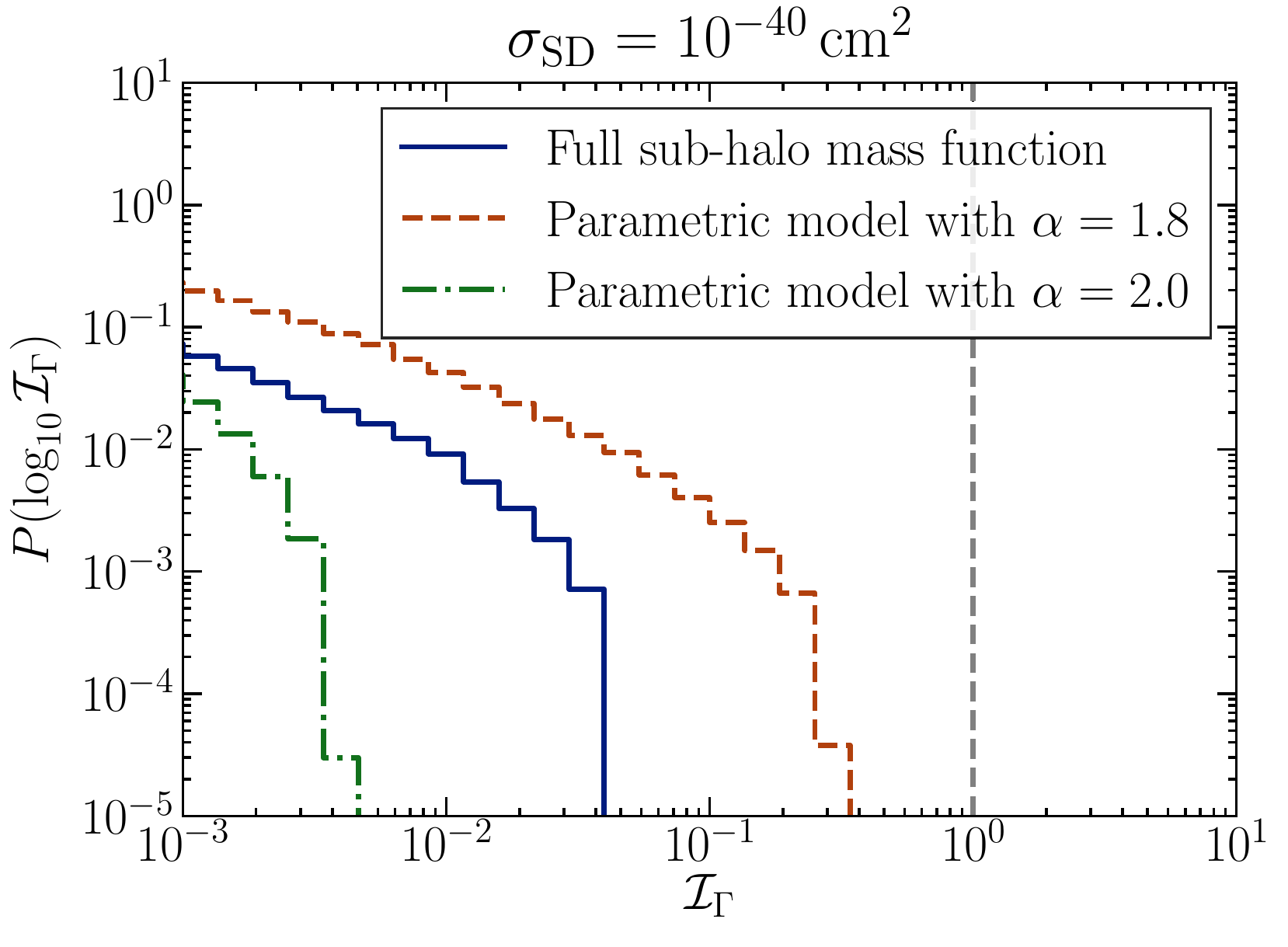}\\
	\hspace{-0.75cm}
	\includegraphics[width=0.49\textwidth]{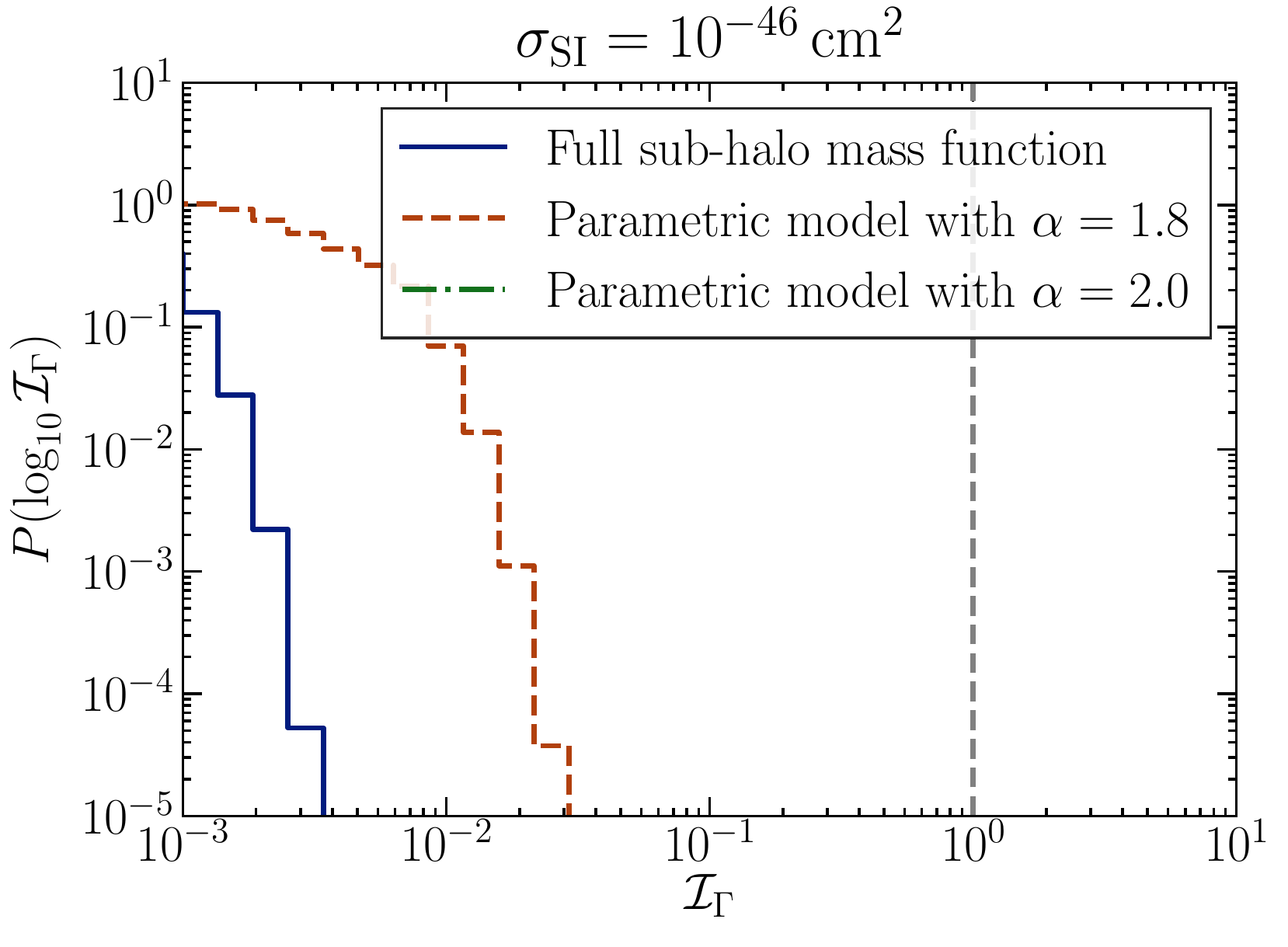}
	\includegraphics[width=0.49\textwidth]{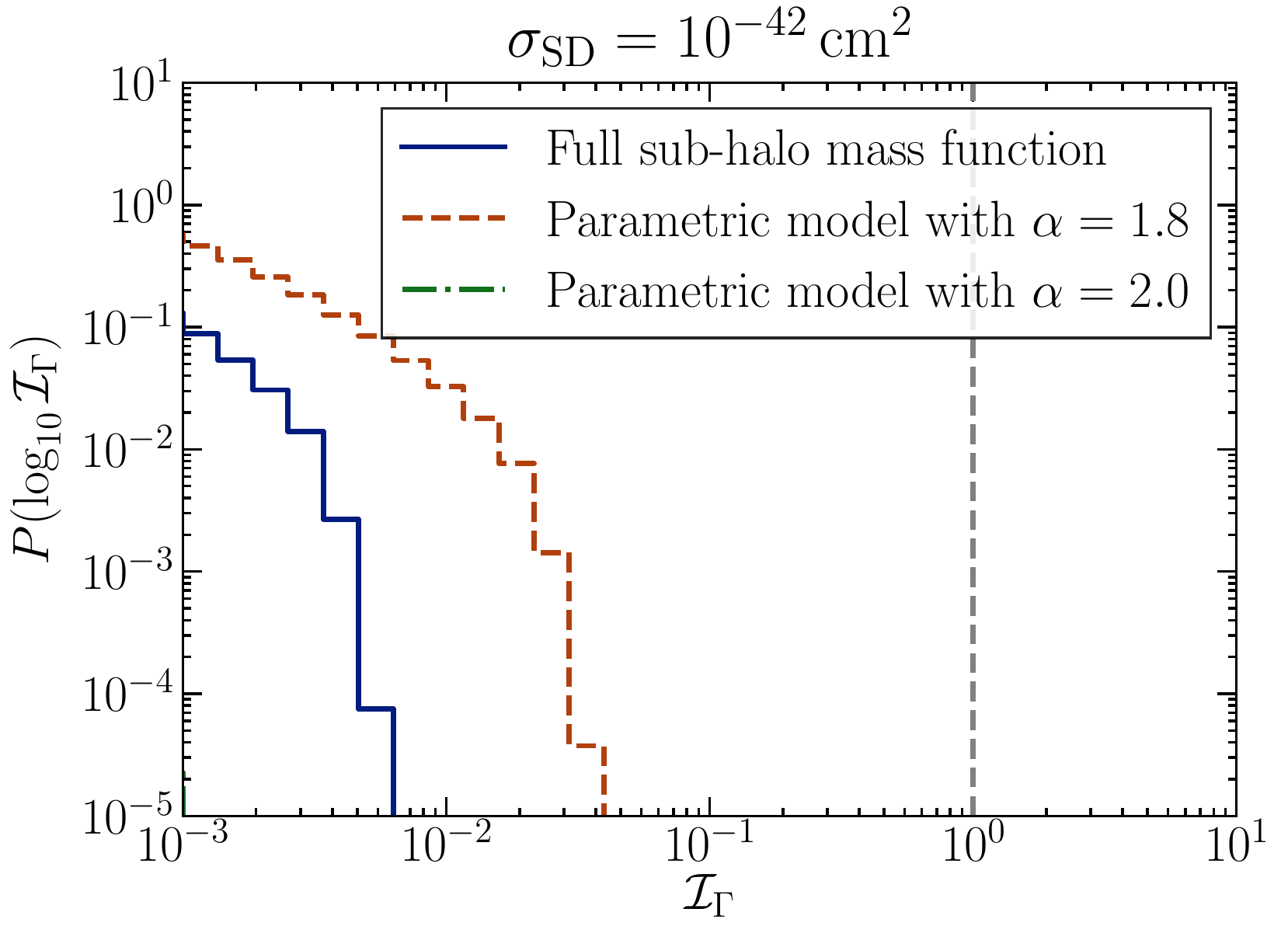}
\end{center}
\caption{\small Probability distributions of the increment in the annihilation rate inside the Sun for a dark matter mass of 5 GeV  for the spin-independent cross sections $\sigma_\text{SI}=10^{-42}\,\text{cm}^2, 10^{-44}\,\text{cm}^2$ and $10^{-46}\,\text{cm}^2$ (left panel, from top to bottom), and for the spin-dependent cross sections $\sigma_\text{SD}=10^{-38}\,\text{cm}^2, 10^{-40}\,\text{cm}^2$ and $10^{-42}\,\text{cm}^2$ (right panel, from top to bottom).
}
\label{fig:PDF_5GeV}
\end{figure}

\begin{figure}[!p]
	\begin{center}
		\hspace{-0.75cm}
		\includegraphics[width=0.49\textwidth]{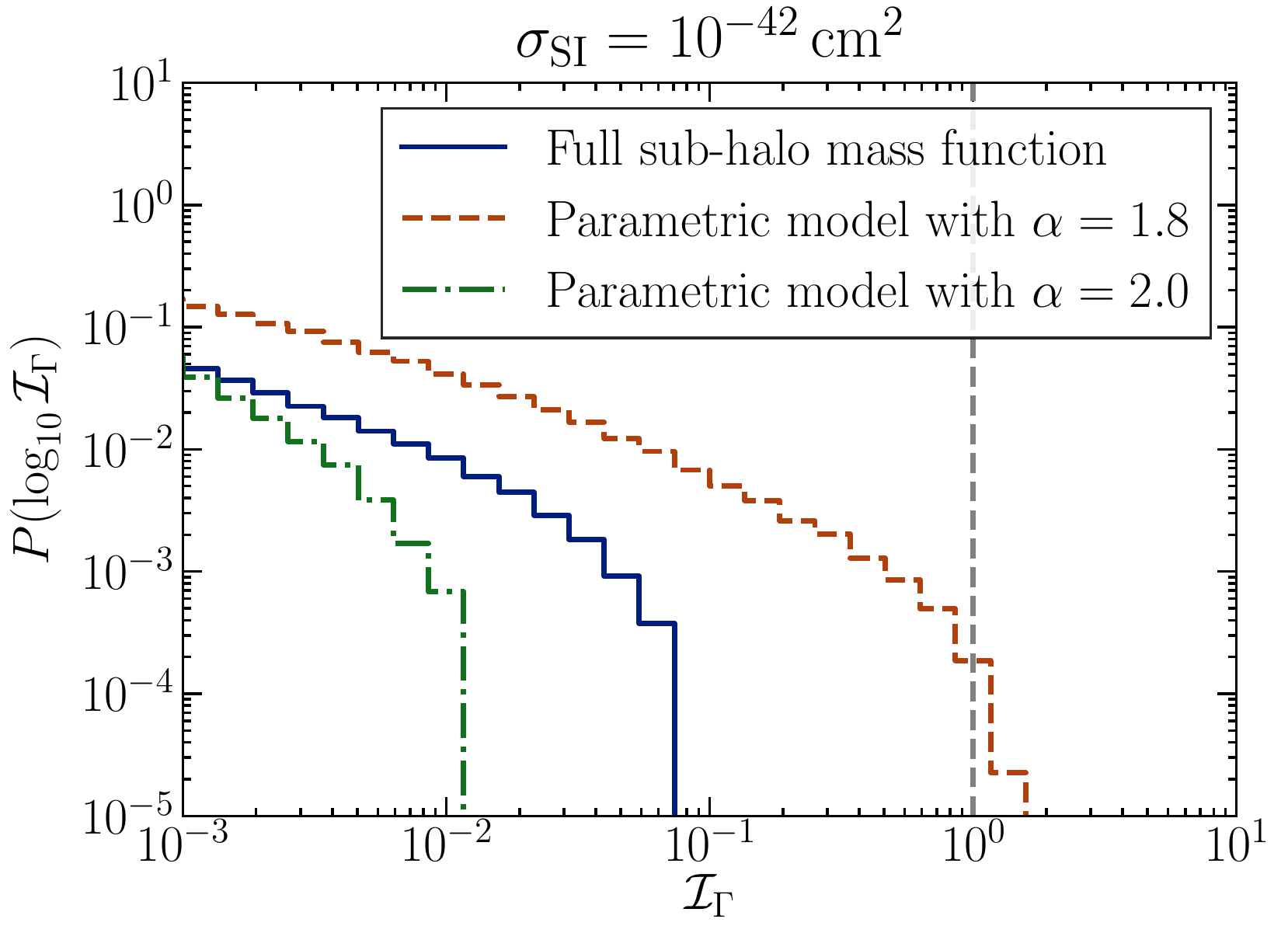}
		\includegraphics[width=0.49\textwidth]{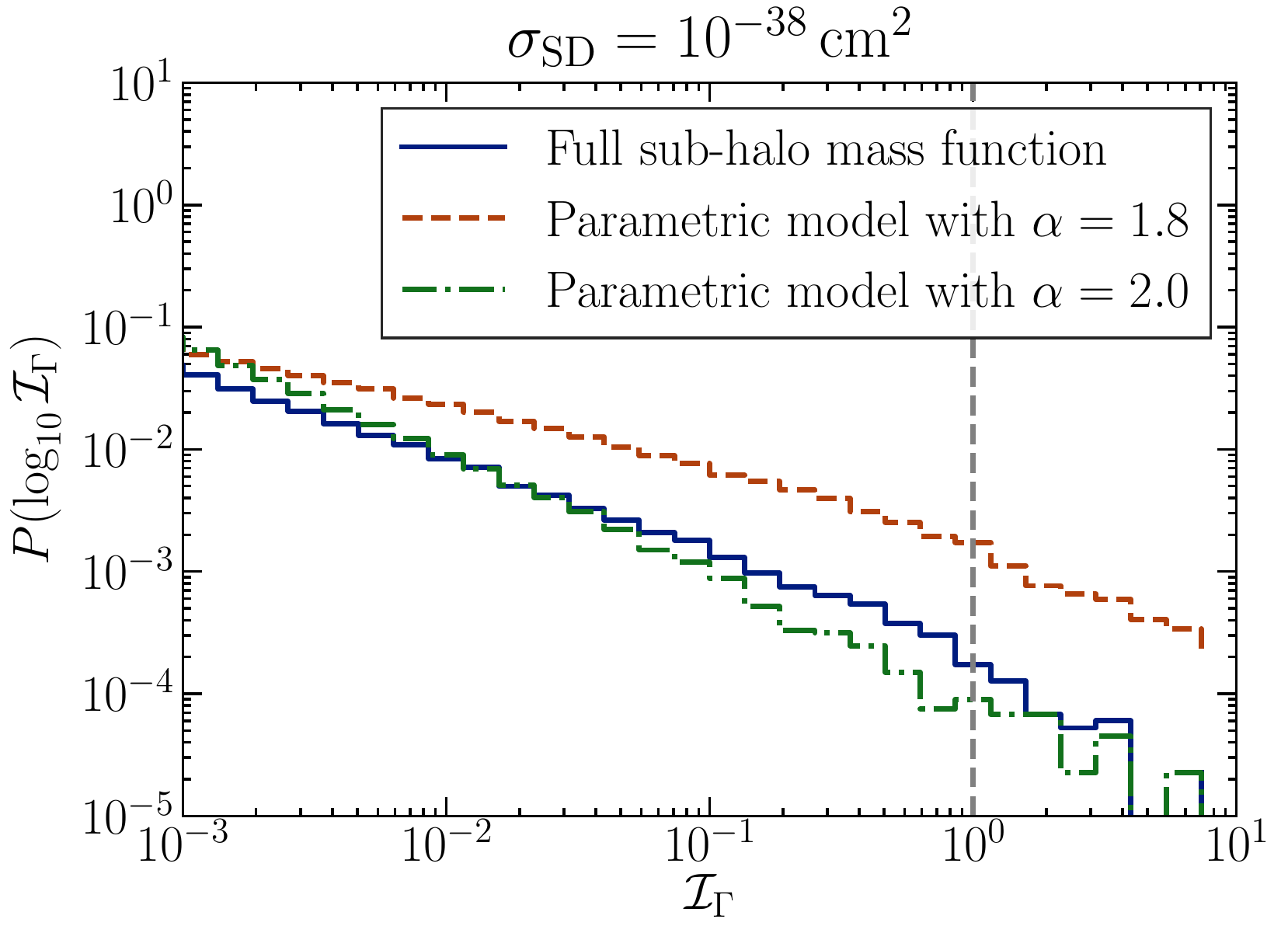}\\
		\hspace{-0.75cm}
		\includegraphics[width=0.49\textwidth]{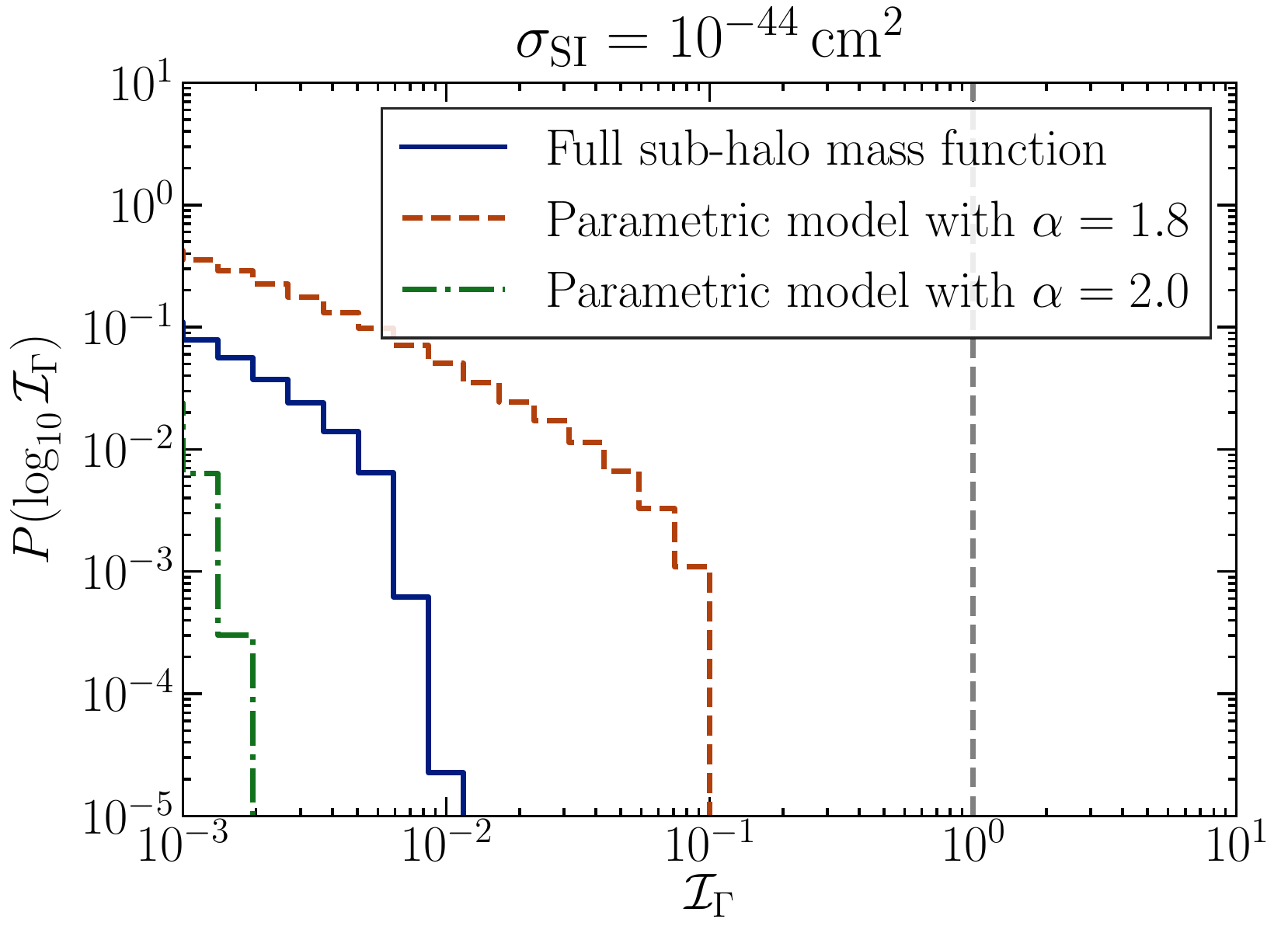}
		\includegraphics[width=0.49\textwidth]{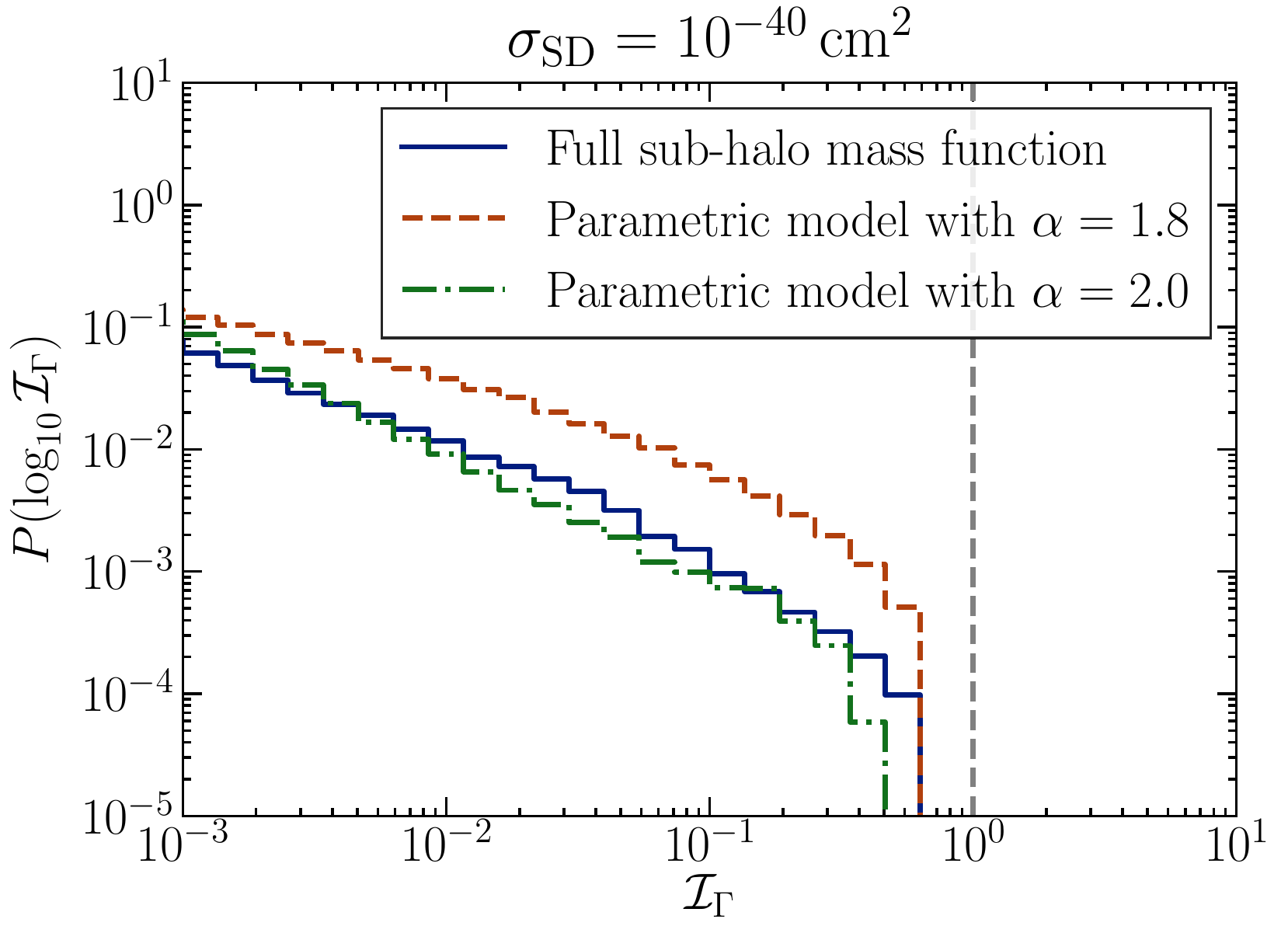}\\
		\hspace{-0.75cm}
		\includegraphics[width=0.49\textwidth]{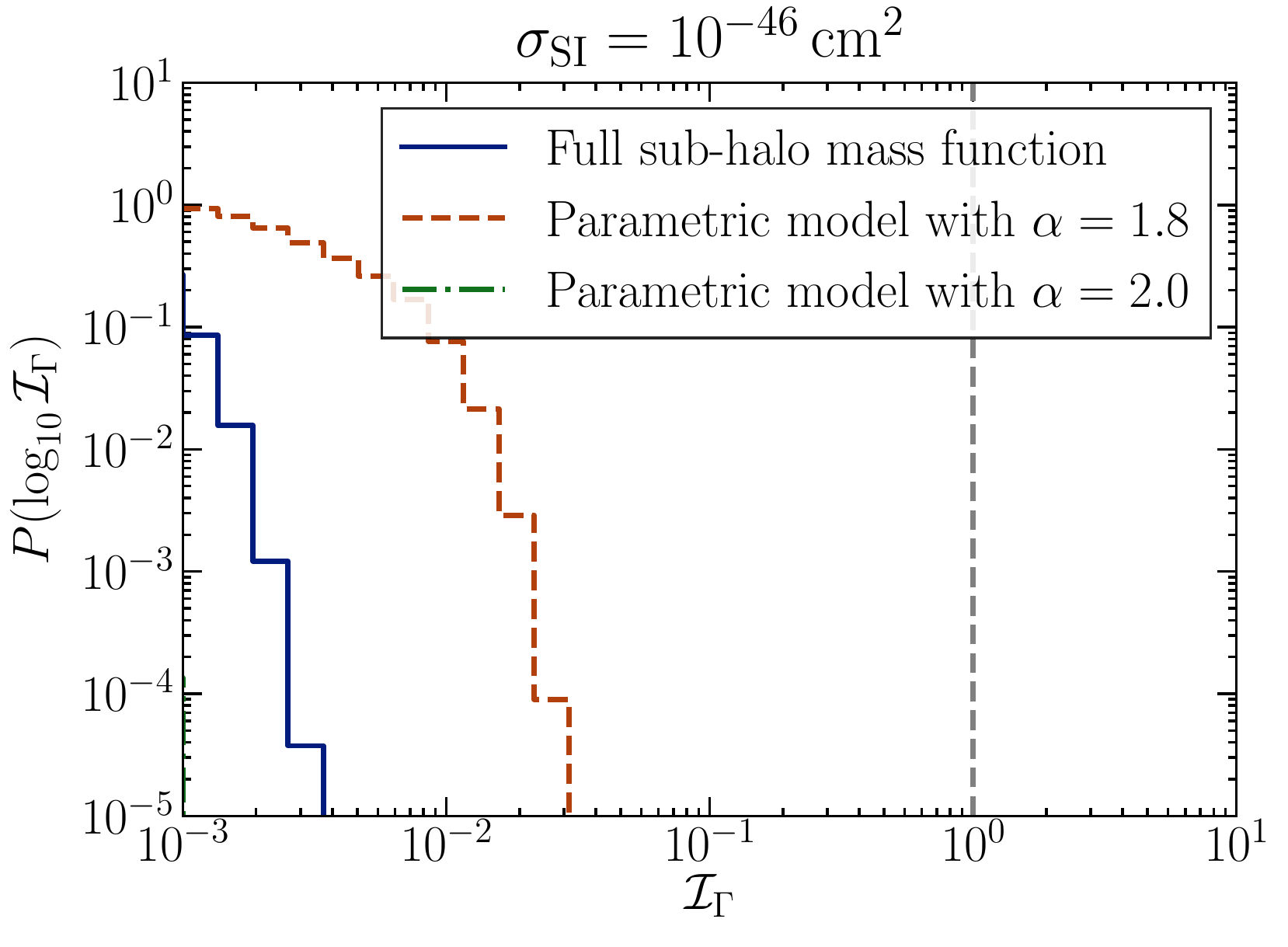}
		\includegraphics[width=0.49\textwidth]{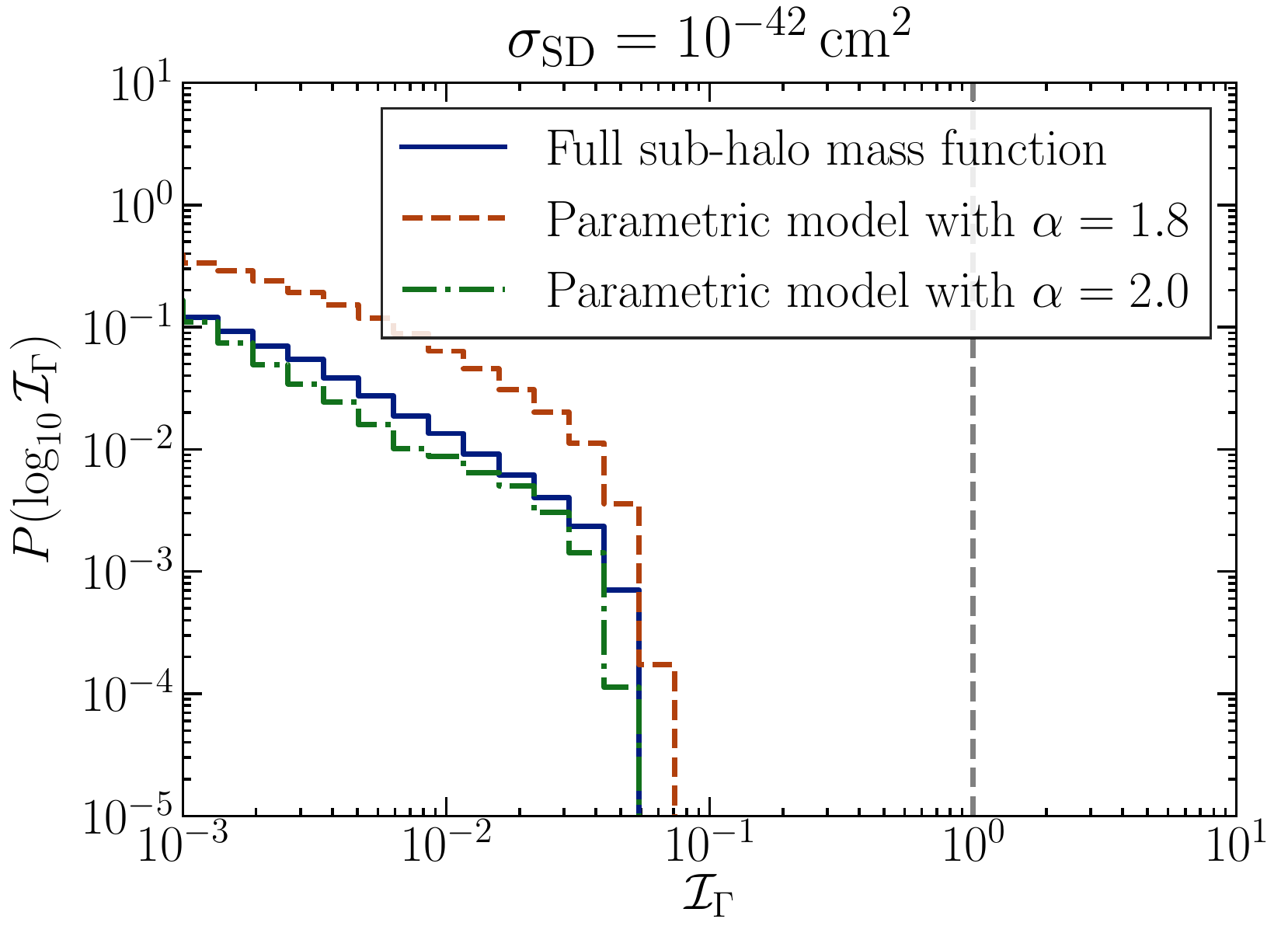}
	\end{center}
	\caption{\small Same as Fig.~\ref{fig:PDF_5GeV}, but for a dark matter mass of 100 GeV.}
	\label{fig:PDF_100GeV}
\end{figure}

\begin{figure}[!p]
	\begin{center}
		\hspace{-0.75cm}
		\includegraphics[width=0.49\textwidth]{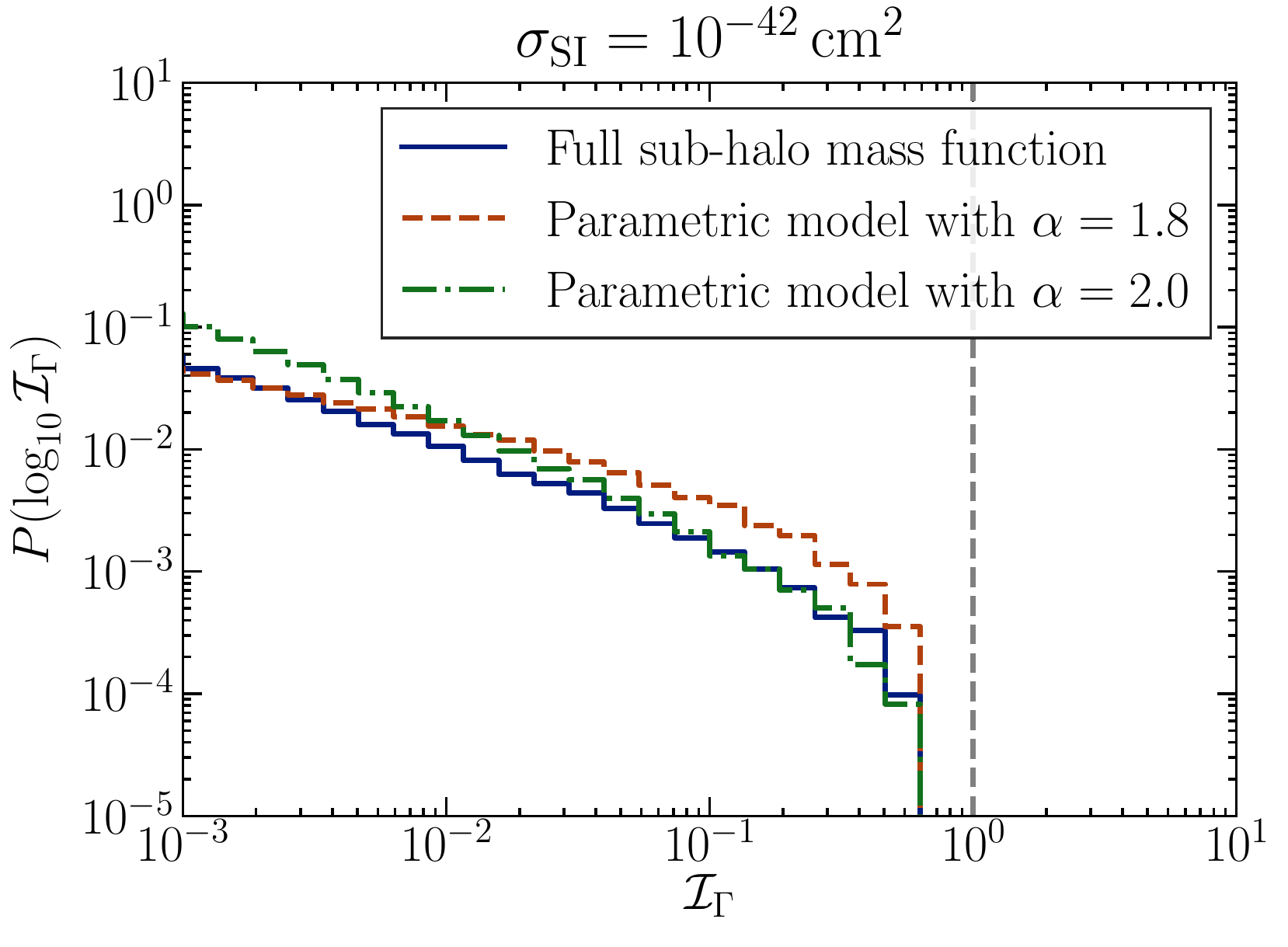}
		\includegraphics[width=0.49\textwidth]{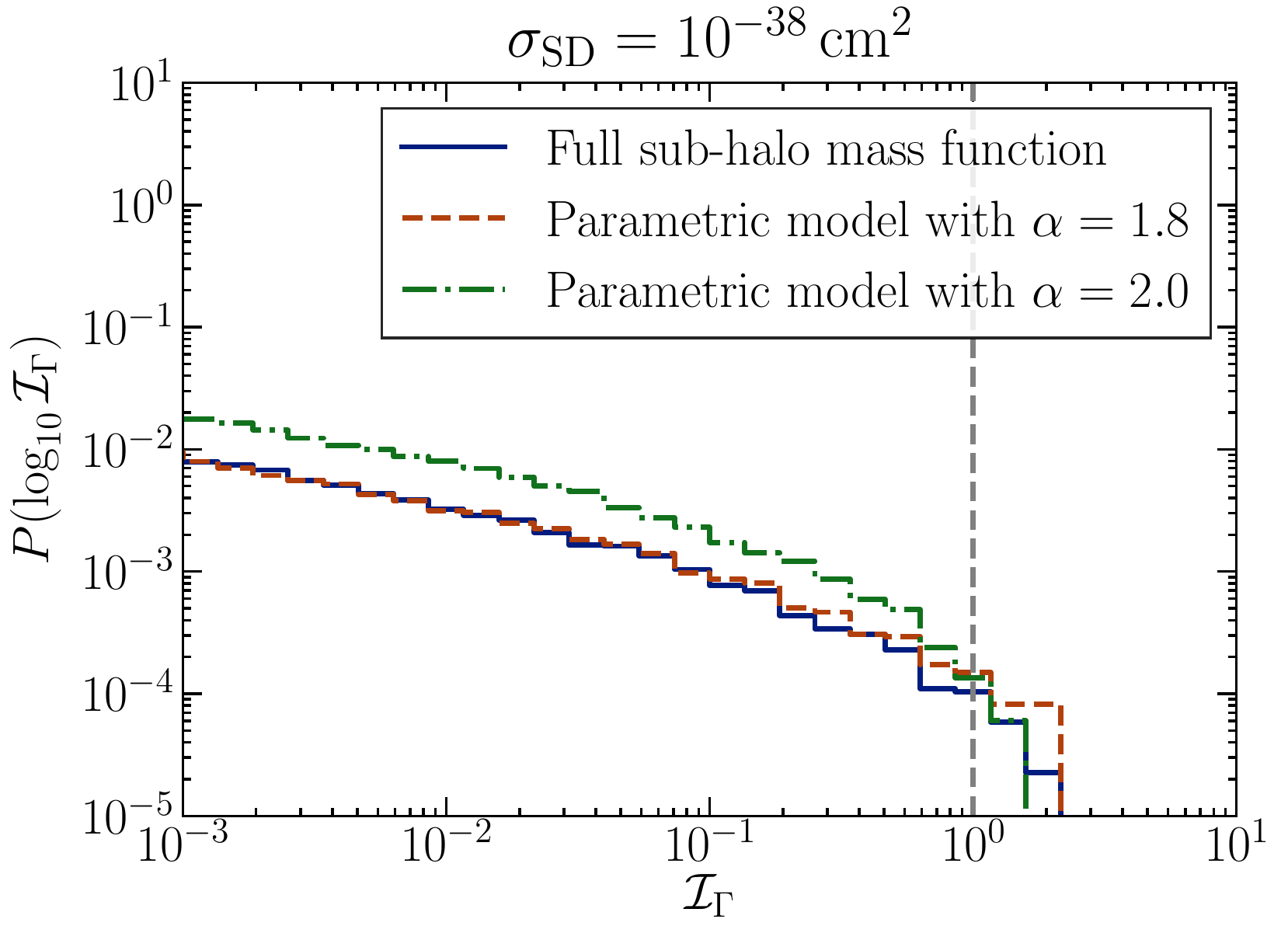}\\
		\hspace{-0.75cm}
		\includegraphics[width=0.49\textwidth]{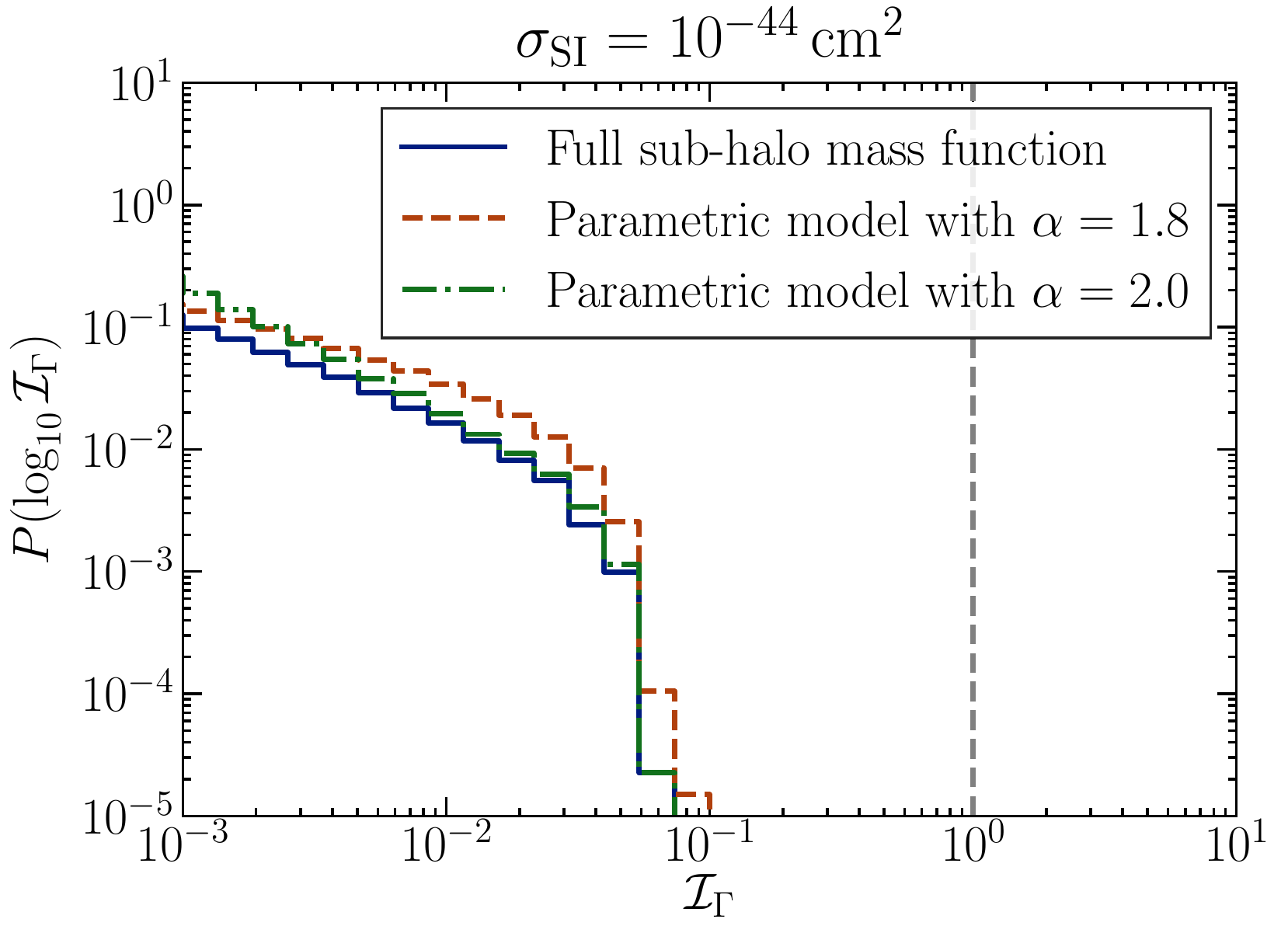}
		\includegraphics[width=0.49\textwidth]{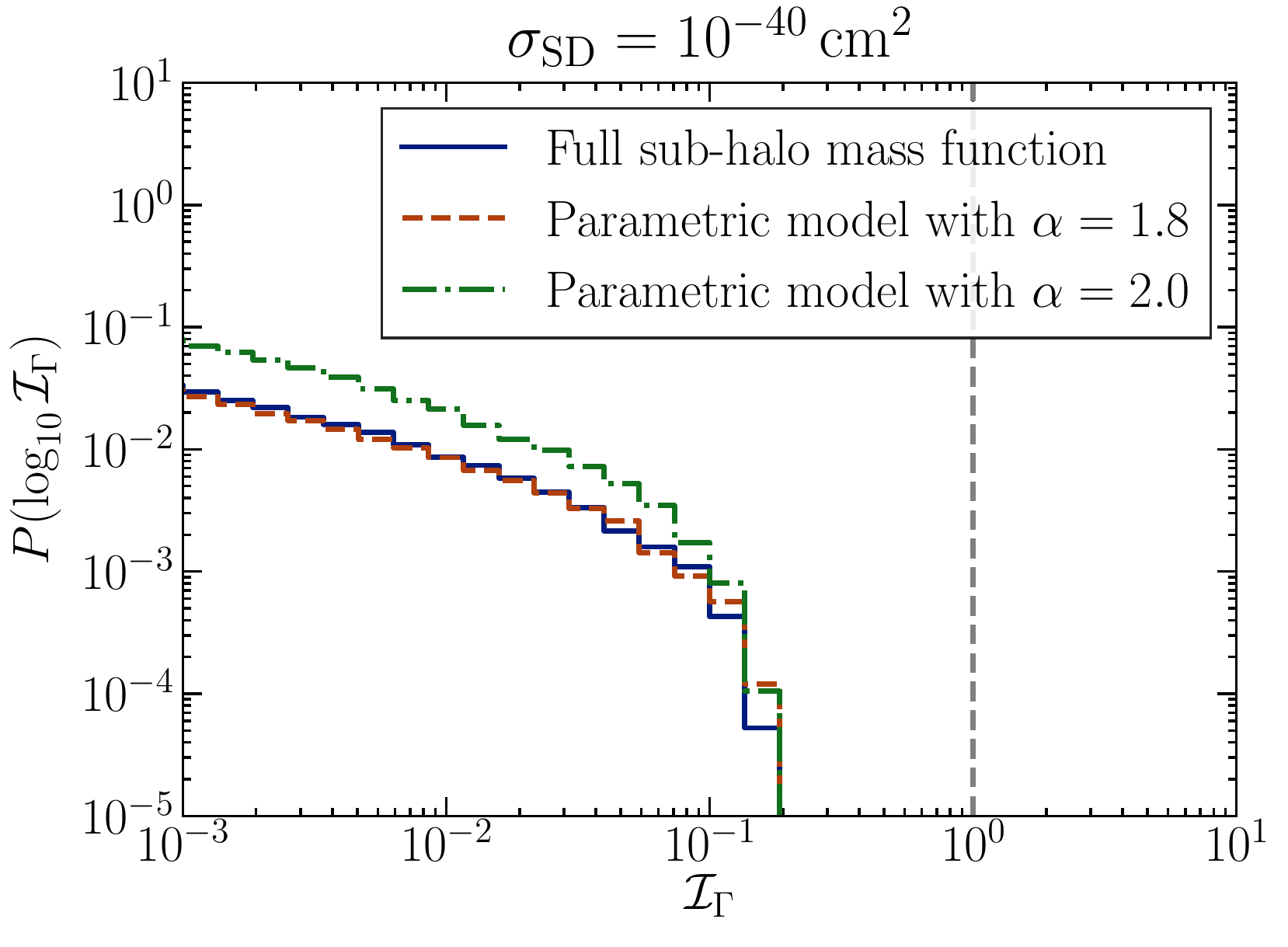}\\
		\hspace{-0.75cm}
		\includegraphics[width=0.49\textwidth]{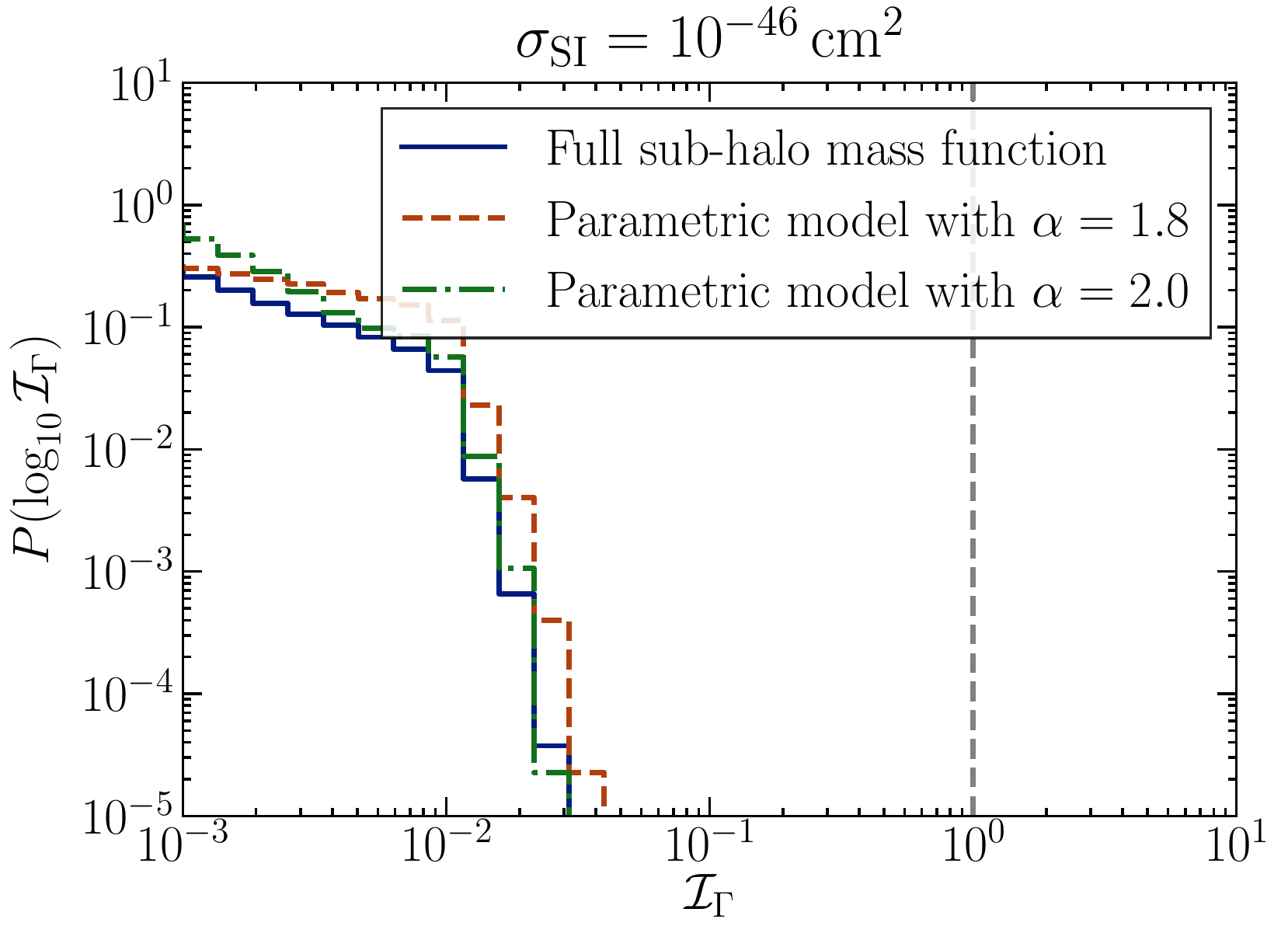}
		\includegraphics[width=0.49\textwidth]{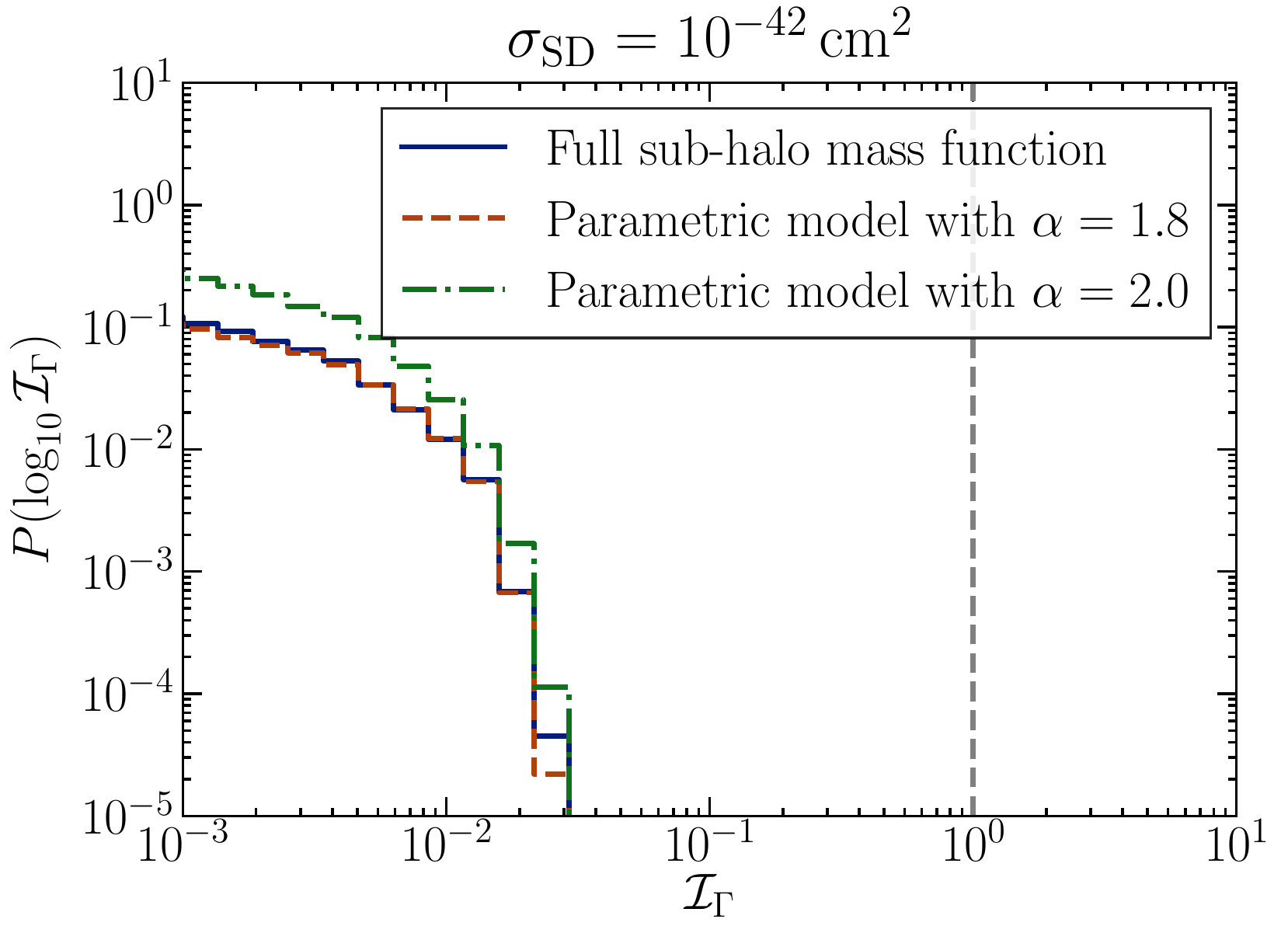}
	\end{center}
	\caption{\small Same as Fig.~\ref{fig:PDF_5GeV}, but for a dark matter mass of 10000 GeV.}
	\label{fig:PDF_10000GeV}
\end{figure}

We show the results of this calculation in figures~\ref{fig:PDF_5GeV}, \ref{fig:PDF_100GeV} and \ref{fig:PDF_10000GeV} for different DM masses.
For each interaction type, we calculate the annihilation boost for three different interaction strengths, ranging down to the interaction cross sections currently probed by direct detection experiments \cite{Schumann:2019eaa}.
Similar to figure \ref{fig:BFPDF}, we additionally include the results for power-law sub-halo mass functions $\mathrm{d}N/\mathrm{d}M \propto M^{-\alpha}$ with exponents $\alpha=1.8$ and $\alpha=2.0$.
Again, the probability to find large increments is usually greatest for $\alpha=1.8$ as this predicts the highest number of heavy sub-halos and smallest for $\alpha=2.0$.
Furthermore, larger values of $\mathcal{I}_\Gamma$ are more likely to be achieved for large cross sections for the discrete examples we studied.

As apparent from Fig. \ref{fig:BFPDF}, we expect larger $\Delta N/N_\text{SHM}$ as the factor $(\tau/t_0)$ decreases with increasing cross section. Therefore, the ratio of captured particles $\Delta N/N_\text{SHM}$ increases while the equilibration time decreases, which implies that the importance of recent encounters rises while the history of sub-halo passages becomes less relevant.
This enhances the probability to find large increments as most of them are due to recent encounters. Ultimately, the distribution of $\mathcal{I}_\Gamma$ will become similar to the results of section \ref{sec:DD} as soon as the cross section gets large enough such that the only relevant sub-halo encounters are those that take place in the very recent past.

When decreasing the cross section even further than in this work, large boost factors are possible again.
In this regime, capture and annihilation processes are far away from equilibrium and the number of particles in the Sun can be dominated by capture from sub-halos.
However, we do not show this here as those cross sections cannot be probed even with future upgrades of neutrino telescopes.
We note that the formalism developed in this work is applicable for all cross sections.

\section{Conclusions}\label{sec:Conclusions}

In this paper, we have studied the impact of sub-halos on the interpretation of direct dark matter searches.
Based on the sub-halo mass functions presented in Ref.~\cite{Hiroshima:2018kfv}, we calculated the probability of the Earth currently being inside a sub-halo, as well as the enhancement of the local dark matter density due to this.
Furthermore, we investigated the sub-halo encounters of the Sun during its lifetime.
For this, we calculated the average number of sub-halo crossings as well as the probability distribution of the enhancement due to one individual encounter.

We find that the impact on direct detection experiments can be large and should be taken into account when analyzing the results of direct detection experiments.
The probability of a sub-halo enhancing the local dark matter density by a factor $\mathcal{O}(1)$ is $\sim10^{-3}$ (figure~\ref{fig:SingleDD}).
Such large enhancements are far less likely (and in some cases impossible) for the neutrino signal from the Sun, which is enhanced only at a sub-percent level (figures~\ref{fig:PDF_5GeV}, \ref{fig:PDF_100GeV} and \ref{fig:PDF_10000GeV}).
We see that the probability of finding a large boost factor of $\mathcal{O}(1)$ is $\lesssim 10^{-5}$, suggesting that single large encounters will average away over the Sun's lifetime.
Thus,  we conclude that perturbations of the dark matter density and velocity distribution due to sub-halos can have a substantial impact on direct detection experiments while the neutrino signal from the Sun is largely unaffected.  

Our results are broadly consistent with those of previous studies on the impact of sub-structure on direct detection. For example, Ref.~\cite{Stiff:2001dq} estimates a roughly 1\% probability of an $\mathcal{O}(1)$ enhancement in the local density from sub-structure. Though our results suggest a somewhat smaller probability, there is perhaps a factor of a few uncertainty coming from the precise sub-halo mass function, as we point out in figure~\ref{fig:DDboost}.  The authors of Ref.~\cite{Kamionkowski:2008vw} also built an analytic model for Milky Way sub-halos, inspired by N-body simulations. Here, we use more recent, state-of-the-art determinations of the properties of the sub-halo population, though we find similar results. In the case of solar capture, Ref.~\cite{Koushiappas:2009ee} demonstrated that sufficiently dense sub-structures, traversed for a sufficiently long time, may have a substantial impact on the neutrino signal from annihilating DM in the Sun. We have demonstrated that such encounters are rare and that a realistic population of sub-structures is unlikely to give a large enhancement.

This work alleviates a major uncertainty of dark matter searches with neutrino telescopes \cite{Choi:2013eda,Nunez-Castineyra:2019odi} and emphasizes the impact of astrophysical uncertainties on direct detection experiments. In particular, given that deviations from the Standard Halo Model are expected due to sub-halos, then it will be necessary to properly account for these uncertainties when producing constraints from direct detection experiments. Of course, given a future signal, it may be possible to directly reconstruct the local DM velocity distribution (as proposed in e.g.~\cite{Peter:2011eu,Kavanagh:2013wba,Kavanagh:2014rya,Feldstein:2014gza,Ferrer:2015bta,Ibarra:2018yxq}). With this, it should be possible to determine whether the local DM density does indeed receive an enhancement due to a sub-halo, as well as hinting at the properties of the sub-halo itself.

While we have made a number of assumptions about the properties and distributions of DM substructure, the formalism we have developed is applicable more generally. For example, it was recently claimed that the density profile of disrupted sub-halos is not well described by an NFW profile but instead by an exponential profile $\rho(r)\propto r^{-\gamma}\,\exp(-r/\text{R}_b)$ \cite{Hooper:2016cld, Kazantzidis:2003hb}.
The parameters $\gamma$ and $\text{R}_b$ are drawn from generalized normal distributions whose mean values are determined from ELVIS \cite{Garrison-Kimmel:2013eoa} and Via Lactea II \cite{Diemand:2008in} simulations for different bins of sub-halo masses.
As these two simulations are able to resolve sub-halos only down to $10^7\,M_\odot$, one has to rely on extrapolations to assess sub-halos as light as $10^{-5}\,M_\odot$.
In this work, we do not use this profile since we found that the largest contribution to the enhancement of the local dark matter density at Earth and to the enhancement of the neutrino signal from the Sun is due to sub-halos in the extrapolated mass range. 
However, we note that using the exponential density profile for our calculation would be straightforward and could be easily implemented as soon as the resolution is sufficient.

We have focused in this work on the sub-halo distribution for standard Cold Dark Matter. Our approach could equally well be applied to other DM candidates, such as Warm Dark Matter \cite{Lovell:2013ola} and Self-interacting Dark Matter \cite{Robles:2019mfq,Moline:2019zev}, as well as DM with more exotic clustering properties \cite{Nussinov:2018ofp,Grabowska:2018lnd}. We have also focused on direct searches for WIMP DM, though our framework may also be applied to axion searches, where axions are expected to form bound `mini-clusters' \cite{Tinyakov:2015cgg}, potentially giving an enhancement to the local density. This work indicates that the observation of DM substructures in direct detection experiments may still be promising, which would shed further light on the nature of the elusive dark matter particle. 

\section*{Acknowledgments}
This work has been supported by the DFG cluster of excellence ORIGINS and by the Collaborative Research Center SFB1258. BJK gratefully acknowledges partial support from the Netherlands Organisation for Scientific Research (NWO) through the VIDI research program ``Probing the Genesis of Dark Matter'' (680-47-532). We thank the Munich Institute for Astro- and Particle Physics (MIAPP) where part of this work was developed. We also thank Ranjan Laha for discussions on the Solar capture of substructure, as well as for pointing us towards some of the relevant literature; Shin'ichiro Ando for sharing the numerical results of Ref.~\cite{Hiroshima:2018kfv}; and Adam Coogan for helpful discussions on the contribution of sub-halos to the local DM density.

\bibliographystyle{JHEP-mod}
\bibliography{paper}
\end{document}